\documentclass[hyper]{JHEP3}
\pdfoutput=1
\usepackage{graphicx}
\newcommand{\Scal}[1]{\Bigl ({#1} \Bigr )}
\newcommand{\scal}[1]{\bigl ({#1} \bigr )}
\def\ie{{\it i.e.}\ }
\def\eg{{\it e.g.}\ }

\def\N{\mathcal{N}}
\DeclareMathAlphabet{\mathpzc}{OT1}{pzc}{m}{it}
\usepackage{dsfont}
\usepackage{verbatim}

\newcommand{\ord}[1]{{\scriptscriptstyle (#1)}}
\newcommand{\Tgp}{{\mathrm T}^+}
\newcommand{\dTgp}{{\mathrm dT}^+}
\newcommand{\Tgm}{{\mathrm T}^-}
\newcommand{\Tgpm}{{\mathrm T}^\pm}
\newcommand{\cTgp}{\overset{\circ}{\mathrm T}{}^+}
\newcommand{\cTgm}{\overset{\circ}{\mathrm T}{}^-}
\newcommand{\cTgpm}{\overset{\circ}{\mathrm T}{}^\pm}

\newcommand{\CR}{\nonumber \\*}
 \usepackage{amssymb}
 \usepackage{amsmath}
 \def\nn{\nonumber}

\def\cF{{\mathcal F}}
\def\cH{{\mathcal H}}
\def\cV{{\mathcal V}}
\def\cK{{\mathcal{K}}}
\def\csK{{\scriptscriptstyle \cK}}
\newcommand{\Iprod}[2]{\langle {#1}, {#2} \rangle}
\newcommand{\J}{\mathsf{J}}
\newcommand{\gz}{\Gamma_{ \mbox{\tiny{0}} } }
\newcommand{\cHz}{ \cH_{ \mbox{\tiny{0}} } }
\newcommand{\Rstz}{ R^*_{ \mbox{\tiny{0}} } }

\newcommand{\gzA}{\Gamma_{\mbox{\tiny{0}} \pA}}
\newcommand{\gzB}{\Gamma_{\mbox{\tiny{0}} \pB}}
\newcommand{\gzC}{\Gamma_{\mbox{\tiny{0}} \pC}}
\newcommand{\kd}{\mathpzc{d}}
\newcommand{\fd}{\mathsf{d}}
\newcommand{\rc}[1]{r_{\text{\tiny #1}}}
\newcommand{\pA}{{\text{\tiny A}}}
\newcommand{\pB}{{\text{\tiny B}}}
\newcommand{\pC}{{\text{\tiny C}}}
\newcommand{\pAB}{{\text{\tiny AB}}}
\newcommand{\pAC}{{\text{\tiny AC}}}
\newcommand{\pBC}{{\text{\tiny BC}}}
\newcommand{\RAB}{{\mbox{\footnotesize{R}}_\pAB}}
\newcommand{\RBC}{{\mbox{\footnotesize{R}}_\pBC}}
\newcommand{\RAC}{{\mbox{\footnotesize{R}}_\pAC}}
\newcommand{\Rot}{{\mbox{\footnotesize{R}}_{\text{\tiny 12}}}}
\newcommand{\RABsq}{{\mbox{\footnotesize{R}}^2_\pAB}}
\newcommand{\RABin}{{\mbox{\footnotesize{R}}^i_\pAB}}
\def\va{{\bf a}}
\def\vb{{\bf b}}
\def\vc{{\bf c}}
\def\ve{{\bf e}}
\def\vk{{\bf k}}
\def\vq{{\bf q}}
\def\vp{{\bf p}}
\def\vl{{\bf l}}
\def\vL{{\bf L}}
\def\vK{{\bf K}}
\def\vt{{\bf t}}
\def\vu{{\bf u}}
\def\vpe{{\bf p_{e}}}

\def\tr{{\rm tr}\, }
\def\det{{\rm det}}

\def\zer{{ \mbox{\tiny{0}} }}
\def\un{{\mathpzc{1}}}
\def\deux{{\mathpzc{2}}}
\def\trois{{\mathpzc{3}}}
\def\Gammamun{{\Gamma^\ord{{\mbox{\tiny -}}1}}}
\def\Gammamtrois{{\Gamma^\ord{{\mbox{\tiny -}}3}}}
\def\kmun{{k^\ord{{\mbox{\tiny -}}1}}}

\title{non-BPS walls of marginal stability}

\author{Guillaume Bossard$^{\sf{a}}$  and Stefanos Katmadas$^{\sf{a}, \sf{b}}$
\\ {$\sf{a}$ Centre de Physique Th\'eorique, \'Ecole Polytechnique, CNRS, 91128
Palaiseau, France}
\\ {$\sf{b}$ Institut de Physique Th\'eorique, CEA Saclay, CNRS-URA 2306, \\ \hspace{.14cm}
  91191 Gif sur Yvette, France }
\\ \email{guillaume.bossard [at] cpht.polytechnique.fr},\\
   \email{stefanos.katmadas  [at] cpht.polytechnique.fr}}

\abstract{ We explore the properties of non-BPS multi-centre extremal black holes in
ungauged $\N=2$ supergravity coupled to vector multiplets, as described by solutions
to the composite non-BPS linear system. After setting up an explicit description that
allows for arbitrary non-BPS charges to be realised at each centre, we study the
structure of the resulting solutions.
Using these results, we prove that the binding energy of the composite is always positive and we show explicitly the existence of walls of marginal stability for
generic choices of charges. The two-centre solutions only exist on a hypersurface of
dimension $n_v+1$ in moduli space, with an $n_v$-dimensional boundary, where the distance
between the centres diverges and the binding energy vanishes. 
}
\preprint{ {CPHT-RR091.0913} }

\keywords{Black Holes in String Theory, Supergravity Models}

\begin{document}

%\tableofcontents
%\setcounter{footnote}{0} \vspace{1.5cm}
\section{Introduction and Overview}
\label{sec:intro}

The description of black holes in supergravity, viewed as a low energy effective
description of string compactifications, has been a useful tool for understanding
the structure and properties of nonperturbative features of the theory. In
particular, the possible bound states of D-branes manifest themselves as multi-centre
supergravity solutions at strong coupling \cite{Denef:2000nb}. In the BPS sector, the properties of the
supergravity solutions, such as the walls of marginal stability and attractor flow
trees \cite{Denef:2000nb,Denef:2007vg}, have been instrumental in uncovering this connection, leading to
remarkable results on the description of D-brane bound states in terms of quiver
quantum mechanics \cite{Denef:2002ru}.

The purpose of this paper is to set the stage for a similar study of a particular
subsector of the non-BPS spectrum. We restrict attention to zero temperature
under-rotating multi-centre black holes, \ie charged and rotating extremal black
holes, for which the extremality bound is saturated by the charges.\footnote{Black
holes for which the extremality bound is saturated by the angular momentum are
similarly called over-rotating.} This class includes BPS solutions, wherein all
charges involved allow for some supersymmetry to be preserved and no local rotation
at the horizons is allowed, even though there is a global angular momentum generically.
In this paper, we study the reverse situation, \ie solutions where only non-BPS charges (of strictly negative quartic invariant)
are allowed at the centres, as described by the composite non-BPS system
\cite{Bossard:2011kz, Bossard:2012ge, Bossard:2013oga}. The mixed situation, in which both BPS and
non-BPS charges are allowed is described by the more complicated almost-BPS system
\cite{Goldstein:2008fq, Bossard:2013oga}, but will not be discussed here.

Using the formalism developed in \cite{Bossard:2013oga}, we are able to solve the system
completely, in a general duality frame. As was
already noted in \cite{Bossard:2011kz, Bossard:2012ge}, the resulting composite
solutions only exist on certain hypersurfaces of the moduli space, unlike the BPS
solutions whose domain of existence is of codimension zero in moduli space. The origin
of this complication can be understood from the property that the phase of the central charge,
which determines the BPS flow in multi-centre solutions, is somehow replaced by the $n_v-1$
flat directions of the individual charges \footnote{For a single center solution of given
non-BPS charge, there are exactly $n_v-1$ scalars that remain constant throughout the flow
and are by definition determined by the $n_v-1$ non-compact generators of the duality group leaving the
charge invariant.} in the composite non-BPS system. It follows that instead of the $N-1$
equations for $N$ centres one finds in the BPS system, one now finds $n_v\times ( N-1)$
equations, which not only fix the distances between the centres, but also constrain the
electromagnetic charges and the asymptotic scalars in general. Nonetheless, restricting
attention to the relevant hypersurface in moduli space where the solution exists, we find
that the situation is essentially the same as for BPS solutions, \ie that this hypersurface
admits a co-dimension one boundary in moduli space corresponding to walls of marginal
stability, where (some of) the distances between the centres diverge.

Furthermore, we study explicitly the binding energy of multi-centre
solutions within the composite non-BPS system. This is based on an extension of the
notion of the fake superpotential, as it has been defined for single-centre solutions
\cite{Ceresole:2007wx,Andrianopoli:2007gt,LopesCardoso:2007ky,Perz:2008kh,Bossard:2009we,Ceresole:2009vp}.
Here, we give the general expression for the single-centre fake superpotential, for
any value of the charge vector, in terms of the scalar fields and the parameters
describing the flat directions mentioned above. The latter correspond precisely to
the auxiliary variables introduced for the ST$^2$ and STU models in
\cite{Ceresole:2009vp}, and can be identified with them. In this paper we prove that the `true fake superpotential' describing the single-centre flow is not only obtained as an extremum of the flat directions dependent potential as defined in \cite{Ceresole:2009vp}, but is in fact always a global maximum. 

Let us stress that the expression of the fake superpotential linear in the charges as defined in \cite{Ceresole:2009vp}, is a rather involved function of the moduli and the flat directions parameters, already for the STU model. Proving that the extrema of the parameters describing flat directions were maxima therefore has been a technical obstacle for some time. Using a parametrisation of the moduli and the flat directions that depends explicitly on the electromagnetic charges, as inspired by the structure of the general single-centre solution, we shall see that the expression of the fake superpotential simplifies drastically such that we are able to prove our results for any cubic model with symmetric K\"{a}hler space. 

Using the property that the energy of a composite bound state is described by the same potential linear in the charges, at non-extremum values of the flat directions parameters \cite{Bossard:2011kz, Bossard:2012ge}, we are able to prove that the binding energy is always positive. Furthermore, we also exhibit that the total energy at the location of a wall of marginal stability in moduli space is equal to the sum of the masses of the constituents that decouple, irrespectively of whether they are single-centre or composite themselves.

In fact we also find that the total mass of a composite solution is
always lower than that of a single-centre solution of the same total charge, so that composite
solutions are actually energetically favored, whenever they exist. This is in contrast with BPS configurations, for which the mass is entirely determined by the total electromagnetic charges and the asymptotic scalars, such that a BPS bound state always has the same mass as the single-centre BPS black hole with identical charges. The existence and structure
of the composite solutions is also shown to be connected to a notion of attractor flow
tree, very similar to the corresponding one for BPS solutions \cite{Denef:2000nb}.

This paper is organised as follows. In section \ref{sec:system} we introduce some notations
and discuss the composite non-BPS system without any restriction on the charges, in a
convenient basis. We then discuss the properties of single-centre and multi-centre non-BPS
solutions in section \ref{sec:explicit-sols}, using the same basis. In particular, we present
the most general single-centre solution in section \ref{sec:sing-cent-expl}, while in section
\ref{sec:superpotential} we define the fake superpotential and consider its properties. These
are then used in section \ref{sec:multi-cent-expl}, where the multi-centre solutions are
presented and the walls of marginal stability and the binding energy of the composites are
studied. Some of our results are illustrated in an explicit two-centre example carrying
D0-D6 and D0-D4-D6 charges, in section \ref{sec:two-cent-ex}. Section \ref{sec:derive-sols}
is devoted to the detailed derivation of several results used in the previous sections for the
single- and multi-centre solutions in a frame independent formulation. We conclude in
section \ref{sec:concl}, where we discuss our results and point to further directions.
Finally, we recall some technicalities about T-dualities derived in \cite{Bossard:2013oga} in Appendix A, we show the appearance of space-dependent K\"{a}hler transformations to identify different sections describing the same solution in Appendix B, and in Appendix C we compute the stabilizer of two generic charges of negative quartic invariant.

\section{Composite non-BPS system}
\label{sec:system}

In this section, we give some basic properties of the supergravity models
we consider, in subsection \ref{sec:preliminaries}, and define the general
composite non-BPS system in a convenient basis in subsection \ref{sec:def-sys}.
Using this basis, we give expressions for the general multi-centre solution,
in terms of harmonic functions, referring to section \ref{sec:derive-sols}
for the details of the derivation in a general basis.

\subsection{Preliminaries}
\label{sec:preliminaries}

In this paper we wish to describe stationary asymptotically flat extremal
black holes in the context of $\N=2$ supergravity coupled to $n_v$ vector
multiplets. The bosonic field content consists of the
metric, $n_v$ complex scalar fields, $t^i$, and $n_v+1$ gauge fields, $A^I$, where
$i=1\dots n_v$ and $I=1\dots n_v+1$.
The bosonic Lagrangian then reads \cite{deWit:1984pk, deWit:1984px}
(see \cite{Bossard:2013oga} for our conventions)
\begin{eqnarray}\label{eq:Poincare-4d}
8\pi\,e^{-1}\, {\cal L} &=&
  - \tfrac12  R - g_{i\bar{\jmath}}\, \partial^{\mu} t^i \partial_{\mu} \bar{t}^{\bar\jmath}
-\tfrac{\mathrm{1}}4\, F^I_{\mu\nu}\, G_I^{\mu\nu} \,.
\end{eqnarray}
Here, the $F^{I}_{\mu\nu} = \partial_\mu A^I_\nu - \partial_\nu A_\mu^I$
encompass the graviphoton and the gauge fields of the vector multiplets,
while $G_I^{\mu\nu}$ are the dual field strengths, defined in terms of the
$F^I_{\mu\nu}$ though the scalar dependent couplings. The explicit form of
these couplings and of the K\"ahler metric, $g_{i\bar{\jmath}}$, will not be
relevant in what follows, but can be computed in terms of the prepotential,
which we will always consider to be cubic
\begin{equation}\label{prep-def}
F=-\frac{1}{6}c_{ijk}\frac{X^i X^j X^k}{X^0} \equiv  -\frac{\det[{\bf{X}}]}{X^0} \,.
\end{equation}
Here, the tensor $c_{ijk}$, $i=1,\dots n_v$, is completely symmetric and we
introduced the cubic determinant $\det[{\bf{X}}]=\frac{1}{6}c_{ijk}\,X^i X^j X^k$
and the shorthand boldface notation for objects carrying an index $i,j,\dots$.

Here, we consider $\N=2$ supergravity theories for which the special
K\"ahler target space, $\mathcal{M}_4$, is a symmetric space and can be
obtained by Kaluza--Klein reduction from the corresponding five dimensional
theories \footnote{This excludes theories with minimally coupled vector
multiplets, which do not contain systems of the type we consider here.}
defined in \cite{Gunaydin:1983bi}. In this case, $\mathcal{M}_4$ is a coset
space, while the symmetric tensor $c_{ijk}$ satisfies special properties.

In order to set up the notation used throughout this paper, we define the cross product
\begin{equation}\label{cross-def}
 (\va\times\vb)_i= \frac12\,c_{ijk} a^j b^k\,,
\end{equation}
where we use boldface notation for vectors, omitting the indices $i\,,j\,,\dots$ for brevity.
Symmetric special target spaces are defined by tensors satisfying the Jordan algebra
identity
\begin{equation}
 (\va\times\va)\times(\va\times \va) = \det\va\, \va\,,
\end{equation}
for any vector $\va$. Taking derivatives of this basic identity, one can easily show
identities involving different vectors, as
\begin{align}
 4\,(\va\times\va)\times(\va\times \vb) =&\, \det\va\, \vb+ \va\,\tr[\va\times\va \,\vb]\,,
\CR
%  4\, \ve \times\left((\ve\times\ve)\times \vp\right) =&\, \det\ve \vp + \ve\times\ve\,\tr[\ve \,\vp]\,,
% \CR
%  4\, \left((\ve\times\ve)\times \vp\right) \times \left((\ve\times\ve)\times \vp\right)
% =&\, -2 \,\ve\times(\vp\times\vp)\,\det\ve + \vp\, \det\ve\,\tr[\ve\, \vp]
% \CR
% &\, + \ve\times\ve\,\tr[(\ve\times\ve)\times(\vp\times\vp)]\,,
% \CR
4 \, (\va\times\vb)\times (\va\times\vb)
=&\, - 2 \, (\va\times\va)\times (\va\times\vb)
\CR
&\, + \va \,\tr[\vb\times\vb\, \va] + \vb \,\tr[\va\times\va\, \vb]\,,
\end{align}
which will be used extensively in what follows. Note that the notation
$\tr[\va \vc]= a^i c_i$ denotes the contraction of two elements with
two different kinds of indices.\footnote{For symmetric models, one can define
a dual tensor $c^{ijk}$, that allows for the cross product \eqref{cross-def}
to be defined for vectors with lower indices.} Similar notation will be used for
vector and scalar fields when writing components, so that we write $\vt$ for the
complex scalars. This notation is rather natural for the so-called magic theories, for which a vector $\va$ can be represented as a three by three Hermitian matrix over a Hurwitz algebra (\ie $\mathds{R},\, \mathds{C},\, \mathds{H},\, \mathds{O}$) \cite{Gunaydin:1983bi}.

Throughout this work, we use objects transforming covariantly under
electric/magnetic duality, in order to naturally parametrise solutions. The
gauge field equations of motion and Bianchi identities can then be cast as a
Bianchi identity on the symplectic vector
\begin{equation}\label{eq:dual-gauge}
 \cF_{\mu\nu}=\begin{pmatrix} G_{I\,\! \mu\nu} \vspace{.3cm} \\ F^I_{\mu\nu}\end{pmatrix}\,,
\end{equation}
whose integral over any two-cycle defines the associated electromagnetic charges
through%\footnote{Note that we use an unconventional normalisation factor of
%$\sqrt{2}$ associated to taking electromagnetic charges as $\sqrt{2} $ times
%integers.}
\begin{equation}\label{ch-def}
\Gamma = \frac{1}{2\pi} \,\int_{S^2} \cF
 =\begin{pmatrix} q_{0} \\ \vq \\ \vp \\ p^0 \end{pmatrix}
 \,,
\end{equation}
where we explicitly show the decomposition of the charge vector in the $n_v+1$ electric
and magnetic components. We use exactly the same decomposition for all other
symplectic vectors. The symplectic inner product in this representation then takes
the form
\begin{equation}\label{inn-prod-def}
\Iprod{\Gamma_1}{\Gamma_2} =
 q_{0\,1} p^0_{\,2}  + \tr[\vq_1\,\vp_2] - p^0_{\,1} q_{0\,2} - \tr[\vp_1\,\vq_2] \,.
\end{equation}

Finally, the physical scalar fields, $\vt$, also appear through a symplectically
covariant object, the so called symplectic section, $\cV$, which is uniquely
determined by the physical scalar fields as
\begin{equation}\label{eq:sec-canon}
 \mathcal{V} = \begin{pmatrix} F_I \\ X^I \end{pmatrix}
 = X^0 \left(\begin{array}{c} \det \vt  \\ -\vt \times \vt  \\ \vt \\ 1 \end{array}\right)\,,
\end{equation}
up to the local $U(1)$ phase $X^0$.

\paragraph{Quartic invariant and charges of restricted rank} \hspace{3cm} \\
The invariance of the cubic norm $\det\va$ can be used to define duality invariants
and restricted charge vectors, a concept that is of central importance for the
applications we consider later in this paper. First, we introduce the quartic
invariant for a charge vector $\Gamma$, as
\begin{eqnarray}\label{I4-basis}
I_4(\Gamma)&=& \frac{1}{4!} t^{MNPQ}\Gamma_M\Gamma_N\Gamma_P\Gamma_Q
\nonumber\\
&=& -4\,q_0\,\det\vp + 4\,p^0\,\det\vq + 4\,\tr[\vp\times\vp\,\vq\times\vq]
    - (p^0 q_0 + \tr \vp \vq)^2 \,,
\label{I4-ch}
\end{eqnarray}
where we also defined the completely symmetric tensor $t^{MNPQ}$ for later reference.
It is also convenient to define a symplectic vector out the first derivative, $I_4^\prime(\Gamma)$,
of the quartic invariant, as
\begin{equation} \label{I4-der-basis}
I_4^\prime(\Gamma) = 4 \left( \begin{array}{c} -\det \vq + \tfrac{1}{2} q_0 ( q_0 p^0 + \tr \vq \vp )\\
 q_0 \vp \times \vp - 2 \vp \times ( \vq \times \vq ) + \tfrac{1}{2} \vq ( q_0 p^0 + \tr \vq \vp ) \\
 p^0 \vq \times \vq + 2 \vq \times ( \vp \times \vp ) - \tfrac{1}{2} \vp ( q_0 p^0 + \tr \vq \vp ) \\
-\det \vp - \tfrac{1}{2} p^0 ( q_0 p^0 + \tr \vq \vp  )
 \end{array}\right) \ ,
\end{equation}
so that the following relations hold
\begin{equation}
 \Iprod{\Gamma}{I^\prime_4(\Gamma)} = 4  I_4(\Gamma)  \ , \qquad I^\prime_4(\Gamma,\Gamma,\Gamma) = 6 I_4^\prime(\Gamma) \ .
\end{equation}
In the following, all instances of $I_4(\Gamma_1,\Gamma_2,\Gamma_3,\Gamma_4)$ will denote
the contraction of the tensor $t^{MNPQ}$ in \eqref{I4-ch} with the four charges, without any symmetry
factors, except for the case with a single argument, as in $I_4(\Gamma)$ and $I^\prime_4(\Gamma)$.

We are now in a position to introduce the concept of charge vectors of restricted rank.
A generic vector leads to a nonvanishing invariant \eqref{I4-ch} and is also referred
to as a rank-four vector, due to the quartic nature of the invariant. Similarly, a
rank-three vector, $\Gamma_\trois$, is a vector for which the quartic invariant vanishes,
but not its derivative. An obvious example is a vector with only $\vp \neq 0$ and all
other charges vanishing, so that the derivative $I_4^\prime(\Gamma_\trois)$ is nonzero and
proportional to the cubic term $\det\vp$.

There are two more classes of restricted vectors, defined analogously as rank-two (small)
and rank-one (very small) vectors. A rank-two vector, $\Gamma_\deux$, is defined such that both
$I_4(\Gamma_\deux)= I^\prime_4(\Gamma_\deux)=0$, and a simple example is provided by
a vector with all entries vanishing except the $\vp$, with the additional constraint
that $\det\vp=0$. Finally, a very small vector, $\Gamma_\un$, is defined such that
\begin{gather}
 I_4(\Gamma_\un)= I^\prime_4(\Gamma_\un)=0\,,
\CR
\frac{1}{4} I_4(\Gamma_\un , \Gamma_\un , \Gamma , \Gamma )
\equiv \frac{1}{4} t^{MNPQ} \Gamma_{\un M} \Gamma_{\un N} \Gamma_P \Gamma_Q
 = - \Iprod{\Gamma_\un}{\Gamma}^2\,, \label{VerySmallI4}
\end{gather}
for any vector $\Gamma$. Examples of very small vectors are given by vectors
where only the $q_0$ or $p^0$ component is nonzero. More generally, we will use
the parametrisation
\begin{equation}\label{eq:R-gen}
 \hat{R} = \frac{ 2 \sqrt{2}}{ \det \ve }  \left( \begin{array}{c} 1 \\ \ve \\ \ve \times \ve \\ - \det \ve \end{array} \right) \ ,
\end{equation}
for a general very small vector, where the choice of normalisation is for later convenience.
Note that a general rank one vector can always be
written in this way up to a possibly singular rescaling. Since the black hole solutions
described in what follows do not depend on the normalisation of $\hat{R}$, this
parametrisation is completely general, although it is singular for specific rank one
vectors. In the discussion of explicit black hole solutions, we will need to define
a second constant very small vector, denoted $\Rstz$, that does not commute with $\hat{R}$,
so that in the parametrisation \eqref{eq:R-gen}, it reads \footnote{We use the particular
notation $\hat R$ and $\Rstz$ for the two vectors in order to simplify comparison with
the notation introduced in \cite{Bossard:2013oga}, as well as with section
\ref{sec:derive-sols} below, which uses the notation of that paper.}
\begin{equation} \label{pre-Rst}
  \Rstz = \frac{\sqrt{2}\, \det \ve }{\det( \ve +\ve^*)} \left( \begin{array}{c} 1 \\ -\ve^* \\ \ve^* \times \ve^* \\ \det \ve^* \end{array} \right) \ ,
\end{equation}
where $\ve^*$ is defined such that $\det (\ve +\ve^*) \ne 0$ and
$\langle \hat{R} , \Rstz \rangle = 4$ by construction. Despite the fact that this
provides a natural parametrisation for $\Rstz$, it turns out it is not the most
convenient, as it obscures the action of T-dualities, which are central to our
construction and we describe next.

\paragraph{T-dualities} \hspace{3cm} \\
A crucial ingredient in the description of black hole solutions in supegravity is
the action of abelian isometries of the scalar manifold in the real basis. These
isometries are defined as including the standard spectral flow transformations,
given by (the notation $\exp(\cTgm_\vk)$ for this action will become clear shortly)
\begin{equation} \exp(\cTgm_\vk) \left(\begin{array}{c} q_0 \\ \vq \\ \vp \\ p^0 \end{array}\right) = \left(\begin{array}{c} q_0 - \tr \vk\vq + \tr \vk \times \vk \, \vp + \det \vk \, p^0  \\ \vq -2 \vk \times \vp - \vk \times \vk \, p^0  \\ \vp + \vk \, p^0  \\ p^0 \end{array}\right) \ , \label{Tminus0} \end{equation}
% \begin{equation} \label{eq:sp-flow-ex}
%  \left(\begin{array}{c} q_0 \\ \vq \\ \vp \\ p^0 \end{array}\right)
% \rightarrow
%  \left(\begin{array}{c} q_0 - \tr \vk\vq + \tr \vk \times \vk \, \vp + \det \vk \, p^0
% \\ \vq -2 \vk \times \vp - \vk \times \vk \, p^0  \\ \vp + \vk \, p^0  \\ p^0 \end{array}\right) \ ,
% \end{equation}
as well as all the abelian isometries dual to \eqref{Tminus0}. An obvious example
are the transformations obtained by S-duality on \eqref{Tminus0}, as 
\begin{equation} \exp(\cTgp_\vk) \left(\begin{array}{c} q_0 \\ \vq \\ \vp \\ p^0 \end{array}\right) =  \left(\begin{array}{c} q_0 \\ \vq - \vk q_0 \\ \vp - 2 \vk \times \vq  + \vk \times \vk\,  q_0 \\ p^0+\tr \vk \vp - \tr \vk \times \vk \, \vq + \det \vk \, q_0  \end{array}\right) \ . \label{Tplus0}  \end{equation}
For the purposes of this paper, we define general T-dualities as
the collection of all abelian subgroups in the duality group, obtained from 
the spectral flows by dualities. These can be described in terms of real vector
parameters in the general case, similar to spectral flows, as shown in
\cite{Bossard:2013oga}. We refer to that work for the details of the
description in the symplectic real basis and concentrate on the results for
the representation of T-dualities that will be used extensively in constructing
black hole solutions.

It is useful to think of T-dualities as raising and lowering operators
$\Tgpm$ on the components in \eqref{Tminus0}-\eqref{Tplus0}. This is clearly
the case for \eg the spectral flow parametrised by $\vk$ in \eqref{Tminus0},
whose generators never generate $p^0$, while the magnetic components, $\vp$, are
only generated by the action on $p^0$ etc. As shown in \cite{Bossard:2013oga},
this structure is general to all T-dualities, which act on four separate
eigenspaces in a similar fashion.
The relevant generator is given by
\begin{equation}
{\bf h}_T\, \Gamma \equiv \Iprod{\hat{R}}{\Rstz}^{-1} \Scal{  
\frac{1}{2} I_4^{\prime}( \hat{R}, \Rstz ,\Gamma)  + \Iprod{\Gamma}{\Rstz} \hat{R}  - \Rstz \Iprod{\hat R}{\Gamma}} \  , 
\label{hTReal-1}
\end{equation}
where $\hat{R}$ and $\Rstz$ are two mutually nonlocal very small vectors.  One can 
verify that ${\bf h}_T$ preserves both the symplectic product and the quartic invariant.
For example, taking $\ve\rightarrow 0$ and $\ve^*\rightarrow \infty$ in \eqref{eq:R-gen}-\eqref{pre-Rst} leads to a pair $\hat{R}$, $\Rstz$ along $p^0$ and $q_0$ respectively and
to the decomposition seen in the spectral flow transformations \eqref{Tminus0}-\eqref{Tplus0}.
In the following, we denote the four eigenspaces of \eqref{hTReal-1} by their corresponding
eigenvalue.\footnote{We refer to appendix \ref{sec:T-duality} for a more
detailed discussion} Indeed, it is simple to show that $\hat{R}$ and $\Rstz$ have eigenvalues
$+3$ and $-3$ respectively, while the remaining charge components are evenly split into
$+1$ and $-1$ eigenvalue vectors. For the spectral flows of \eqref{Tminus0}, the magnetic
components $\vp$ are of eigenvalue $+1$, while the electric components $\vq$ are of
eigenvalue $-1$.

In the general case, one should use the parametrisation \eqref{eq:R-gen} for $\hat R$, which
by using \eqref{Tminus0}, can be written as
\begin{equation} \label{SimilarlyR}
 \hat{R} = \exp\Scal{ \cTgm_{ - \frac{ \ve \times \ve}{\det \ve}}} \left( \begin{array}{c} 0 \\ 0 \\ 0 \\ - 2 \sqrt{2}  \end{array} \right)
=  \exp\Scal{ \cTgm_{ - \frac{ \ve \times \ve}{\det \ve}}}  \exp\scal{ \cTgp_{ \vk}} \left( \begin{array}{c} 0 \\ 0 \\ 0 \\ - 2 \sqrt{2}  \end{array} \right) \ .
\end{equation}
Here we used the property that the vector is invariant with respect to $\cTgp$ in the second line.
This way it is straightforward to write another parametrisation for $\Rstz$ in \eqref{pre-Rst},
where such a T-duality parameter appears polynomially, as
\begin{eqnarray}
  \Rstz &=& \exp\Scal{ \cTgm_{ - \frac{ \ve \times \ve}{\det \ve}}}  \exp\scal{ \cTgp_{\vk}}  \left( \begin{array}{c} \sqrt{2}  \\ 0 \\ 0 \\ 0 \end{array} \right) \CR
&=&  \frac{\sqrt{2}}{ \det \ve} \left( \begin{array}{c} \det(\ve - \vk)    \\  -  \vk\,  \det \ve + 2 ( \ve \times \ve ) \times ( \vk \times \vk ) - \ve \, \det \vk  \\   \vk  \times   \vk   \, \det \ve-  \ve \times \ve\,  \det \vk  \\ \det \ve \, \det\vk   \end{array} \right) \ .\label{RsK}
\end{eqnarray}
Of course this base will be rather singular when $\det \ve = 0 $, but this is only the case for
isolated points in the moduli space of pairs of rank one vectors with a fixed symplectic product.

Using the relations above, we can obtain an explicit representation for general T-dualities,
denoted $\Tgpm$, that will be useful in what follows, especially in section \ref{sec:explicit-sols}.
As explained in Appendix \ref{sec:T-duality}, the representation \eqref{SimilarlyR} and \eqref{RsK}
allows one to define the generic $\Tgp$ from the spectral flows and their S-dual through
\eqref{eq:T-conj}, or explicitly
\begin{equation} \label{T-similar}
\Tgp_{\vk} = \exp\Scal{ \cTgm_{ - \frac{ \ve \times \ve}{\det \ve}}} \cTgp_{\vk}  \exp\Scal{ \cTgm_{ \frac{ \ve \times \ve}{\det \ve}}}  \ . \end{equation}
Similarly, one can define the dual T-dualities $\Tgm_{\vk_-}$ through \eqref{eq:T-conj-dual},
but these do not appear in the composite non-BPS system studied here.

We emphasise that all explicit formulae above are fully duality covariant,
despite the fact that we use spectral flows as preffered transformations in order to
define a representation. On the contrary, our parametrisation identifies the correct
combinations of a general charge vector that transform under the simple spectral flows
\eqref{Tplus0}-\eqref{Tminus0}. To be precise, we record the following rewriting of
the charge vector in the preffered basis,
\begin{equation}\label{eq:ch-dec-basis-gen}
\Gamma =  \exp\Scal{ \cTgm_{ - \frac{ \ve \times \ve}{\det \ve}}}  \exp\scal{ \cTgp_{ \vk} }
\left( \begin{array}{c} P \\  \vl_\Gamma + \vk\, P\\
\vp + \frac{\ve \times \ve}{\det \ve}  \, p^0 + 2 \vk \times \vl_\Gamma  + \vk\times \vk \, P\\
p^0 - \tr \vk \Scal{\vp + \frac{\ve \times \ve}{\det \ve}  \, p^0  }
- \tr \vk \times \vk \, \vl_\Gamma - \det\vk\,P
 \end{array} \right)  \ ,
\end{equation}
where
\begin{align}
P = &\, \frac1{2\,\sqrt{2}}\Iprod{\hat{R}}{\Gamma}
= \frac1{\det\ve}\,(p^0 + \tr \ve \vp - \tr \ve \times \ve \, \vq)  + q_0\,,
\CR
 \vl_\Gamma = &\, \vq -  2 \frac{ \ve \times \ve}{\det \ve }  \times \vp - p^0  \frac{\ve}{\det \ve }  \ .
\label{lGamma-gen}
\end{align}
It is straightforward to verify that the action of the general T-duality \eqref{T-similar}
on $\Gamma$ is equivalent to the action of $\cTgp$ in \eqref{Tplus0} on the combinations
$P$, $\vl_\Gamma$, $\vp + \frac{\ve \times \ve}{\det \ve}  \, p^0$ and $p^0$, in the order
they appear in \eqref{eq:ch-dec-basis-gen}. Therefore, $P$ is the charge of grade $-3$,
$\vl_\Gamma$ is of grade $-1$, while $\vp + \frac{\ve \times \ve}{\det \ve}  \, p^0$ and
$p^0$ are of grade $+1$ and $+3$ respectively.

Similarly, we use the definition in \eqref{T-similar} to act on the moduli, given the
known action of $\cTgpm$. By definition, the spectral flow $\cTgm$ is the T-duality
shifting the axions as
\begin{equation}\label{Tm-scal}
 \exp\scal{ \cTgm_{\vk}} \vt = \vt + \vk\,,
\end{equation}
which is exactly the action of $\cTgm$ on the physical scalar following by application of
\eqref{Tminus0} on the section in \eqref{eq:sec-canon}.
Finally, the action of $\cTgp$ on \eqref{eq:sec-canon} leads to the transformation
\begin{equation}\label{Tp-scal}
 \exp\scal{ \cTgp_{\vk}} \vt = \left( \vt^{-1} + \vk \right)^{-1} =
\frac{\vt + 2 \vk\times(\vt\times\vt) + \vk\times\vk\,\det\vt }
     {1 + \tr\vt\times\vt\,\vk + \tr\vk\times\vk\,\vt + \det\vk \,\det\vt}\,.
\end{equation}
Here, the inverse is the Jordan inverse $\vt^{-1}\equiv\frac{\vt\times\vt}{\det\vt}$ and the first
equality expresses the fact that $\cTgp$ is related to $\cTgm$ by an S-duality.

\subsection{Definition of the system}
\label{sec:def-sys}

We are now ready to introduce the composite non-BPS system for constructing
multi-centre black hole solutions. We assume stationary backgrounds
and restrict ourselves to the solutions with a flat $\mathbb{R}^3$ base space. We
therefore introduce the standard Ansatz for the metric
\begin{equation}
 \label{metricMltc}
 ds^2=- e^{2U}( d t+\omega )^2 + e^{-2U}   d \vec{x} \cdot  d\vec{x} \,,
\end{equation}
in terms of a scale function $U(x)$ and the Kaluza--Klein one-form
$\omega(x)$ (with spatial components only), which are both required to asymptote to zero at
spatial infinity. Here and henceforth, all quantities are independent of
time, so that all scalars and forms are defined on the flat three-dimensional
base.

For a background as in \eqref{metricMltc}, the $n_v+1$ gauge fields of
the theory, together with their magnetic duals, as arranged in the symplectic
vector, $\cF$,  in \eqref{eq:dual-gauge} are decomposed as
\begin{equation}
 \label{gauge-decop}
2  \cF= d \zeta \, ( d t+\omega) + F   \,, \qquad
   F= \zeta\, d  \omega + d w \,.
\end{equation}
Here, we defined the gauge field scalars $\zeta$, arising as the time component
of the corresponding gauge fields, and the one-forms $w$ describing the charges.
Of these components, only the vector fields $w$ are indepedent, while the $\zeta$
can straightforwardly be constructed once the solution for the scalars is known.

In order to describe a solution, one therefore needs to specify the spatial
part of the gauge fields, $dw$, the scalar section $\cV$ (or the physical
moduli $t^i$ directly), as well as the metric components $e^U$ and $\omega$.
The composite non-BPS system can be described by introducing two constant,
mutually nonlocal very small vectors, $\hat R$ and $\Rstz$, as above and two
vectors of functions, denoted $\cHz$ and $\cK$. The former is contains
eigenvectors of eigenvalues $(-1)\oplus(+3)$ with respect to the grading
\eqref{eq:vec-decomp} and will be parameterised as
\begin{equation} \label{V-L-def}
\cHz= \frac{1}{\sqrt{2} }\,
\exp\Scal{ \cTgm_{ - \frac{ \ve \times \ve}{\det \ve}}}  \exp\scal{ \cTgp_{ \vk} }
\left(\begin{array}{c} 0 \\ \vL \\ 0 \\ -V \end{array}\right) \,.
\end{equation}
Here, $\vL$ and $V$ are the two functions parametrising the $(-1)$ and $(+3)$
components respectively.\footnote{Note that in \eqref{V-L-def}
we rescaled these functions by factors of $\det \ve$ with respect to their
definition in terms of $\hat{R}$ and $\Rstz$ in \eqref{SimilarlyR}-\eqref{RsK},
for simplicity (this can be reabsorbed by a rescaling of these two very small
vectors).}
The second vector of functions, $\cK$, appears only as a parameter of T-dualities
that vary in space. We therefore do not need its explicit covariant form, but only
the corresponding parameter in the chosen representation, which we denote by $\vK$.

The two vectors, $\vK$ and $\vL$, are harmonic on the flat
$\mathbb{R}^3$ base, as
\begin{equation}
 d\star d \vK = d\star d \vL=0\,,
\end{equation}
while the function $V$ is specified by the Poisson equation
\begin{equation}\label{V-Poiss-b}
 d\star d V = \tr [\vL\, d\star d (\vK \times \vK)] \,.
\end{equation}
The final dynamical equation required is the one for the angular momentum
vector $\omega$, which is given by
\begin{align}\label{eq:dom-5-b}
\star d \omega - d M = \tr [\vL\times \vL \,d\vK] \,,
\end{align}
where $M$ is a new local function that appears explicitly in the
solutions. Taking the divergence of \eqref{eq:dom-5-b}, we obtain the Poisson
equation
\begin{align}\label{eq:Poiss-M-b}
d \star d M = -d\,\tr [\vL\times \vL \,\star d\vK] \,,
\end{align}
in terms of $\vK$ and $\vL$.

The solutions are then given by the above functions, as follows. The scalars
are given by
\begin{align}
 2\,e^{-U}\mbox{Im}(e^{-i\alpha}\cV)
=&\, -\exp\Scal{ \cTgm_{ - \frac{ \ve \times \ve}{\det \ve}}}
\exp\scal{ \cTgp_{\vK-\vk} } \exp\Scal{ \cTgm_{ \frac{ \ve \times \ve}{\det \ve}}}
\left(\cHz - \tfrac12\, V\,\hat{R} - \tfrac{M}{V}\,\Rstz \right) \,
\CR
=&\, -\frac{1}{\sqrt{2} }  \exp\Scal{ \cTgm_{ - \frac{ \ve \times \ve}{\det \ve}}}  \exp\scal{ \cTgp_{\vK}}
\left( \begin{array}{c} -2 \frac{M}{V}  \\ \vL   \\  0   \\ V  \end{array} \right) \,,
\label{scals-5-b}
\end{align}
where we have used  in the second line the explicit form of $\cHz$ and the very small
vectors. The physical scalars do not depend on the K\"{a}hler phase $\alpha$. Note that
the vector of harmonic functions, $\vK$, appears in place of the constant parameter of
the basis, $\vk$, which can be viewed as the asymptotic value of $\vK$, parametrising
$\Rstz$ (cf. also the discussion below \eqref{Rstar0-real}).
Similarly, the vector fields are defined from the first order equation
\begin{align}
  \star dw = &\, \frac{1}{\sqrt{2}} 
\exp\Scal{ \cTgm_{ - \frac{ \ve \times \ve}{\det \ve}}}  \exp\scal{ \cTgp_{\vK}}
\left(  d  \left( \begin{array}{c} 0 \\ \vL  \\  0   \\ -V \end{array} \right)
      - \cTgp_{d \vK}  \left( \begin{array}{c} 0 \\ \vL  \\  0   \\ -V \end{array} \right)\right)
\CR
= &\,\frac{1}{\sqrt{2}}  \exp\Scal{ \cTgm_{ - \frac{ \ve \times \ve}{\det \ve}}}
 \left( \begin{array}{c} 0 \\ d \vL  \\   2 \vL \times d \vK - 2 \vK \times d \vL   \\ -dV - \tr \vK \times \vK \, d \vL + \tr \vL \, d ( \vK \times \vK )  \end{array} \right)\,,
\label{dw-bas}
\end{align}
so that the additional harmonic functions $\vK$ modify the charges explicitly. One computes the gauge fields scalars according to \cite{Bossard:2013oga}
\begin{equation} \zeta = -\frac{1}{\sqrt{2}}e^{4U} \exp\Scal{ \cTgm_{ - \frac{ \ve \times \ve}{\det \ve}}}  \exp\scal{ \cTgp_{\vK}}  \left( \begin{array}{c} \det \vL  \\ - M \vL   \\  V \vL \times \vL    \\ M V \end{array} \right) \ . \end{equation}
We note that \eqref{scals-5-b} can be solved in exactly the same
way as for the BPS solutions \cite{Bates:2003vx}, which in our basis gives
\begin{eqnarray}
 \vt &=& \exp\Scal{ \cTgm_{ - \frac{ \ve \times \ve}{\det \ve}}}  \exp\scal{ \cTgp_{\vK}} \frac{ \vL \times \vL}{M-ie^{-2U} } \CR
&=&  \frac{ \scal{ \frac{ M - i e^{-2U}}{\det \vL} \vL +\vK } \times \scal{ \frac{ M - i e^{-2U}}{\det \vL} \vL +\vK }}{\det\scal{ \frac{ M - i e^{-2U}}{\det \vL} \vL +\vK }} - \frac{\ve \times \ve}{\det \ve }  \ . \label{Gen-Moduli}
\end{eqnarray}
Similarly, the metric scale factor is given by
\begin{equation}
e^{-4U} = V\, \det\vL - M^2 \ . \label{ScallingFactorInvariant}
\end{equation}
Regularity implies that the $n_v$ harmonic functions $\vL$ must correspond to a strictly positive Jordan algebra element,  so that \eqref{ScallingFactorInvariant} leads to a non-degenerate metric and the scalar fields \eqref{Gen-Moduli} lie in the K\"{a}hler cone.  Strictly positive  means that the three eigen values of  $\vL$ must be strictly positive for a classical Jordan algebra, and equivalently in the STU truncation that the three functions $L_i$ are strictly positive.

Explicit solutions to this system where derived in a particular frame in \cite{Bossard:2011kz, Bossard:2012ge}
while in the next section we discuss the general solution carrying arbitrary charges in the specific base above.
The general manifestly duality covariant solution is derived in section \ref{sec:derive-sols}, independently of any specific frame.

\section{Composite non-BPS solutions}
\label{sec:explicit-sols}

In this section we discuss the general properties of composite non-BPS solutions,
in the explicit parametrisation of the previous section. This representation
is useful in studying the properties of solutions, since it provides explicit formulae
for all quantities, as explained above. In particular, the natural parametrisation of
the moduli in terms of integration constants in \eqref{Gen-Moduli} allows us to study
the behaviour of solutions as a function of the asymptotic scalars for fixed
electromagnetic charges.

We find that all regular composite solutions only exist for moduli constrained to a
$(n_v+1)$-dimensional hypersurface with an $n_v$-dimensional boundary defining a wall
of marginal stability. The solution admits a non-zero binding energy that tends to zero
at the wall, while the distance between the centres diverges, in complete analogy to
BPS composite solutions. Somewhat surprisingly, we find that a single-centre solution
always has a greater energy compared to the total energy of a composite solution of
the same total charge, at points in moduli space where it exists. Finally, we
show that one can introduce a notion of attractor tree flow, similar to the
existing one for BPS solutions \cite{Denef:2000nb}. 

In section \ref{sec:sing-cent-expl} we first discuss the general single-centre solution,
while in section \ref{sec:superpotential} we give a detailed presentation of the properties
of the fake superpotential for single-centre solutions, in the basis introduced in the
previous section. This completes a longstanding discussion in the literature
\cite{Ceresole:2007wx,Andrianopoli:2007gt,LopesCardoso:2007ky,Perz:2008kh,Bossard:2009we,Ceresole:2009vp,Yeranyan:2012au,Ferrara:2012qm}
and at the same time establishes various relations that are crucial in our treatment
of multi-centre solutions. Indeed, as it turns out, this same function describes the
total mass of the multi-centre solutions, and satisfy to a generalisation of the triangular identity that permits to prove the positivity of the binding energy. Finally, in section
\ref{sec:two-cent-ex} we present an explicit example including two centres, for which
we make all relations fully explicit, including a numerical treatment of some aspects
of the solution.

\subsection{Revisiting the single-centre solution}
\label{sec:sing-cent-expl}

We now turn to an explicit description of the general single-centre solution in the representation
introduced above. The general solution was constructed in \cite{Bossard:2012xsa}, but it has not
been given in a fully explicit form, while the mass formula and the properties of the relevant
fake superpotential were only briefly discussed in that paper. In addition, a precise description
of these properties will prove crucial in the discussion of composite solutions in what follows.

For a single-centre solution, the functions $\vK$ can be consistently set a to a specific
constant $\vk_\Gamma$,\footnote{This need not be the case, but allowing $\vK$ to be a harmonic
function leads to exactly the same physical results, as we will discuss in
\eqref{sing-complicated}.} which depends on the charge vector $\Gamma$.
We then find from \eqref{dw-bas} that the charge is defined by the poles
of $\vL$ and $V$ \ie of grade $(-1)\oplus(+3)$. This is not a constraint, but rather a choice of
basis, as a single charge can be always brought to this form by choosing $\hat R$ and $\Rstz$ appropriately.
Indeed, consider a general charge vector and constrain the vector $\ve$ as satisfying
\begin{equation} \frac{ \det \ve }{2 \sqrt{2} } \Iprod{\hat{R}}{\Gamma} = p^0 + \tr \ve \vp - \tr \ve \times \ve \, \vq  + \det \ve\, q_0 = 0 \  \label{NoRstar}\,, \end{equation}
which from \eqref{lGamma-gen} sets the grade $(-3)$ charge to zero.
One then finds
\begin{equation}\label{eq:ch-dec-basis}
\Gamma =  \exp\Scal{ \cTgm_{ - \frac{ \ve \times \ve}{\det \ve}}}  \exp\scal{ \cTgp_{ \vk} } \left( \begin{array}{c} 0 \\  \vl_\Gamma \\  \vp + \frac{\ve \times \ve}{\det \ve}  \, p^0 + 2 \vk \times \vl_\Gamma  \\ p^0 - \tr \vk \Scal{\vp + \frac{\ve \times \ve}{\det \ve}  \, p^0  } - \tr \vk \times \vk \, \vl_\Gamma  \end{array} \right)  \ ,
\end{equation}
for
\begin{equation} \vl_\Gamma =  \vq -  2 \frac{ \ve \times \ve}{\det \ve }  \times \vp - p^0  \frac{\ve}{\det \ve }  \ . \label{lGamma}  \end{equation}
For a single-centre solution, one may additionally choose the grade $(+1)$ component
of the charge to vanish, by choosing $\vk$ appropriately. The appropriate value, $\vk_\Gamma$,
is found by setting the third row in \eqref{eq:ch-dec-basis} to zero, as
\begin{equation} 2 \vl_\Gamma \times \vk_\Gamma = - \vp - \frac{ \ve \times \ve }{\det \ve}  \, p^0 \  .\label{k-sing} \end{equation}
One then obtains the charge
\begin{equation}\label{sing-charge}
 \Gamma =  \exp\Scal{ \cTgm_{ - \frac{ \ve \times \ve}{\det \ve}}}  \exp\scal{ \cTgp_{\vk_\Gamma}} \left( \begin{array}{c} 0 \\ \vl_\Gamma  \\  0   \\ \frac{I_4(\Gamma)}{4 \det \vl_\Gamma }  \end{array} \right)  \  ,  
\end{equation}
which is indeed a general vector of grade $(-1)\oplus(+3)$ for $\vk=\vk_\Gamma$. Note that we used the parametrisation
$\frac{I_4(\Gamma)}{4 \det \vl_\Gamma }$ for the charge of grade $(+3)$, instead of the equivalent
expression in the last line of \eqref{eq:ch-dec-basis}.
The general solution of \eqref{k-sing} for $\vk_\Gamma$ is
\begin{eqnarray}
 \vk_\Gamma &=&  -\frac{ 2 ( \vl_\Gamma \times \vl_\Gamma ) \times \Scal{\vp + \frac{\ve \times \ve}{\det\ve} \, p^0 } - \tfrac{1}{2} \vl_\Gamma \, \tr \vl_\Gamma \Scal{\vp + \frac{\ve \times \ve}{\det\ve} \, p^0 }}{\det \vl_\Gamma}  \label{kGamma} \\
&=&  \frac1{\det \ve \, \det \vl_\Gamma}\biggl(  \det \ve \, q_0 \, \vp \times \vp + \ve \, \det \vp - 2\, p^0 \, ( \ve \times \ve ) \times ( \vq \times \vq )
\CR
&& \hspace{1.5cm}
 - 2 \,\det \ve \, ( \vq \times \vq ) \times \vp
- 4 \,( \ve \times \ve ) \times \scal{  \vq \times ( \vp \times \vp ) }
\CR
&&  \hspace{1.5cm}
+ \frac{1}{2} \scal{ \det \ve \, \vq + 2 ( \ve \times \ve ) \times \vp + \ve \, p^0 } ( q_0 p^0 + \tr \vq \vp ) \biggr)  \ , \nonumber
\end{eqnarray}
where the determinant of $\vl_\Gamma$ is given explicitly by
\begin{align}
\det \ve \, \det \vl_\Gamma = &\, \det \vp + q_0 \tr \ve \times \ve \, \vp \times \vp + \det \ve \, \det \vq
+ \scal{  - \det \ve \, q_0 + \tr \ve \times \ve \, \vq } ( q_0 p^0 + \tr \vq \vp )
\CR
&\,- p^0 \, \tr \ve \, \vq \times \vq
- 2 \tr ( \ve \times \ve)\times ( \vq \times \vq)\, \vp - 2 \tr \ve \times \vq\, \vp \times \vp \ .
\end{align}

It is then straightforward to define the general single-centre solution from these data.
One chooses the vector of harmonic functions
\begin{equation} \cHz = \exp\Scal{ \cTgm_{ - \frac{ \ve \times \ve}{\det \ve}}}  \exp\scal{ \cTgp_{\vk_\Gamma}} \left( \begin{array}{c} 0 \\ \frac{1}{\sqrt{2}}  \vl+  \frac{1}{r} \vl_\Gamma  \\  0   \\ -\frac{1}{\sqrt{2}} \frac{1+m^2}{\det \vl}  + \frac{1}{r} \frac{I_4(\Gamma)}{4\det \vl_\Gamma }  \end{array} \right)  \ , \end{equation}
and the corresponding section reads
\begin{equation}  \label{eq:sec-expli-sing}
2 \mbox{Im}( e^{-U-i\alpha} \cV ) = -\frac{1}{\sqrt{2}}  \exp\Scal{ \cTgm_{ - \frac{ \ve \times \ve}{\det \ve}}}  \exp\scal{ \cTgp_{\vk_\Gamma}} \left( \begin{array}{c} -2 \frac{M}{V}  \\ \vL   \\  0   \\ V  \end{array} \right)  \ , \end{equation}
for
\begin{eqnarray}
 V &=& \frac{1+m^2}{\det \vl } - \frac{1}{2 \sqrt{2}} \,  \frac{I_4(\Gamma)}{r \det \vl_\Gamma}\ ,  \CR
\vL &=&  \vl +\sqrt{2}  \,  \frac{\vl_\Gamma }{r} \ , \CR
M &=& m + \J \frac{\cos \theta}{r^2} \ .
\end{eqnarray}
One then obtains the scaling factor
\begin{equation} e^{-4U}  = V \det \vL - M^2 \, \end{equation}
and the scalar fields
\begin{eqnarray}
 \vt &=& \exp\Scal{ \cTgm_{ - \frac{ \ve \times \ve}{\det \ve}}}  \exp\scal{ \cTgp_{\vk_\Gamma}} \frac{ \vL \times \vL}{M-ie^{-2U} } \CR
&=&  \frac{ \scal{ \frac{ M - i e^{-2U}}{\det \vL} \vL +\vk_\Gamma } \times \scal{ \frac{ M - i e^{-2U}}{\det \vL} \vL +\vk_\Gamma }}{\det\scal{ \frac{ M - i e^{-2U}}{\det \vL} \vL +\vk_\Gamma }} - \frac{\ve \times \ve}{\det \ve }  \ , \label{SingleCentreModuli}
\end{eqnarray}
where we used \eqref{Tp-scal} and \eqref{Tm-scal}. The solution will be regular provided
\begin{equation} - I_4(\Gamma) - \J^2 > 0\,, \end{equation}
and the vector $\vL \times \vL$ is a positive Jordan algebra element everywhere.\footnote{\eg $L_{i+1} L_{i+2} > 0$ within the STU truncation.} For $\vl$ positive,
this requires that $\vl_\Gamma$ be positive, which fixes some conditions on the vector  $\ve$ that
parametrises partially the asymptotic scalars. Note that, in principle, we should consider regularity of
$\vk_\Gamma$ as well, but since the denominator of the explicit solution in \eqref{kGamma} is
$\det\ve\,\det \vl_\Gamma$, this condition is already implied by the regularity of $\vl_\Gamma$, $\ve$.

Before concluding our discussion of the single-centre solution, let us return to the choice
made above \eqref{NoRstar} and note that one can write the same solution with non constant $\vK$.
This function is however quite restricted, since the requirement of regularity at the horizon
implies that its poles are proportional to those of $\vL$. The relevant expressions for the various
functions then follow from \eqref{V-Poiss-b}-\eqref{eq:Poiss-M-b} as
\begin{eqnarray}
 \vL &= &\vl + \sqrt{2} \frac{\vl_\Gamma}{r} \ ,
\quad \ V = \frac{1+m^2}{\det \vl } + \gamma^2 \scal{ \det \vL - \det \vl } - \frac{1}{2\sqrt{2}}\frac{1}{r}  \frac{I_4(\Gamma)}{\det \vl_\Gamma} \ , \CR
\vK &=& \vk_{\Gamma} + \gamma \vL  \ , \qquad
M = m + \gamma \scal{ \det \vl - \det \vL } + \frac{ \J \cos \theta}{r^2} \ , \label{sing-complicated}
\end{eqnarray}
where $\gamma$ is the proportionality constant relating the poles of $\vL$ and $\vK$.
The asymptotic scalars are then parametrised according to
\begin{eqnarray}
 \vt_{\infty}  &=& \exp\Scal{ \cTgm_{ - \frac{ \ve \times \ve}{\det \ve}}}  \exp\scal{ \cTgp_{  \vk}} \frac{ \vl \times \vl}{m-i} \CR
&=&  \frac{ \scal{ \frac{ m - i }{\det \vl} \vl + \vk  } \times \scal{ \frac{ m - i }{\det \vl} \vl +  \vk  }}{\det\scal{ \frac{ m - i }{\det \vl} \vl +\vk }} - \frac{\ve \times \ve}{\det \ve }  \ , \label{AsymptoticModuli}
\end{eqnarray}
for $\vk = \vk_\Gamma +\gamma \vl$. The full expression for the moduli follows from \eqref{Gen-Moduli}
and, as it turns out, is equivalent to the one in \eqref{SingleCentreModuli}, where all functions are
harmonic. One can easily check that \eqref{sing-complicated} is only a rewriting of the simple single
centre solution, since $\gamma$ can be absorbed in a redefinition of the parameters, as
$m \rightarrow m + \gamma \det \vl$. The proof in the general frame independent case is given in
\ref{app:kahler-trans}.  This redefinition defines a different set of coordinates in
moduli space \eqref{AsymptoticModuli}, which will prove useful in various settings below.

\subsection{The fake superpotential}
\label{sec:superpotential}
The mass formula for single-centre solutions is crucial for the applications that follow, especially
in comparing the mass of multi-centre solutions to that of their constituents. We therefore wish to
rewrite the explicit expression of the mass in the representation used in this paper, in terms of
the fake superpotential proposed in \cite{Ceresole:2007wx} and defined in
\cite{Bossard:2009we,Ceresole:2009vp}.
Using the parametrisation \eqref{SingleCentreModuli} for the moduli, the non-BPS mass formula takes the
rather simple form
\begin{equation} \label{W-single}
W(\Gamma) = \frac{1}{2\sqrt{2}} \Scal{ ( 1 + m^2 ) \frac{ \tr \vl \times \vl \, \vl_\Gamma }{\det \vl} -\frac{1}{4}  \frac{ \det \vl}{\det \vl_\Gamma} I_4(\Gamma) } \ . \end{equation}
Similarly, the asymptotic central charge in this basis is
\begin{equation} |Z(\Gamma)| = \frac{1}{2\sqrt{2}} \Bigl| ( m - i  )^2  \frac{ \tr \vl \times \vl \, \vl_\Gamma }{\det \vl} - \frac{1}{4}  \frac{ \det \vl}{\det \vl_\Gamma} I_4(\Gamma) \Bigr|  \ . \end{equation}
Noting that the constant $m$ is finite for regular values of the moduli,
it is simple to verify that $W(\Gamma) > |Z(\Gamma)|$, provided $I_4(\Gamma) < 0 $. In contrast, one would
have  $W(\Gamma) < |Z(\Gamma)|$ for $I_4(\Gamma)>0$. This proves that such a regular non-BPS extremal black
hole always satisfies to the BPS bound. We emphasise that our formula \eqref{W-single} is not linear in
the charge $\Gamma$, as the parametrisation \eqref{SingleCentreModuli} of the asymptotic scalars we use
depends implicitly on the charge (through $\vk_\Gamma$ and $\ve$ via the condition
$\Iprod{\hat{R}}{\Gamma}=0$). In order to understand this property, it is convenient to rewrite the mass
formula in a form that only depends on the asymptotic scalars, the charges, and the auxiliary vector $\ve$.
This vector, defined such that it satisfies to \eqref{NoRstar}, can then be understood as parametrising the
$n_v-1$ flat directions associated to the charge vector $\Gamma$.

To this end, it is convenient to introduce some shorthand notation, that will be used in the remainder
of this section. First, we define one complex and one real variable
\begin{equation} \label{u-defs}
\vu \equiv ( \vt + \ve^{-1} )^{-1} \ ,
\qquad
\vpe \equiv  \vp  + \ve^{-1} p^0\,,
\end{equation}
which appear in all expressions involving $W$. Here,  the inverse is the Jordan inverse
$\ve^{-1} = \frac{ \ve \times \ve}{\det \ve}$ (and similarly for $(\vt + \ve^{-1})^{-1}$). It is important to note that these variables only depend on the moduli $\vt$, the electromagnetic charge $\Gamma$ and the parameter $\ve$.
Using these objects, one computes indeed that $W(\Gamma)$ can be rewritten as
\begin{align}
 W(\Gamma) =&\,
\frac{|\det(\vt + \ve^{-1})|}{ \sqrt{i \det (\vt - \bar \vt) } } \biggl(
\tr\bigl[ \vu \times \bar{\vu}\, \vl_\Gamma \bigr]
- p^0
+ \frac12\,\tr\bigl[ \scal{ \vu + \bar{\vu} } \vpe  \bigr]
 \biggr)\ ,  \label{FakeSuperpotential}
\end{align}
where $\vl_\Gamma$ is given by \eqref{lGamma}. Note that this expression is linear in the charge $\Gamma$. In this form, the fake superpotential reproduces
the formula derived in \cite{Ceresole:2009vp}, where the vector $\ve$ satisfying to \eqref{NoRstar}
parametrises the $n_v-1$ flat directions associated to the charge $\Gamma$. Moreover, this
parametrisation of the flat directions exhibits the similarity of the fake superpotential and the
central charge in this basis. The latter can be shown to take the form
\begin{align}
 |Z(\Gamma)| =&\, \biggl|
\frac{\det(\vt + \ve^{-1})}{ \sqrt{ i \det (\vt - \bar \vt) } } \biggl(
\tr\bigl[ \vu \times \vu\, \vl_\Gamma \bigr]
- p^0
+ \,\tr \vu \,\vpe
 \biggr) \biggr| \ ,  \label{eCentralCharge}
\end{align}
in this basis, using the constraint $\Iprod{\Gamma}{\hat R}=0$. Note that the explicit dependence
of \eqref{eCentralCharge} on $\ve$ is due to exactly this constraint, and can be eliminated by
rewriting $\vl_\Gamma$ and $\vpe$ in terms of the charges.

According to \cite{Ceresole:2009vp}, $\ve$ must be such that it extremises $W(\Gamma,\ve)$, with respect to variations preserving \eqref{NoRstar}.
In order to check this property we compute the variation of $W$ with respect to $\ve^{-1}$
while keeping the charge and the moduli fixed. The variation of $\vu$ following from
\eqref{u-defs} reads
\begin{equation}
 \delta \vu = 2 ( \vu \times \vu ) \times \delta \ve^{-1} - \vu \tr \vu \delta \ve^{-1}\,,
\end{equation}
so that we obtain
\begin{eqnarray}
 \delta W &=& \frac{|\det(\vt + \ve^{-1})|}{ \sqrt{i \det (\vt - \bar \vt) } } \biggl( 2 \tr \Scal{ \scal{ ( \vu \times \vu ) \times \delta\ve^{-1} } \bar \vu \times \vl_\Gamma } + 2 \tr \Scal{ \scal{ ( \bar \vu \times \bar \vu ) \times \delta\ve^{-1} } \vu \times \vl_\Gamma } \biggr . \CR
&& \hspace{40mm}  - \frac{1}{2} \tr \scal{ ( \vu + \bar \vu ) \delta \ve^{-1} } \ \tr \scal{ \vu \times \bar \vu \, \vl_\Gamma } \\
&& \hspace{10mm}  \biggl . + \tr\scal{  ( \vu - \bar \vu ) \times ( \vu - \bar \vu )\, \vpe \times \delta \ve^{-1} } - \frac{1}{4} \tr \scal{ ( \vu - \bar \vu ) \delta \ve^{-1}}  \tr \scal{ ( \vu - \bar \vu ) \vpe}  \biggr)\,. \nn
\end{eqnarray}
For the single-centre solution, one computes that this variation reduces to
\begin{equation}\label{W-first}
 \delta W = \frac{\det \vl }{2\sqrt{2}} \biggl(  \det \Scal{ \vk_\Gamma + m \frac{\vl}{\det \vl}} + \tr \frac{ \vl\times\vl}{(\det\vl)^2 } \Scal{ \vk_\Gamma + m \frac{\vl}{\det \vl}} \biggr) \ \tr \vl_\Gamma \delta\ve^{-1} \ ,
\end{equation}
which indeed vanishes for $\delta \ve^{-1}$ preserving the condition \eqref{NoRstar}, that $\hat{R}$
mutually commutes with the charge, \ie
\begin{equation}
 \delta\Iprod{\hat R}{\Gamma}=0\quad \Rightarrow \quad \tr \vl_\Gamma \delta\ve^{-1} =0\,,
\end{equation}
in agreement with  \cite{Ceresole:2009vp}. Note that it is important \cite{Ceresole:2009vp}, that the flat directions parameter extremises the fake superpotential for arbitrary moduli, and so the reader may worry that we only check this variation within the solution. But note that the asymptotic scalars are completely arbitrary in this solution, and so this is perfectly consistent.

In the construction of \cite{Ceresole:2009vp} it is also important that these extrema are unique, so as to fix unambiguously the expression of the fake superpotential in terms of the charge $\Gamma$ and the moduli. To check this, we can simply consider the moduli to be parametrised by \eqref{AsymptoticModuli} with $\vk$ arbitrary and not necessarily equal to $\vk_\Gamma$, in which case $\ve$ would extremise $W$ as we just explained. One computes that the condition that $\delta W$ is proportional to $\tr \vl_\Gamma \delta \ve $ gives 
\begin{equation} \Scal{ 4 ( \vl \times \vl ) \times \scal{ ( \vk - \vk_{\Gamma})  \times \vl_{\Gamma}  }
- \vl\, \tr \vl \,  ( \vk - \vk_{\Gamma})  \times \vl_{\Gamma}}
=     \det \vl \ \gamma \vl_{\Gamma} \ , \end{equation}
for some arbitrary Lagrange multipliers $\gamma$.  Using the property that
$\vl$ is positive, one can simplify this equation to
\begin{equation}  ( \vk - \vk_{\Gamma} - \gamma \vl ) \times \vl_{\Gamma} = 0 \  , \end{equation}
which because $\vl_\Gamma$ is also positive, reduces to
\begin{equation} \vk =  \vk_{\Gamma} + \gamma \vl  \ . \end{equation}
Since the term in $\gamma$ can always be reabsorbed in a redefinition of $m$ as in \eqref{sing-complicated}
without affecting $\ve$, we find that the unique solution for $\ve$ is indeed the expression it takes for
a single-centre solution.

Beyond the first order variation \eqref{W-first}, it is important for the multi-centre applications that follow
to consider the second variation of $W$ as well, as it turns out to be crucial in comparing the mass of a
composite to that of its constituents. In the remainder of this section, we compute explicitly the Hessian of
$W$ at its extremum, viewed as a function of $\ve$, imposing the constraint that this vector is such that
$\Iprod{\Gamma}{\hat R}=0$. We find that the resulting quadratic form is negative definite along
all directions preserving the constraint, in an open set in moduli space for general charges, so that one can extend the
result to the full moduli space by duality. Because the extremum is unique, it follows that it is moreover a global maximum. The result that the extremum of $W$ is moreover a global maximum is
crucial in defining a generalisation of the triangular identity for BPS black holes, which states that
the mass of a composite is always lower than the masses of its constituents.
However, the details of the proof are technical and not directly relevant for the
remainder of this paper, so that they can be skipped by a hasty reader. 

In order to prove that the second derivative of $W$ is a negative definite quadratic form, we consider the explicit form of the latter, which reads
\begin{eqnarray}
 \delta^2 W &=&  \frac{|\det(\vt + \ve^{-1})|}{ \sqrt{ i \det (\vt - \bar \vt) } } \biggl( 2 \tr \Bigl[ ( \delta \ve^{-1} \times \delta \ve^{-1} ) \times \vl_\Gamma \, \Scal{ \det \vu \, \bar \vu - 2 ( \vu \times \vu ) \times ( \bar \vu \times \bar \vu ) + \det \bar \vu \, \vu }\Bigr]  \biggr . \CR
&& \hspace{10mm} + \tr \bigl[ \delta \ve^{-1} \times \delta \ve^{-1} \, \vu \times \vu \bigr] \ \tr \bar \vu \times \bar \vu \, \vl_\Gamma+ \tr \bigl[  \delta \ve^{-1} \times \delta \ve^{-1} \, \bar \vu \times \bar \vu \bigr] \ \tr  \vu \times  \vu \, \vl_\Gamma  \CR
&& \hspace{35mm} - \tr \bigl[  \delta \ve^{-1} \times \delta \ve^{-1} \, \ ( \vu \times \vu + \bar \vu \times \bar \vu ) \bigr]  \ \tr \vu \times \bar \vu \, \vl_\Gamma \CR
&& \hspace{10mm} - 2 \tr\bigl[  ( \vu + \bar \vu ) \delta \ve^{-1}\bigr]  \,\tr \bigl[ \delta \ve^{-1}  \scal{ ( \vu \times \vu ) \times ( \bar \vu \times \vl_\Gamma) + ( \bar  \vu \times\bar  \vu  ) \times (   \vu \times \vl_\Gamma ) } \bigr] \CR
&& \hspace{10mm} + \tr [\vu \times \bar \vu \, \vl_\Gamma] \, \Scal{ \frac{1}{2} \scal{ \tr \vu \delta \ve^{-1} }^2 + \frac{1}{4} \scal{ \tr ( \vu + \bar \vu ) \delta \ve^{-1} }^2 + \frac{1}{2}  \scal{ \tr \bar \vu \delta \ve^{-1} }^2} \CR
&& \hspace{5mm} + 4 \tr \bigl[ ( \delta \ve^{-1} \times \delta \ve^{-1} ) \times \bar \vu \ ( \vu \times \vu ) \times \vpe \bigr] + 4 \tr\bigl[  ( \delta \ve^{-1} \times \delta \ve^{-1} ) \times  \vu \ ( \bar \vu \times \bar \vu ) \times \vpe\bigr]   \CR
&& \hspace{10mm} + \tr \bigl[ \delta \ve^{-1} \times \delta \ve^{-1}\, \vpe\bigr]  \ \Scal{ \det \vu - \tr \vu \times \vu \, \bar \vu - \tr \vu \, \bar \vu \times \bar \vu + \det \bar \vu } \CR
&& \hspace{10mm} - \frac{1}{2} \tr  \bigl[ \delta \ve^{-1} \times \delta \ve^{-1}\, ( \vu \times \vu + \bar \vu \times \bar \vu ) \bigr] \ \tr ( \vu + \bar \vu ) \vpe \CR
&& \hspace{10mm} - \tr\bigl[  ( \vu - \bar \vu ) \times ( \vu -\bar \vu ) \ \vpe \times \delta \ve^{-1}\bigr]  \ \tr ( \vu + \bar \vu ) \delta \ve^{-1} \CR
&& \hspace{5mm}  + \frac{1}{2} \scal{ \tr \vu \delta \ve^{-1} }^2 \, \tr \vu \vpe - \frac{1}{8} \scal{\tr (\vu + \bar \vu )  \delta \ve^{-1} }^2 \, \tr (\vu + \bar \vu ) \vpe+ \frac{1}{2} \scal{\tr \bar  \vu \delta \ve^{-1} }^2 \, \tr \bar \vu \vpe \CR
&& \hspace{10mm}  \biggl . + p^0  \Scal{ \tr \bigl[ ( \vu - \bar \vu ) \times ( \vu - \bar \vu ) \ \delta\ve^{-1} \times \delta \ve^{-1}\bigr]  - \frac{1}{4} \scal{ \tr ( \vu - \bar \vu ) \delta \ve^{-1} }^2 } \biggr)  \ .
\end{eqnarray}
Substituting the single-centre expression one obtains
\begin{eqnarray}
 \delta^2 W &=&  - 2 \sqrt{2} \frac{1+m^2 } { ( \det \vl )^2 } \tr \vl \times \vl_\Gamma \,  \delta \ve^{-1} \times \delta \ve^{-1} +\sqrt{2}  \frac{m}{\det \vl} \tr \vpe \,  \delta \ve^{-1} \times \delta \ve^{-1} \CR&& \qquad + \frac{W}{( \det \vl)^2 } \Scal{ 4 \tr \vl \times \vl \,  \delta  \ve^{-1} \times \delta \ve^{-1} - \scal{ \tr \vl \delta \ve^{-1}}^2 }  \CR
&& -  \frac{1}{ \sqrt{ i \det (\vt - \bar \vt) } } \frac{ \tr ( \vu + \bar \vu ) \vu \times \bar \vu }{|\det \vu|}  \ \delta^2 \frac{ \det \ve  \, q_0 - \tr \ve \times \ve\,  \vq +  \tr\ve   \vp + p^0}{\det \ve} \CR
&\approx&  - 2\sqrt{2}  \frac{1+m^2 } { ( \det \vl )^2 } \tr \vl \times \vl_\Gamma \, \delta \ve^{-1} \times \delta \ve^{-1} + \sqrt{2} \frac{m}{\det \vl} \tr \vpe \,  \delta \ve^{-1} \times \delta \ve^{-1} \CR&& \qquad + \frac{W}{( \det \vl)^2 } \Scal{ 4 \tr \vl \times \vl \,  \delta  \ve^{-1} \times \delta \ve^{-1} - \scal{ \tr \vl \delta \ve^{-1}}^2 }  \ , \label{SecondDerW}
\end{eqnarray}
where in the second equality we neglected the component that vanishes assuming that the variation of
$\ve$ preserves $\Iprod{\hat{R}}{\Gamma}=0$. We shall prove that the above defined quadratic form is negative definite for appropriate variations of $\ve^{-1}$ preserving this constraint. However it is generally not negative definite for arbitrary variations $\delta \ve^{-1}$, therefore it is important to take the constraint into account.

In order to proceed, it turns out that a change of variable from $\ve$ to $\vk$ is useful,
where $\vk$ is the arbitrary vector parametrising the asymptotic scalars as in \eqref{AsymptoticModuli}.
Because we consider the variation of $\ve^{-1}$ at fixed moduli, the variation $\delta \ve^{-1}$ is
determined by the corresponding variation of $\vk$ such that \eqref{AsymptoticModuli} is kept constant.
For $\vk \ne \vk_\Gamma$, one can always find the corresponding $\vl^\prime,\, \ve^\prime,\, m^\prime$ such that
\begin{equation}  \Scal{ \frac{ m - i }{\det \vl} \vl + \vk  }^{-1} - \ve^{-1} =  \Scal{ \frac{ m^\prime - i }{\det \vl^\prime} \vl^\prime + \vk_{\Gamma}(\ve^\prime)  }^{-1} - \ve^{\prime\, -1} \ . \end{equation}
For infinitesimal variations of the parameters in the vicinity of $\vk = \vk_\Gamma$, one obtains
\begin{equation} 2 (   \vu^{-1} \times \vu^{-1} )  \times \delta \vk - \vu^{-1}   \tr \vu^{-1}   \delta \vk  =  2 ( \vu^{-1} \times \vu^{-1} ) \times \delta \vu - \vu^{-1} \tr \vu^{-1}  \, \delta \vu - \delta \ve^{-1} \ , \end{equation}
where
\begin{equation} \delta \vu =  \frac{\vl}{\det \vl } \delta m + ( m-i ) \delta \frac{ \vl }{\det \vl } + \frac{ \partial \vk_\Gamma}{\partial \ve^{-1}} \cdot \delta \ve^{-1}  \ . \end{equation}
Note that the variation of $\vk_\Gamma$ is required because the correct $\vk_\Gamma$ as a function of $\ve$ is evaluated at  $\ve + \delta \ve$ at this order. It is convenient to rewrite this equation as
\begin{equation} \delta \vk -  \frac{\vl}{\det \vl } \delta m- ( m-i ) \delta \frac{ \vl }{\det \vl } = - 2 ( \vu \times \vu ) \times \delta \ve^{-1} + \vu  \tr \vu  \, \delta \ve^{-1} + \frac{ \partial \vk_\Gamma}{\partial \ve^{-1}} \cdot \delta \ve^{-1}  \ . \label{ConstantModuli} \end{equation}
One can compute the variation of $\vk_\Gamma$ as
\begin{equation}  \frac{ \partial \vk_\Gamma}{\partial \ve^{-1}} \cdot \delta \ve^{-1}  = 2 ( \vk_\Gamma \times \vk_\Gamma ) \times \delta \ve^{-1}  - \vk_\Gamma \tr \vk_\Gamma \, \delta \ve^{-1} - \frac{1}{2}  \frac{I_4(\Gamma)}{(\det \vl_\Gamma)^2 } ( \vl_\Gamma \times \vl_\Gamma ) \times \delta \ve^{-1} \ , \end{equation}
such that the terms quadratic in $\vk_\Gamma$ cancel in \eqref{ConstantModuli}  after substituting $\vu = \frac{m-i}{\det \vl} \vl + \vk_\Gamma$.
Decomposing \eqref{ConstantModuli} into its imaginary and real components one obtains
\begin{multline}  \delta \frac{ \vl }{\det \vl} = \frac{ 4}{\det \vl } ( \vl \times \vk_\Gamma) \times \delta \ve^{-1} - \frac{ \vl}{\det \vl } \tr \vk_\Gamma \, \delta \ve^{-1} - \frac{ \vk_\Gamma}{\det \vl } \tr \vl \, \delta \ve^{-1}\\ + \frac{ 2 m }{ (\det \vl)^2 } \scal{ 2 ( \vl \times \vl ) \times \delta \ve^{-1} - \vl \tr \vl \, \delta \ve^{-1}} \ , \end{multline}
and
\begin{equation} \delta \vk - \frac{\vl}{\det \vl } \delta m = - \frac{1}{2}   \frac{I_4(\Gamma)}{(\det \vl_\Gamma)^2 } ( \vl_\Gamma \times \vl_\Gamma ) \times \delta \ve^{-1} + \frac{1  + m^2 }{( \det \vl )^2 } \scal{ 2 ( \vl \times \vl ) \times \delta \ve^{-1} - \vl \tr \vl \, \delta \ve^{-1}} \ ,\label{kgivese} \end{equation}
where $\delta m$ is itself determined such that $\tr \vl_\Gamma \, \delta \ve^{-1} = 0 $.
The last two expressions exhibit that $\delta \ve^{-1}$ and $\delta \vl$ are completely determined by the
variation $\delta \vk$. The second formula in particular provides the required change of variable from
$\delta \ve^{-1}$ to $\delta \vk$. As a variation of $\delta \vk$ proportional to $\vl$ can be reabsorbed
in a redefinition of $m$ (cf. \eqref{sing-complicated}), one can restrict attention to the variations
of $\vk$ linearly independent of $\vl$.  This parametrisation of the flat directions in terms of $\vk$
is more useful because we can forget about the constraint, which now simply determines the decomposition
of \eqref{kgivese} in the variations $\delta\ve^{-1}$ and $\delta m$. Moreover, it is the variation
$\delta \vk$ rather than $\delta \ve^{-1}$ that will appear explicitly in the two-centre solutions as 
will be shown shortly.

The quadratic form written in terms of $\delta \vk$ is a rather complicated expression in general, and we shall only consider the limit of small and large $\vl$. These limits both correspond to asymptotic scalars that have order one axions and very large dilatons. Note however that we do not consider any restriction on the electromagnetic charges, so it is enough to prove the result on an open set in moduli space to ensure that this property holds in general, given that the mass formula is duality invariant and the duality group $G_4$ acts transitively on the moduli space. The only restriction on our representation arises from singularities at regions where $\det \ve=0$, but the corresponding configurations define isolated points in moduli space. Therefore, showing that the extrema of $W(\Gamma,\ve)$ are maxima on an open set in moduli space for all charges, is enough to circumvent this issue. The reader might note that the domain of large dilatons in moduli space corresponds to small volume moduli in string
theory, and the supergravity approximation cannot be trusted in this limit. However this is only a technical detail at this level, since the proof extends by duality to all values of the moduli.

\subsubsection*{Small $\vl$}

For very small $\vl$, the expression of  $\delta \ve^{-1}$ simplifies to
\begin{equation}  \delta \ve^{-1}  = 2  \frac{ \det \vl }{1 + m^2 } \Scal{ \vl \times \delta \vk - \vl \times \vl \frac{ \tr \vl  \, \vl_\Gamma \times \delta \vk }{\tr \vl \times \vl \, \vl_\Gamma}} + \mathcal{O}(\vl^8)  \ , \end{equation}
and the quadratic form reads
\begin{eqnarray}
 \delta^2 W &=& \frac{1}{2\sqrt{2}} \frac{1 + m^2}{( \det \vl)^3} \biggl( \tr \vl \times \vl \, \vl_\Gamma \Scal{  4 \tr \vl \times \vl\, \delta\ve^{-1} \times \delta \ve^{-1} - \scal{ \tr \vl \delta \ve^{-1} }^2 } \biggr . \CR
 && \hspace{55mm} \biggl . - 8\,  \det \vl \, \tr \vl \times \vl_\Gamma \, \delta \ve^{-1} \times \delta \ve^{-1} \biggr) + \mathcal{O}(\vl^{-3})   \CR
&=&\sqrt{2}  \frac{ \det \vl }{1 + m^2} \biggl( \tr \vl_\Gamma \, \delta \vk \times \delta \vk - \frac{ (\tr \vl \, \vl_\Gamma \times \delta \vk )^2}{\tr \vl \times \vl \, \vl_\Gamma}  \biggr)   + \mathcal{O}(\vl^5) \ ,
\end{eqnarray}
which can be shown to be negative for all $\delta \vk$. To prove this one can use the $G_5$ invariance of this equation to chose both $\vl$ and $\vl_\Gamma$ to be diagonal Jordan algebra elements sitting in the STU truncation. Regularity of the solution then requires all components of $\vl$ and $\vl_\Gamma$ to be strictly positive. This permits to rewrite
\begin{eqnarray}
 \delta^2 W &=& -\frac{1 }{ 2 W(l_2 l_{\Gamma 1} + l_1 l_{\Gamma 2} )^2 }  \biggl(    \det \vl_\Gamma \, \tr \vl \times \vl \,  \vl_\Gamma  \, ( l_2 \delta k_1 - l_1 \delta k_2)^2 \CR
& &  + \frac{1}{2\sqrt{2}} \big[(l_2 l_{\Gamma 1} + l_1 l_{\Gamma 2} )^2  \delta k_3 + \tr \vl \times \vl \, \vl_\Gamma \, ( l_{\Gamma 1 } \delta k_2 + l_{\Gamma 2} \delta k_1 )
 \CR
&& \hspace{6cm} 
- l_{\Gamma 3} ( l_1^{\; 2} l_{\Gamma 2 } \delta k_2  + l_2^{\; 2} l_{\Gamma 1} \delta k_1 ) \big]^2 \biggr)
\CR
&& - \frac{1}{\sqrt{2}} \frac{\det \vl}{1 + m^2 } \Scal{\tr \vl_\Gamma (\delta \vk )^2 - \sum_i l_{\Gamma i}  (\delta k_i)^2 }  + \mathcal{O}(\vl^5)  \ ,
\end{eqnarray}
which is manifestly negative. The last line shows that the non-diagonal components of $\delta \vk$ necessarily contribute negatively to $\delta^2 W$ and $(\delta \vk)^2$ is the Jordan square of $\delta \vk$, \ie as a matrix in an explicit basis.
For completeness, we note that this region corresponds to small volume moduli and finite axions
\begin{equation} \vt  =  \frac{ \vl \times \vl}{m-i} - \ve^{-1}  + \mathcal{O}(\vl^2) \ . \end{equation}

\subsubsection*{Large $\vl$}

 In this case $\delta \ve^{-1}$ reduces to
\begin{equation} \delta \ve^{-1} =  - 8 \frac{\det \vl_\Gamma}{I_4(\Gamma)} \, \vl_\Gamma \times \Scal{ \delta \vk - \vl \frac{ \tr \vl_\Gamma \times \vl_\Gamma \, \delta \vk}{\tr \vl\, \vl_\Gamma \times \vl_\Gamma} }  + \mathcal{O}(\vl^{-4})  \ . \end{equation}

After some algebra one obtains the perturbation of the mass to be
\begin{eqnarray} \label{d2WlargeL}
\delta^2 W &=&4 \sqrt{2}  \frac{ \det \vl_\Gamma}{-  I_4(\Gamma_\pA) \det \vl} \biggl( - 2 \tr \vl \times \vl  \,  ( \vl_\Gamma \times  \vl_\Gamma  ) \times ( \delta \vk \times  \delta \vk  ) - \det \vl \frac{ ( \tr \vl_\Gamma \times \vl_\Gamma \, \delta \vk )^2 }{\tr \vl\, \vl_\Gamma \times \vl_\Gamma} \biggr . \\ && \biggl . + \tr \vl \times \vl \, \delta \vk \ \tr   \vl_\Gamma \times \vl_\Gamma \, \delta \vk+ \tr \vl \times \vl \, \vl_\Gamma \ \tr   \delta \vk \times \delta \vk \, \vl_\Gamma - ( \tr \vl \, \vl_\Gamma \times \delta \vk)^2 \biggr)  + \mathcal{O}(\vl^{-3}) \ .  \nn \end{eqnarray} 
Although this is not manifest in this equation, it is possible to show that \eqref{d2WlargeL} is always strictly negative, in a similar way as shown for the case of very small $\vl$, provided that $\delta \vk$ is not proportional to $\vl$ (in which case  $\delta \ve^{-1}$ itself vanishes). To do this we restricted ourselves to the STU truncation, requiring that all components of $\vl, \vl_\Gamma$ are strictly positive.
This region also corresponds to small volume moduli with finite axions
\begin{equation} \vt  =  \vk_\Gamma^{\;-1} - \ve^{-1} + \frac{-m+i}{\det \vl (\det \vk_\Gamma)^2 } \Scal{ \vk_\Gamma \times \vk_\Gamma \, \tr \vk_\Gamma\times \vk_\Gamma \, \vl -  2 \vk_\Gamma \times \vl \, \det \vk_\Gamma }  + \mathcal{O}(\vl^{-4})\ . \end{equation}

\subsection{The multi-centre solutions}
\label{sec:multi-cent-expl}

We now turn to multi-centre solutions, using the same parametrisation of charges and moduli
as for the single-centre solutions above. The scalar section and the gauge fields are now
given by \eqref{scals-5-b}-\eqref{dw-bas}, where the harmonic functions $\vK$ parametrising the
T-dualities are now nontrivial. In the preferred basis of the previous section, we consider
a system of $N$ centres, labeled by an index $\mathrm{A}$. Then, \eqref{dw-bas} implies that
all charges commute with the vector $\hat R$, so that all $P_\pA=0$ in the decomposition
\eqref{eq:ch-dec-basis-gen}. Of the remaining components, it turns out that only the
$\vl_{\Gamma_\pA}$ appear in the various expressions, since we have
\begin{eqnarray}
 \vL &=&  \vl + \sqrt{2} \sum_\pA \frac{\vl_{\Gamma_\pA} }{r_\pA} \ , \CR
\vK &=& \vk + \sqrt{2}  \sum_\pA \gamma_\pA \frac{\vl_{\Gamma_\pA} }{r_\pA}\ ,\label{LKdef}
\end{eqnarray}
where $\vk$, $\vl$ and $\gamma_\pA$ are constants and $r_\pA=|x-x_\pA|$ is the distance from
centre $\mathrm{A}$. We stress here that regularity imposes that the poles of $\vK$ and $\vL$
be linearly dependent at each centre (see the constraint \eqref{keql-5-multi} ).
Given these expressions, the solutions to \eqref{V-Poiss-b}-\eqref{eq:Poiss-M-b}
are given by \eqref{V-gen}, \eqref{M-gen} and \eqref{omega-gen} upon substituting the
explicit expression \eqref{V-L-def} for $\cHz$, leading to
  \begin{multline} \label{V-frame}
V = \frac{1  + m^2}{\det \vl }  +  \sqrt{2} \sum_\pA   \frac{ - p^0_A  + \tr \bigl[ (  {\bf k}- 2 {\bf k}_{\Gamma_A}  ) ({\bf k} \times {\bf l}_{\Gamma_\pA} )\bigr]   }{r_\pA}
- 2 \sum_\pA \gamma_\pA \frac{ \J_{\pA i } r_\pA^i}{ r_\pA^3}\\  
+ \sum_\pA   \gamma_\pA^{\; 2} \biggl(2 \sqrt{2}  \frac{ \det[\vl_{\Gamma_\pA}] }{r_\pA^3}  +2  \frac{\tr {\bf l} \, \vl_{\Gamma_\pA} \times \vl_{\Gamma_\pA} }{r_\pA^2}  \biggr )
+2  \sum_{\pA \ne \pB} \gamma_\pA \gamma_\pB \frac{\tr  {\bf l} \, \vl_{\Gamma_\pA} \times \vl_{\Gamma_\pB} }{r_\pA r_\pB}  \\
 +2 \sqrt{2}  \sum_{\pA\ne \pB} \gamma_\pA \tr  \vl_{\Gamma_\pB} \, \vl_{\Gamma_\pA} \times \vl_{\Gamma_\pA} \biggl( \frac{ \gamma_\pB}{r_\pA^2 r_\pB} + \frac{\gamma_\pA-\gamma_\pB}{\RABsq} \Bigl( \frac{ r_\pB}{r_\pA^2} - \frac{1}{r_\pB}\Bigr) \biggr) 
\\
-4 \sqrt{2}   \sum_{\pA\ne \pB \ne \pC} \gamma_\pA \gamma_\pB  \tr  \vl_{\Gamma_\pA} \, \vl_{\Gamma_\pB} \times \vl_{\Gamma_\pC} \Bigl( F_{\pA,\pBC} + \frac{1}{\RAC\,\RBC\, r_\pC} \Bigr)
  \\
  + 2 \sqrt{2}   \sum_{\pA\ne \pB \ne \pC} \gamma_\pA \gamma_\pB  \frac{\tr  \vl_{\Gamma_\pA} \, \vl_{\Gamma_\pB} \times \vl_{\Gamma_C}} {r_\pA r_\pB r_\pC}
 \end{multline}
and
 \begin{multline} \label{M-frame}
M = m  +  \sum_\pA \gamma_\pA \biggl( \det \vl - \det \Scal{ \vl + \sqrt{2 } \frac{ \vl_{\Gamma_\pA}}{r_\pA}} \biggr)  + \sum_\pA \frac{ \J_{\pA i } r_\pA^i}{ r_\pA^3}
\\ 
-2  \sum_{\pA \ne \pB} \gamma_\pA \tr  {\bf l} \, \vl_{\Gamma_\pA} \times \vl_{\Gamma_\pB} \biggl( \frac{ 1}{r_\pA r_\pB} + \frac{1}{\RAB\, r_\pA} - \frac{1}{\RAB\,r_\pB} \biggr) \\
 -\sqrt{2}  \sum_{\pA\ne \pB} \tr  \vl_{\Gamma_\pB} \, \vl_{\Gamma_\pA} \times \vl_{\Gamma_\pA} \biggl( \frac{ \gamma_\pA + \gamma_\pB}{r_\pA^2 r_\pB} + \frac{\gamma_\pA-\gamma_\pB}{\RABsq} \Bigl( \frac{ r_\pB}{r_\pA^2} - \frac{1}{r_\pB}\Bigr) \biggr) 
\\
 -2 \sqrt{2}  \sum_{\pA\ne \pB \ne \pC} \gamma_\pC \tr  \vl_{\Gamma_\pA} \, \vl_{\Gamma_\pB} \times \vl_{\Gamma_\pC} \Bigl( F_{\pA,\pBC} + \frac{1}{\RAC\, \RBC\,r_\pC} \Bigr)
 \end{multline}
In these expressions, $\J_{\pA i }$ is the intrinsic `under-rotating' angular momentum at each centre, $\RAB$ is the distance
between the centres labeled by $A$ and $B$, while the function $F_{\pA,\pBC}$ was defined in \cite{Bossard:2012ge} as
the everywhere regular solution to \eqref{F-triple-def}. We refer to that work for the properties of this function.

Using these explicit functions, one can now write down the scalar fields and the metric, using 
\eqref{Gen-Moduli}-\eqref{ScallingFactorInvariant}. In addition, one can readily understand the
property that the poles of $\vK$ and $\vL$ must be linearly dependent at each centre, by considering
the explicit expression for the scalar fields at the horizons
\begin{equation}
 \vt(x_\pA)   = \exp\Scal{ \cTgm_{ - \frac{ \ve \times \ve}{\det \ve}}}  \exp\scal{ \cTgp_{\vK(x_\pA)-\gamma_\pA \vL(x_\pA)   }} \frac{ 2 \, \vl_{\Gamma_\pA} \times \vl_{\Gamma_\pA}}{\J_\pA \cos \theta_\pA - i \sqrt{- I_4(\Gamma_\pA) - \J_\pA^2 }} \ , \label{AttractorModuli}   
\end{equation}
which leads to the requirement that $\vK-\gamma_\pA \vL$ must be regular at $x_\pA$.
The finite values of these functions at each horizon define a set of T-duality parameters that play an
important role in the definition of the electromagnetic charges.

In the multi-centre case, the charges at the various centres are allowed to have nontrivial
grade $(+1)$ components, but are still constrained to have a vanishing grade $(-3)$ component,
as \eqref{dw-bas} commutes with $\hat R$. It follows that the most general charge allowed in each
centre is given by \eqref{eq:ch-dec-basis}. In the explicit parametrisation of the previous section, the additional
components can be computed by decomposing all charges as in \eqref{ch-full-5-0}, \ie by viewing
each charge as the result of a T-duality acting on the poles of $\vL$ and $V$. This is conveniently
realised in terms of the functions above, since the expression for the charges at a given centre,
obtained by integrating \eqref{dw-bas}, is indeed given as a T-duality of parameter
$\vK(x_\pA)-\gamma_\pA \vL(x_\pA)$ acting on an underlying vector defined from the poles of
$\vL$ and $V$ at that centre, as
\begin{equation}\label{ch-full-5-0}
 \Gamma_\pA = \exp\Scal{ \cTgm_{ - \frac{ \ve \times \ve}{\det \ve}}}
 \exp\Scal{ \cTgp_{ \vK(x_{\pA})-\gamma_\pA \vL(x_\pA)} }
\left( \begin{array}{c} 0 \\  \vl_{\Gamma_\pA} \\  0 \\
 \frac{I_4(\Gamma_\pA)}{4 \det \vl_{\Gamma_\pA}}
 \end{array} \right) \,,
\end{equation}
which is exactly of the form \eqref{eq:ch-dec-basis} for a vanishing
grade $(+1)$ charge.
Note that, while this equation simply defines the charge for given harmonic functions
$\vK$ and $\vL$, it becomes a nontrivial constraint on the parameters of the solutions if the charges are kept fixed.

The parameters of the T-dualities in \eqref{ch-full-5-0} are the central objects governing
the structure of multi-centre solutions. In order to obtain their value, one can compare
\eqref{ch-full-5-0} to \eqref{sing-charge}, to find that 
\begin{align}
  \vk_{\Gamma_\pA}=&\, \vK(x_\pA)-\gamma_\pA \vL(x_\pA) 
\CR
=&\,
 \vk - \gamma_\pA \vl + \sqrt{2} \sum_{\pB\ne \pA} \frac{\gamma_\pB - \gamma_\pA}{\RAB} \vl_{\Gamma_\pB}   \ , \label{BubbleTframe}
\end{align} 
because the action of T-dualities is faithful on charges carrying a non-zero grad 3 component as does \eqref{ch-full-5-0}. Alternatively, the same result is obtained
by use of the general formula derived in \eqref{d-expr-mult} below, which can be
used in any other basis as well.

Note that \eqref{BubbleTframe} is consistent with the property that $\vk = \vk_\Gamma$ for a single
centre solution, due to \eqref{k-sing} and \eqref{sing-complicated}, since the single-centre limit
of the multi-centre solution naturally leads to a nontrivial $\vK$. We stress that although this
formula is identical to the ones displayed in \cite{Bossard:2011kz,Bossard:2012ge} in a specific
duality frame, the dependence of the vectors $\vl,\ \vk,\ \vk_\Gamma$ and $\vl_\Gamma$ in terms of
the charges and the asymptotic scalars is here manifest. Within the formulation of this paper, we
can therefore keep the charges fixed and rather consider \eqref{BubbleTframe} as a constraint on
the asymptotic scalars.

Although there is no solution for generic charge configurations with more than three centres,
the problem generally admits a solution for two centres. In this case, one can
easily solve \eqref{BubbleTframe} as
\begin{eqnarray}
 \vl &=& -\frac{\sqrt{2}}{\Rot} \scal{  \vl_{\Gamma_1} + \vl_{\Gamma_{2}} } + \frac{1}{\gamma_1-\gamma_2} \scal{  \vk_{\Gamma_2} - \vk_{\Gamma_1}}\ , \label{AsympL}  \\
\vk &=&- \frac{\sqrt{2}}{ \Rot} \scal{ \gamma_1 \vl_{\Gamma_1} + \gamma_2 \vl_{\Gamma_{2}} } + \frac{1}{\gamma_1-\gamma_2} \scal{ \gamma_1 \vk_{\Gamma_2} -\gamma_2 \vk_{\Gamma_1}} \ ,
\label{eq:lk-two-c}
\end{eqnarray}
so that the asymptotic scalars are parametrised by the vector $\ve$ satisfying both
$\Iprod{\hat{R}}{\Gamma_\pA}=0$ for $A=1,2$ (\ie \eqref{NoRstar}), the two proportionality
constants, $\gamma_1,\ \gamma_2$, in \eqref{LKdef} and the distance between the two centres,
denoted by $\Rot$. This sums up to a total of $n_v+1$ parameters for the $2 n_v$ asymptotic
moduli (this holds for $n_v\ge 3$). Note that the parameter $m$ does not count, because it can
always be reabsorbed in a redefinition of $\vk$. This is a more general property that persists
when adding more centres, so that composite non-BPS solutions only exist for moduli constrained
to an (at most) $(n_v+1)$-dimensional subsurface, specified by the charges at the centres.

For the asymptotic scalar fields to be well defined, $\vl$ in \eqref{AsympL} must moreover define
a positive Jordan algebra element. As both $\vl_{\Gamma_1}$ and $ \vl_{\Gamma_{2}}$ must be
positive for the solution to be well behaved at the two horizons, the positive contribution must come
from $\vk_{\Gamma_2}-\vk_{\Gamma_1}$. Regularity therefore requires that
$\vk_{\Gamma_2}-\vk_{\Gamma_1}$ is a strictly positive Jordan algebra element (or strictly negative,
depending on the sign of $\gamma_1-\gamma_2$). Indeed, one can always find a $G_5$ element that
rotates $\vl_{\Gamma_1}+ \vl_{\Gamma_{2}}$ to a Jordan algebra element proportional to the identity.
The group $K_5$ defined as leaving the identity element invariant then permits to rotate
$\vk_{\Gamma_2}-\vk_{\Gamma_1}$ to a diagonal Jordan algebra element. For the three eigenvalues of
$\vl$ to all be positive, it is then clear that the three eigen values of
$(\gamma_1-\gamma_2)^{-1}(\vk_{\Gamma_2}-\vk_{\Gamma_1})$ must themselves be strictly positive.

A final aspect of the multi-centre solutions worth discussing is the issue of flat directions.
As is well known, the scalar fields of single-centre solutions admit $n_v-1$ flat directions, in
the sense that $n_v-1$ of the $2\,n_v$ scalar fields are undetermined constants throughout the flow.
In the description of the previous subsection, one can easily check that the $n_v-1$ parameters in
$\ve$ account for exactly these flat directions. These directions can be also viewed as the invariance
group of the non-BPS charge vector, embedded in the duality group \cite{Bellucci:2006xz}. These are the flat directions of the individual centres, whereas the flat directions associated to the common stabilizer of the two charges are much more restricted if not trivial. The latter define the actual flat directions of the solution, and we shall discuss them latter.

The scalar fields at the horizon $x_\pA$, \eqref{AttractorModuli}, are still
localised at the non-BPS attractor of the corresponding charge $\Gamma_\pA$, and therefore only
know about the global structure of the solution through the explicit expression of their own
flat directions parametrised by $\ve$. In this case, $\ve$ is not determined by the
charge $\Gamma_\pA$ of the centre at hand and the asymptotic moduli only, as it would be in the single
centre solution, but is determined by the property that $\ve$ extremises the fake superpotential
$W(\sum_\pA \Gamma_\pA,\ve)$ with respect to the variations leaving invariant all the constraints
$\Iprod{\hat{R}}{\Gamma_\pA}=0$ for all centres $A=1,\, N$. Through this property, $\ve$ is in fact
a function of the asymptotic scalars and all the charges $\Gamma_\pA$. This vector does
not depend explicitly on the distances between the centres, although the latter are eventually
determined in terms of the asymptotic moduli and the individual charges themselves.

Given the above, the possibility of genuine flat directions for multi-centre solutions is not excluded.
This turns out to depend on the values of the charges at the centres, as we show explicitly in
section \ref{sec:two-cent-ex} for a two-centre example of restricted charges, while in appendix
\ref{app:stabilizer} we discuss the classification of the allowed flat directions for a two-centre
configuration of generic charges.

\subsubsection{Binding energy of composite states}

One of the most important advantages of obtaining explicit general multi-centre black hole solutions,
as we have done in this paper, is the possibility of studying the binding energy of the constituents.
Indeed, showing that the solutions obtained are genuine bound states, rather than collections of
marginally interacting objects, could be useful in the study of the non-BPS bound states at the microscopic level.

The energy of a composite solution is defined by the standard ADM expansion of the metric at infinity,
which can be carried out in the general multi-centre case. For the solutions in the previous section,
this involves expanding the metric scale function in \eqref{ScallingFactorInvariant} near infinity,
using the expressions \eqref{V-frame}-\eqref{M-frame}. The resulting expression, in terms of the
parameters introduced above, takes the form
\begin{multline}  W\scal{ \Gamma_\pA|_{\pA=1}^{N} } = \frac{1}{2\sqrt{2}} \sum_{\pA=1}^N \Bigl(  ( 1 + m^2 )  \frac{\tr \vl \times \vl \, \vl_{\Gamma_\pA} }{\det \vl} -\frac{1}{4}  \frac{ \det \vl}{\det \vl_{\Gamma_\pA}} I_4({\Gamma_\pA})   \\ + \det \vl \, \tr \vl_{\Gamma_\pA} \scal{ 2 \gamma_\pA \vl \times \vk - \vk \times \vk  + \vk_{\Gamma_\pA} \times \vk_{\Gamma_\pA} } + 2 m \gamma_\pA \tr \vl \times \vl \, \vl_{\Gamma_\pA} \Bigr)\,. \label{EnergyMulticentre}   \end{multline}
Note that the first line of this expression is simply the sum of the single-centre mass formulae
\eqref{W-single} for each centre. In the non-interacting limit, where all charges mutually
commute, all $\gamma_\pA$ are equal and  $\vk=\vk_\Gamma + \gamma \vl$, so that the second
line can be reabsorbed into a redefinition of $m$, as in \eqref{sing-complicated}.

While the structure of \eqref{EnergyMulticentre} is suggestive of a nontrivial binding energy, verifying
this directly is rather complicated in general. Using the parametrisation \eqref{AsymptoticModuli}, one shows that \eqref{EnergyMulticentre} can still be rewritten in terms of the fake superpotential  \eqref{FakeSuperpotential},
\begin{equation} W\scal{ \Gamma_\pA|_{\pA=1}^{N} }  = \sum_\pA W(\Gamma_\pA,\ve) \ , \end{equation}
where the value of $\ve$ does not however extremise each of its components in the sum. This simplifies the
comparison to the mass of the constituent centres, since the single-centre mass, $M_\pA$ is found by
extremising the fake superpotential, $W(\Gamma_\pA,\ve)$, with respect to $\ve$, to obtain a vector
$\ve_\pA$. As we have seen in section \ref{sec:superpotential}, the resulting value $W(\Gamma_\pA,\ve_\pA)$
is a global maximum of the fake superpotential, and therefore one finds
\begin{equation} W(\Gamma_\pA,\ve ) \le W(\Gamma_\pA,\ve_\pA) \equiv M_\pA \ , \end{equation}
for each centre separately. This directly implies that the binding energy of any composite solution
is necessarily positive, since
\begin{equation}
 W\scal{ \Gamma_\pA|_{\pA=1}^{N} }  = \sum_\pA W(\Gamma_\pA,\ve)
 \le \sum_\pA W(\Gamma_\pA,\ve_\pA) = \sum_\pA M_\pA\,.
\end{equation}
Moreover, by exactly the same argument, one finds that the energy of a composite solution is lower
than the the energy of a single-centre solution of the same total charge, as
\begin{equation}\label{single-comp}
 W\scal{ \Gamma_\pA|_{\pA=1}^{N} }  = W(\Gamma_{\sf t} ,\ve)  \le M_{\Gamma_{\sf t}}\,,
\end{equation}
where $\Gamma_{\sf t}=\sum_\pA\Gamma_\pA$. We then conclude that single-centre non-BPS solutions
are energetically disfavored over multi-centre solutions with the same total charge. Note that
this comparison can only be done on the relevant hypersurface in moduli space where the composite
solutions exist. This property is in contrast with the BPS case, in which the mass is uniquely
determined by the total charge and the moduli. However, we should mention that single-centre black
holes are generically entropically favored over multi-centre black holes, as can be computed using the area law and the properties of the quartic invariant. We shall prove that this is always the case in the two-centre configurations we consider in section \ref{sec:two-cent-ex}.

The discussion above illustrates that the parameter $\ve$, describing flat directions for non-BPS solutions,
plays a role analogous to the one of the K\"{a}hler phase $\alpha$ of the central charge in the corresponding
BPS solutions. Indeed, the linear mass formula for a BPS black hole also follows from a superpotential, given
by
\begin{equation}
W_{\sf BPS} = \mbox{Re}[  e^{-i\alpha} Z(\Gamma)] \ ,
\end{equation}
where $\alpha$ is chosen such that $ W_{\sf BPS} $ is maximised, \ie to be the phase of the central charge
$Z(\Gamma)$. The resulting mass formula is of course
\begin{equation}
M = |Z(\Gamma)| \ .
\end{equation}
The energy of a two-centre solution then satisfies the triangular identity
\begin{gather}
 \mbox{Re}[ e^{-i\alpha} Z(\Gamma_1+\Gamma_2)] \le
\mbox{Re}[e^{-i\alpha_1} Z(\Gamma_1)] +   \mbox{Re}[e^{-i\alpha_2} Z(\Gamma_2)] \Rightarrow
\CR
M_{\sf tot} \le M_1 + M_2\,,
\end{gather}
where $\alpha$ is the phase of $Z(\Gamma_1 + \Gamma_2)$ and $\alpha_{\pA}$ are the phases of
$Z(\Gamma_{\pA})$ for $A=1,2$. Positivity of the binding energy between the two
centres then follows from the property that $\alpha_\pA$ is a maximum of $ \mbox{Re}[e^{-i\alpha} Z(\Gamma_\pA)]$.
As seen above, the composite non-BPS mass formula obeys the same property, but with the subtle difference
that, if a single-centre non-BPS solution of charge $\Gamma_1 + \Gamma_2$ exists, it has a mass bigger than
that of the bound states of constituent charges $\Gamma_1$ and $\Gamma_2$, due to \eqref{single-comp}.

For a two-centre solution, one can make these properties more explicit in the regime of large separation
of centres, using the expression \eqref{BubbleTframe},
\begin{equation}\label{BubbleTframe-2}
  \vk = \gamma_\pA \vl + \vk_{\Gamma_\pA} + \sqrt{2}\frac{ \gamma_\pA - \gamma_\pB}{\RAB} \vl_{\Gamma_\pB} \ ,
\end{equation}
to obtain the variation from the single-centre solution. In this case, the term
proportional to $\gamma_\pA \vl $ can be reabsorbed into a redefinition of $m$, so that the relevant
small perturbation $\delta \vk$ that determines $\delta \ve$ in \eqref{kgivese} is
\begin{equation}
 \delta \vk = \sqrt{2} \frac{ \gamma_\pA - \gamma_\pB}{\RAB} \vl_{\Gamma_\pB} \ ,
\end{equation}
whenever the solution exists, provided $\vl_{\Gamma_B}$ is linearly independent of $\vl$.
This can be used in \eqref{SecondDerW} to verify that the two-centre binding energy is indeed nontrivial
whenever the solution exists.  Moreover, \eqref{EnergyMulticentre} in the two-centre case simplifies to
\begin{multline}  W(\Gamma_1,\Gamma_2) = \frac{1}{2\sqrt{2}} \sum_{\pA=1}^2 \Bigl(  ( 1 + ( m + \gamma_\pA \det \vl )^2 )  \frac{\tr \vl \times \vl \, \vl_{\Gamma_\pA} }{\det \vl} -\frac{1}{4}  \frac{ \det \vl}{\det \vl_{\Gamma_\pA}} I_4({\Gamma_\pA}) \Bigr)    \\ +\frac{1}{\sqrt{2}} \frac{(\gamma_1-\gamma_2)^2}{ \Rot}  \det \vl \, \tr \vl_{\Gamma_1} \times \vl_{\Gamma_2} \, \Scal{ \sqrt{2}\,  \vl + \tfrac{1}{\Rot} ( \vl_{\Gamma_1} + \vl_{\Gamma_2} )}   \ , \end{multline}
where we used the explicit form of $\vk$ in \eqref{BubbleTframe-2}. This expression of the mass looks naively like the sum of the individual masses plus a manifestly positive quantity on the second line, which would be in contradiction with a positive binding energy. It is important to point out however that this is not the case because these expressions of the individual masses do not correspond to the individual masses at the same moduli whenever $\Rot < \infty$, because then $\vk \ne \vk_{\Gamma_A} +\gamma_A \vl$. When $\Rot \rightarrow \infty$, one gets instead that $\vk = \vk_{\Gamma_A} +\gamma_A \vl$ and the second line vanishes, which shows that the binding energy vanishes in the limit of large radius. This limit corresponds to a wall of marginal stability in moduli space, as we shall discuss in more detail in the next subsection.

It is interesting to compare this expression of the mass to the central charge, which reads
\begin{multline}  |Z(\Gamma_1+\Gamma_2)| = \frac{1}{2\sqrt{2}} \biggl|  \sum_{\pA=1}^2 \Bigl(  ( m + \gamma_\pA \det \vl  - i  )^2   \frac{\tr \vl \times \vl \, \vl_{\Gamma_\pA} }{\det \vl} - \frac{1}{4} \frac{ \det \vl}{\det \vl_{\Gamma_\pA}} I_4({\Gamma_\pA}) \biggr)    \Bigr . \\ \biggl .  + 2 \frac{(\gamma_1-\gamma_2)^2}{ \Rot}  \det \vl \, \tr \vl_{\Gamma_1} \times \vl_{\Gamma_2} \, \Scal{ \sqrt{2}\,  \vl + \tfrac{1}{\Rot} ( \vl_{\Gamma_1} + \vl_{\Gamma_2} )}  \biggr| \ . \end{multline}
Now, for a regular solution, one has to demand both that
$\frac{\tr \vl \times \vl \, \vl_{\Gamma_\pA} }{\det \vl} > 0 $ and that
\begin{equation}
- \sum_{\pA=1,2} \frac{ \det \vl}{\det \vl_{\Gamma_\pA}} I_4({\Gamma_\pA})
+ 8 \frac{(\gamma_1-\gamma_2)^2}{ \Rot}  \det \vl \, \tr \vl_{\Gamma_1} \times \vl_{\Gamma_2} \, \Scal{ \sqrt{2}  \vl
+ \tfrac{1}{\Rot} ( \vl_{\Gamma_1} + \vl_{\Gamma_2} )} > 0 \,,
\end{equation}
where all three terms are separately positive. These formulae show explicitly that the energy is always strictly above the BPS bound, as expected.
We stress that this result holds everywhere in moduli space, where the non-BPS
solution exists.

It is interesting to note that using \eqref{AsympL} one finds that there is always a critical radius
$R_{\rm c}$ at which the vector $\vl$ becomes degenerate (\ie $\det\vl=0$), and the asymptotic
scalars are singular, as ${\rm Im}(\vt) \times {\rm Im}(\vt) = \mathcal{O}(R-R_{\rm c})$ for finite value of the other parameters. One
computes that both the energy and the central
charge diverge as $(R-R_{\rm c})^{-1}$ in this limit, while still consistent with the BPS bound
up to order $\mathcal{O}(R-R_{\rm c})$
\begin{equation}
 W(\Gamma_1,\Gamma_2) = |Z(\Gamma_1 + \Gamma_2)| + \mathcal{O}(R-R_{\rm c})  \ .
\end{equation}
This limit should not be considered as a boundary of the $(n_v+1)$-dimensional surface in moduli space on which the solution exists, since it is itself at the boundary of moduli space, consistently with the property that the BPS bound is saturated in this limit.

\subsubsection{Attractor tree and walls of marginal stability}
\label{sec:attr-tree}

Given the results of the previous subsection on the positivity of the binding energy, the natural
next step is to consider the possibility of decay of composite solutions at regular points in
moduli space, \ie the existence of walls of marginal stability. Before turning to the corresponding
analysis, we emphasise that the question at hand is in principle more subtle for non-BPS composites,
which only exist on appropriate hypersurfaces in moduli space, compared to the BPS solutions,
which exist in codimension zero subspaces of the moduli space. In practice, this means that
BPS solutions exist in codimension zero domains in moduli space, which boundaries define walls of marginal stabilities where some of the distance $\RAB$ diverge. For non-BPS composites one finds exactly the same situation, but now restricted on the relevant
hypersurface where the given solution exists, as discussed above. It then follows that the walls
of marginal stability are only defined on the appropriate hypersurfaces as their boundaries in
moduli space and do not extend outside of them, as we discuss in more detail now.

Consider a general two-centre solution, as described by \eqref{BubbleTframe}-\eqref{eq:lk-two-c}
above. Assuming that $\vk_{\Gamma_2}-\vk_{\Gamma_1}$ is indeed a strictly positive Jordan algebra element,
the solution clearly exists for arbitrary large distance between the centres, $\Rot$, since the distance
dependent term in these relations becomes irrelevant at this limit. Taking the limit $\Rot\rightarrow \infty$,
the moduli are regular, so that one finds a wall of marginal stability for finite moduli.

For a marginally bounded solution, one expects that the energy of the composite solution
is equal to the sum of the masses of the constituent black holes. In order to verify that,
we consider \eqref{EnergyMulticentre} in the limit of marginal stability, where \eqref{BubbleTframe}
becomes simply
\begin{equation}\label{Bubble-triv}
 \vk_{\Gamma_\pA} = \vk - \gamma_\pA \vl   \ .
\end{equation}
This is identical to the value of $\vk$ for which \eqref{EnergyMulticentre} describes non-interacting
centres, as explained below that equation. It follows that the total energy decouples
\begin{equation} \lim_{\Rot \rightarrow \infty}  W( \Gamma_1,\Gamma_2) = W(\Gamma_1)  + W(\Gamma_2) \ . \end{equation}
This formula generalises to any number of centres, as long as the solution exists. Suppose that we
have a solution to \eqref{BubbleTframe} for $N+1$ centres, such that $\vl$ is strictly positive, in
the limit where $x_\pA$ goes to spatial infinity for some $A_0$, then
\eqref{Bubble-triv} again applies for that value $A=A_0$, while \eqref{BubbleTframe} for
$A, B \ne A_0$ define the corresponding equations for $N$ centres. Therefore, one finds that the
mass formula satisfies
\begin{equation}\label{wall-dec-gen}
 \lim_{|x_{N+1}| \rightarrow \infty}   W\scal{ \Gamma_\pA|_{A=1}^{N+1} }
=   W\scal{ \Gamma_\pA|_{A=1}^{N} }  + W(\Gamma_{N+1}) \ .
\end{equation}

The property that the binding energy is finite at finite radius relies on the property that $\ve$ is not an extremum of each $W(\Gamma_\pA,\ve)$ for generic variations preserving $\Iprod{\hat{R}}{\Gamma_\pA}=0$ separately. Nonetheless, given a total charge $\Gamma_{\sf t}$ and a set of charges, $\Gamma_\pA$,
such that $\Gamma_{\sf t}=\sum_\pA\Gamma_\pA$, one checks that $\ve$
extremises $W$ on the subvariety satisfying $\Iprod{\hat{R}}{\Gamma_\pA}$ for all $A$. Indeed, one computes using \eqref{BubbleTframe} that
\begin{align}\label{del-W-tot}
\delta W = &\,
\frac{\det \vl }{2\sqrt{2}} \biggl(  \det \Scal{ \vk  + m \frac{\vl}{\det \vl}} + \tr \frac{ \vl\times\vl}{(\det\vl)^2 } \Scal{ \vk  + m \frac{\vl}{\det \vl}} \biggr) \ \tr \vl_{\sum_\pA\Gamma_\pA} \delta\ve^{-1}
\CR
&\, - \frac{1}{\sqrt{2}} \tr \vl_{\sum_\pA \gamma_\pA \Gamma_\pA}  \delta \ve^{-1}  \ .
\end{align}
Here one can interpret the coefficients $\gamma_A$ as Lagrange multipliers for the conditions 
\begin{equation}
 \Iprod{\hat{R}}{ \Gamma_\pA} = 0\quad  \forall A \  . \label{Hypersurface} 
\end{equation}
The existence of a well defined extremum then constrains $\ve$ and the moduli. For a regular multi-centre solution $\ve$ must in fact satisfy \eqref{Hypersurface}, but for a single centre of charge $\sum_\pA \Gamma_\pA$
the $\ve$ following from \eqref{del-W-tot} does not extremise correctly the function
$W\scal{\Gamma_{\sf t},\ve}$ in one direction, and would not reproduce the same mass, as
discussed in \eqref{single-comp}. In fact it follows that unlike for the BPS solutions, a multi-centre non-BPS solution admits an energy strictly lower than the energy of the single-centre solution with the same total charge and asymptotic scalars, so the composite configuration is energetically favored whenever it exists.

Given this extremisation problem, we may now ask the reverse question and consider a top-down approach where one seeks to
infer criteria for the existence of solutions from the function $W(\Gamma_\pA, \ve)$ alone,
instead of using the explicit knowledge of solutions to derive the properties of the fake
superpotential. One may wonder if the condition that $\ve$
extremises $W$ on the subvariety satisfying \eqref{Hypersurface} is strong enough to ensure the existence of a solution. One computes in general that for $\ve$ to be such an extremum, the condition 
\begin{equation} \sum_\pA \Scal{ 4 ( \vl \times \vl ) \times \scal{ ( \vk - \vk_{\Gamma_\pA})  \times \vl_{\Gamma_\pA}  }
- \vl\, \tr \vl \,  ( \vk - \vk_{\Gamma_\pA})  \times \vl_{\Gamma_\pA}}
= \det \vl  \sum_\pA \gamma_\pA  \vl_{\Gamma_\pA} \ , \end{equation}
must hold for some arbitrary Lagrange multipliers $\gamma_A$. 
Using the property that
$\vl$ is positive, one can simplify this to
\begin{equation} \sum_\pA ( \vk - \vk_{\Gamma_\pA} - \gamma_\pA \vl ) \times \vl_{\Gamma_\pA} = 0 \  , \end{equation}
whose general solution is
\begin{equation} \vk =  \vk_{\Gamma_\pA} + \gamma_\pA \vl + \sum_{\pB\ne \pA} A_{\pA\pB} \vl_{\Gamma_\pB} \ , \end{equation}
where $A_{\pA\pB}$ is an antisymmetric matrix. Upon identifying the $\gamma_\pA$ as the
proportionality constants in \eqref{LKdef}, this solution would reproduce equation
\eqref{BubbleTframe} for
\begin{equation} A_{\pA\pB} = \sqrt{2} \frac{ \gamma_\pA - \gamma_\pB}{\RAB} \ . \end{equation}
However, this is not simply a particular parametrisation of $A_{\pA\pB}$, since an $N\times N$
antisymmetric matrix comprises $\frac{N(N-1)}{2}$ independent components, whereas there are
only $3(N-2)$ independent distances $\RAB$ in three dimensions for $N\ge 3$ (so that the solution is not general for $N\ge 5$). Therefore the condition that $\ve$ extremises $W$ on the appropriate subvariety satisfying \eqref{Hypersurface} does not ensure the existence of a solution in general.

Nevertheless, this shows that there is a natural generalisation of the existence of an
attractor flow tree associated to BPS composite solutions. Indeed, any solution to
\eqref{BubbleTframe} with $N+1$ centres  admits a large radius limit in which the solution
decouples in a single-centre solution and an $N$-centre solution to \eqref{BubbleTframe},
as in \eqref{wall-dec-gen}. Therefore, one can solve \eqref{BubbleTframe} by adding each
centre one after the other by following the inverse procedure that ensures \eqref{Bubble-triv}
to be satisfied at each addition. Moreover, the existence of a limit of marginal stability
implies that for any two-centre solution of charges $\Gamma_1$ and $\Gamma_2$, there exist
moduli for which the solution $\ve$ extremising both $W(\Gamma_1,\ve)$ and $W(\Gamma_2,\ve)$
is the same. Then, adding a third centre requires that there exist moduli such that
$\ve$ extremise $W(\Gamma_1+\Gamma_2,\ve)$ with respect to variations preserving both
$\Iprod{\hat{R}}{\Gamma_1}$ and $\Iprod{\hat{R}}{\Gamma_2}$, and extremise $W(\Gamma_3,\ve)$
with respect to variations preserving  $ \Iprod{\hat{R}}{ \Gamma_3} = 0 $. Each new centre
or cluster of centres already satisfying \eqref{BubbleTframe} requires similarly the
existence of a common extremising vector $\ve$ for some moduli.

This construction clearly extends Denef's attractor tree, where the role of the
central charge phase is now played by the Jordan algebra element $\ve$ parametrising
the flat directions of the individual centres. The existence of such a non-BPS attractor
tree is clearly required for the solution to exist, and it is therefore natural to wonder
if it provides a sufficient condition. However, the construction of such a tree is a rather
difficult task in practice. An obvious obstacle is that, unlike for BPS solutions, where a closed
form formula for the phase of the central charge is available, there is no such formula for
the vector $\ve$ extremising the fake superpotential.

It is important to point out that the restrictions on $\ve$ can rapidly become overconstraining.
In particular, for $n_v$ centres one finds configurations of charges such that
$\Iprod{\hat{R}}{\Gamma_\pA}$ determines $\ve$ completely. Then there is no freedom in tuning
$\ve$ to accommodate a new centre anymore, and although the solution may still exist, it can
only be marginally stable. For example, this is the case for solutions within the one modulus
model, for which $\ve$ is necessarily fixed already for a single-centre solution. In this
case we have found that there exist two-centre solutions, but the total energy is independent
of the distance and is equal to sum of the constituent masses.

\subsection{Explicit two-centre example}
\label{sec:two-cent-ex}

In this final subsection, we make the above considerations fully explicit for a two-centre
example. While the structure and properties of the solution are exactly those discussed in
the general case above, having an explicit example allows for an even more detailed description
of the possible bound states. We display most of the formulae within the STU truncation, which does not constitute a restriction for the charge configurations we consider. We also perform numerical estimates of the binding energy.

We consider a system of two black holes, carrying charges $\Gamma_1$ and $\Gamma_2$ respectively,
which we choose as follows. Up to an electric/magnetic duality, one can always bring one of the
charges to be a D0-D6 with $q_0 = - p^0 \equiv Q_0 > 0 $, so we choose for simplicity
\begin{equation}\label{Gamma1}
 \Gamma_1 = \left( \begin{array}{c} Q_0 \\ 0 \\ 0 \\ -Q_0 \end{array}\right)\ .
\end{equation}
Given that we used duality covariance to restrict one of the charges, the second charge
is a priori unrestricted for the most general two-centre solution up to dualities.
Here, we shall nevertheless restrict the second charge to be a D0-D4-D4-D4-D6 for simplicity. We do not
expect a significant change in the physical properties of the solution by adding a D2 charge.
We therefore take
\begin{equation}\label{Gamma2}
 \Gamma_2 = \left( \begin{array}{c} q_0 \\ 0 \\ \vp \\ p^0 \end{array}\right)\ ,
\end{equation}
where we assume that $I_4(\Gamma_2) = - 4 q_0 \det \vp - (p^0 q_0)^2 <  0 $, so that we are indeed dealing
with two non-BPS charges. Note that the inner product $\Iprod{\Gamma_1}{\Gamma_2}=Q_0\,(q_0 + p^0)$,
does not depend on the magnetic charges, $\vp$, and the interaction of the two centres vanishes
in the limit $q_0=-p^0$. It is also worth mentioning that the uplift of this configuration to
five dimensions describes a pair of doubly spinning extremal Kerr black holes \footnote{There
are indeed four independent angular momenta, since we allow for arbitrary under-rotation at both centres
and $q_0\neq Q_0$ in general.} located at the two tips of a two-centre Taub-NUT geometry, while
the magnetic fluxes $\vp$ are threading the two-cycle between the centres.

We start by considering the relevant constraints coming from the two charges on the
auxiliary vector $\ve$. First, the condition $\Iprod{\hat{R}}{\Gamma_1}=0$ simply implies
that $\det \ve = 1$, and \eqref{lGamma}, \eqref{kGamma} for $\Gamma_1$ also simplify dramatically,
to give
\begin{equation} \vl_{\Gamma_1} = Q_0 \ve \  , \qquad \vk_{\Gamma_1} = \frac{1}{2} \ve \ . \end{equation}
The regularity condition at the horizon requires that $\ve$ is a positive Jordan algebra element (\ie all three $e_i$ are positive numbers within the STU truncation) 
and we shall always assume that this condition is satisfied. Using the results of subsection
\ref{sec:sing-cent-expl}, it is straightforward to construct the general single center solution
carrying the particular D0-D6 charge in \eqref{Gamma1}.\footnote{Note that this example is in fact the simplest in our
framework, while the corresponding single-centre solution has not yet been described explicitly in the literature.}
Turning to the second charge \eqref{Gamma2}, the equations \eqref{NoRstar} and \eqref{lGamma} become respectively
\begin{equation} \label{ch2-gr}
q_0 + \tr \ve \vp  + p^0 =0\ ,
\qquad
\vl_{\Gamma_2} = - 2 ( \ve \times \ve ) \times \vp - \ve \, p^0 \ .
\end{equation}
Solving for $p^0$ using the first relation, one finds
\begin{eqnarray}
 \vl_{\Gamma_2} &=& ( q_0 + \tr \ve \vp)  \, \ve - 2 ( \ve \times \ve ) \times \vp \CR
(\vl_{\Gamma_2})_i &=& e_i \scal{ q_0 + e_i p_i } \ ,
\end{eqnarray}
as well as
\begin{eqnarray}
 \vk_{\Gamma_2}  - \vk_{\Gamma_1} &=& \frac{1}{2} \frac{ \ve \scal{ \det \vp - q_0 \tr \ve \times \ve \, \vp \times \vp +  q_0 p^0  ( q_0 + p^0 ) } + 2 q_0 \, \vp \times ( \vp + p^0\,\ve \times \ve )}{ \det \vp + q_0 \tr \ve \times \ve \, \vp \times \vp - q_0^{\; 2} p^0 }  \CR
( \vk_{\Gamma_2}  - \vk_{\Gamma_1})_i  &=& \frac{e_i}{2} \frac{ q_0 \, e_i p_i + e_{i+1} e_{i+2} p_{i+1} p_{i+2}}{( q_0 + e_{i+1} p_{i+1} ) ( q_0 + e_{i+2} p_{i+2} )} \ ,
\label{k-comp-2}
\end{eqnarray}
where we wrote the expressions within the STU truncation in the second lines. These are relevant
in what follows, because we can always use the $G_5$ symmetry of the problem to diagonalise both $\ve$ and $\vp$
and solve these equations within the STU truncation without loss of generality.

\subsubsection*{Existence of solution}

The conditions above can clearly be solved by requiring that $q_0$ and $\vp$ are both positive,
and $p^0$ is negative, but this is not the only solution. Without loss of generality, we can
assume that
\begin{equation} e_1 p_1 \le e_2 p_2 \le e_3 p_3 \ . \end{equation}
One can then verify that regularity requires $q_0>0$, while $\vp$ is not necessarily positive
and satisfies
\begin{equation} - e_2 p_2  \min\Scal{ \frac{ q_0}{e_3 p_3} , \frac{e_3 p_3}{q_0}  }  < e_1 p_1 \le e_2 p_2 \le e_3 p_3 \ . \end{equation}
It follows that $q_0,\, p_2,\, p_3$ must all be strictly positive, but $p_1$ can possibly be negative.
Note that the presence of a nonzero $p^0$ is required for the latter possibility so that $I_4(\Gamma_2)<0$ (even though we have solved
for $p^0$ in all equations).
Indeed, we have by construction that $ \min\scal{ \frac{ q_0}{e_3 p_3} , \frac{e_3 p_3}{q_0}  } \le 1$ and therefore 
\begin{equation} \sum_{i=1}^3 e_i p_i  > e_3 p_3 \ , \end{equation}
so that $-p^0 > q_0 + e_3 p_3$. This also guarantees that the two charges are never mutually commuting, as
\begin{equation} \Iprod{\Gamma_1}{\Gamma_2} = Q_0 ( p^0 + q_0 ) < - Q_0 e_3 p_3 \ . \end{equation}
The existence of a regular solution for $\ve$ to the first of \eqref{ch2-gr} leads to a bound on $p^0$.  For $p_1 \le 0$, it turns out there are solutions for $-p^0 > q_0$ arbitrary
close to $q_0$, whereas it must satisfy to \footnote{The function
$f(e_1,e_2) \equiv e_1 p_1 + e_2 p_2 + \frac{p_3}{e_1 e_2}$ is minimum at
$ \ve = \frac{ \vp \times \vp}{(\det \vp)^{2/3}}$.}
\begin{equation} -p^0 >  q_0 + 3 \sqrt[3]{\det \vp} \ , \label{p0Bound}  \end{equation}
when $p_1 > 0$. The limit case $-p^0 = q_0 + 3 \sqrt[3]{\det \vp} $ is physical but degenerate,
and will be discussed separately.

We shall now concentrate on the example without flat directions, for which $\vp$ is strictly
positive.  In the general situation described by \eqref{p0Bound}, it is simple
to consider the STU truncation, where one can solve for $\ve$ explicitly as
\begin{eqnarray} 
e_2 &=&  \frac{-e_1 p^0 - e_1^{\; 2} p_1 - e_1 q_0
   \pm \sqrt{-4 e_1 p_2 p_3 + (e_1 p^0 + e_1^{\; 2} p_1 + e_1 q_0)^2}}{2 e_1 p_2}
\CR
e_3 &=&  \frac{-e_1 p^0 - e_1^{\; 2} p_1 - e_1 q_0
   \mp \sqrt{-4 e_1 p_2 p_3 + (e_1 p^0 + e_1^{\; 2} p_1 + e_1 q_0)^2}}{2 e_1 p_3}
 \ ,
\end{eqnarray} 
where the signs appearing in the two expressions must be opposite. Defining
\begin{equation} {\rm x} \equiv  - \frac{ q_0 + p^0}{3 \sqrt[3]{\det \vp }} \ , \end{equation}
one obtains that in the generic situation, for which ${\rm x} > 1 $, a positive solution of
\eqref{ch2-gr} for $\ve$ exists, provided that $e_1$ satisfies
\begin{equation}  \scal{ 2 {\rm x}    + a_1  + \bar a_1 }  \sqrt[3]{\frac{p_2 p_3}{p_1^{\; 2} }}  < e_1 < \scal{ 2 {\rm  x}    + a_2  + \bar a_2 }   \sqrt[3]{\frac{p_2 p_3}{p_1^{\; 2} }} \label{ebound} \end{equation}
where $a_i$ are the three cubic roots of $2 - {\rm x}^3  + 2 i \sqrt{  {\rm x}^3 -1}$ ordered such that $\mbox{Re}[a_1] \le \mbox{Re}[a_2] \le  \mbox{Re}[a_3]$, and that $\gamma_1 - \gamma_2 $ satisfies to
\begin{equation} 0 <  \gamma_1 - \gamma_2 < \frac{\Rot}{2\sqrt{2}} \inf_i \left[ \frac{  q_0 \, e_i p_i + e_{i+1} e_{i+2} p_{i+1} p_{i+2}}{( Q_0 + q_0 + e_i p_i ) ( q_0 + e_{i+1} p_{i+1} ) ( q_0 + e_{i+2} p_{i+2} )} \right]  \ . \label{gammabound} \end{equation}
This inequality shows explicitly that one can reach infinite radius at finite values of the moduli,
and therefore exhibits the existence of a wall of marginal stability.

Within the STU truncation, the six moduli are parametrised by the four free parameters $\Rot$,
$e_3,\, \gamma_1,\, \gamma_2$, up to the conditions \eqref{ebound}, \eqref{gammabound}. Therefore, the
wall defined at $\Rot \rightarrow \infty$ is clearly of co-dimension one in the 4-dimensional hypersurface
of allowed moduli. Note that $\Rot$ can be arbitrarily small, but the limit $\Rot\rightarrow 0$ is located at
the boundary of moduli space. A similar analysis leads to the same conclusions for $p_1\le 0$. 

It is also interesting to compare the entropy of such a two-centre configuration with the entropy of the single-centre solution that would carry the same total electromagnetic charges and total angular momentum. For simplicity, we consider the case in which the intrinsic angular momenta of the two individual centres vanish. In this case the total angular momentum of the solution is the symplectic product of the two charges. Using the regularity conditions that $Q_0 > 0 , \, q_0 > 0$ and, either $\det \vp > 0 $ and \eqref{p0Bound} is satisfied, or $\det \vp < 0 $ and $p^0 < - \mbox{max}\bigl[ q^0, \sqrt{ \frac{ - \det \vp}{q_0}}\bigr]$, one proves that
\begin{equation}
\scal{ 2 \det \vp - q_0 p^0 ( 3 Q_0 + q_0 - p^0 ) + Q_0^{\; 2} ( q_0 - p^0 ) }^2 > Q_0^{\; 2} \scal{ 4 q_0 \det \vp + ( q_0 p^0 )^2 } \ ,
\end{equation}
and therefore  
\begin{equation} \sqrt{ - I_4(\Gamma_1 + \Gamma_2) - \Iprod{\Gamma_1}{\Gamma_2}^2 } > \sqrt{ - I_4(\Gamma_1) } + \sqrt{ - I_4(\Gamma_2) } \  . \end{equation}
We conclude that for all the regular solutions we consider in this section, the entropy of the single-centre solution carrying the same total electromagnetic charges and angular momentum is always strictly greater than the entropy of the two-centre solution. 

\FIGURE{
 \centering
 \includegraphics[scale=0.15]{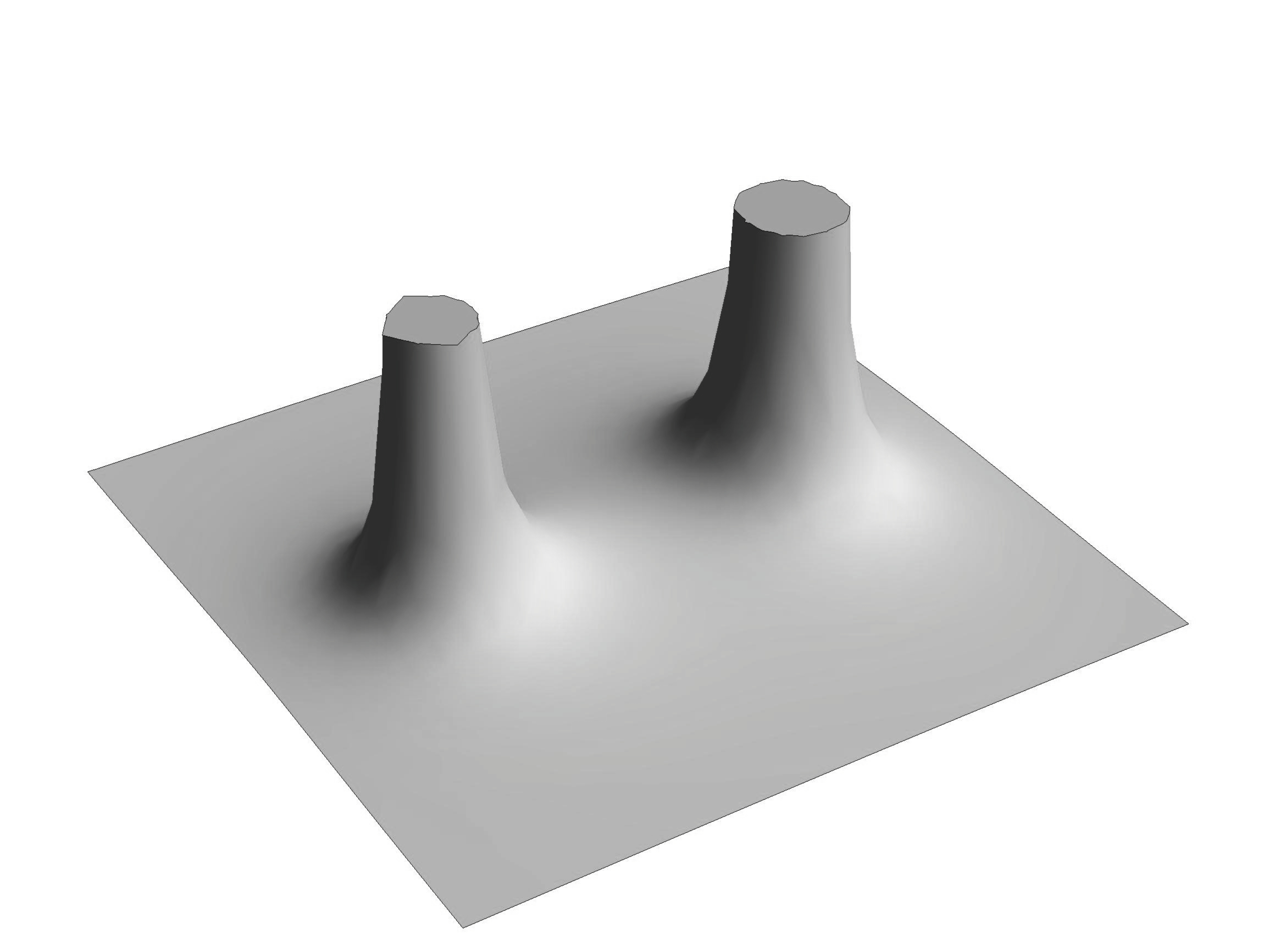}
%  \includegraphics[scale=0.6]{./2centerAX.jpeg}
 % 2center.jpeg: 2340x1732 pixel, 72dpi, 82.55x61.10 cm, bb=0 0 2340 1732
\caption{A plot of the function $e^{-4U} - \frac{ \omega_\varphi^{\; 2}}{r^2 \sin^2 \theta}$,
for an example two-centre solution with $p_1<0$, in cylindrical coordinates $\rho,z$, centered
along the axis between the centres. 
The uneventful behaviour seen in this plot seems to be universal for all examples we considered.}
\label{fig:two-cen-plot}
}

We did several numerical checks of this solution in the parametrisation above, unfortunately excluding
the region of large volume in moduli space. The reason is that reaching large imaginary values requires
a severe fine tuning of the parameters, such that $\vl$ is very large and at the same time
\begin{equation} \vk = - m \frac{\vl}{\det \vl} + \mathcal{O}( \vl^{-3}) \ ,\end{equation}
which is complicated to obtain in practice. We checked for several axisymmetric examples, including different signs of $p_1$, that not only the metric is regular, but there is no closed time-like curves outside the horizons, \ie 
\begin{equation} e^{-4U} - \frac{ \omega_\varphi^{\; 2}}{r^2 \sin^2 \theta} > 0 \ . \end{equation}
An example plot of this function for $p_1<0$ is shown in figure \ref{fig:two-cen-plot}, which 
immediately shows that it falls off monotonically as one approaches asymptotic infinity.
Here, we choose not to display any further details, since this behaviour seems to be universal for
all examples we considered.
The numerical simulations we have been doing all manifest
the behaviour that
\begin{equation} W(\Gamma_1) + W(\Gamma_2) -W(\Gamma_1,\Gamma_2) <\hspace{-1mm}< W(\Gamma_1,\Gamma_2) - | Z(\Gamma_1 + \Gamma_2)| <\hspace{-1mm}< W(\Gamma_1,\Gamma_2) \ ,   \end{equation}
so that the binding energy is extremely small compared to both the total energy of the system and the energy
above the BPS bound. In connection to this, it is interesting to note that given a two-centre non-BPS
composite, generically one can define a BPS solution with the same total charges and moduli, \eg as two
centre configurations. Such a solution is then largely favored energetically and one may expect the
non-BPS composite to decay into the BPS composite by tunneling relatively rapidly. Although they carry
the same total charges, these solutions are however extremely different on large scales, and such a decay
would require rather non-local quantum effects, and it is by no means clear that this can occur in
string theory.

Let us comment finally on the limiting case $-p^0 = q_0 + 3 \sqrt[3]{\det \vp} $, when there is a unique
solution $\ve = \frac{ \vp \times \vp}{(\det \vp)^{2/3}}$ for the vector parametrising $\hat R$. In this
case, the solution is always equivalent up to duality transformations to a solution of the one modulus
model (or $t^3$ model). Since a non-BPS charge has no flat directions in this model, the binding energy
always vanishes in this case, and the energy of the solution is independent of the radius. This solution
is therefore only marginally stable for all values of the radius. This behaviour is peculiar to the
$T^3$ model and it is worth mentioning that the composite non-BPS system is associated
to a nilpotent orbit that in fact does not exist in $\mathfrak{g}_{2(2)}$, the three-dimensional duality group of this model. The existence of non-BPS composite solutions within the $T^3$ model despite the absence of the relevant nilpotent orbit in $\mathfrak{g}_{2(2)}$ was pointed out in \cite{Yeranyan:2012au}.

\subsubsection*{Flat directions}

One can distinguish three cases among these examples, depending on the sign of the minimum eigen value $p_1$ of $\vp$.
The latter determines the flat directions associated to the solution. The D0-D6 charge is by construction
left invariant by a subgroup $G_5 \subset G_4$, whereas the common stabiliser of the two charges is the
stabiliser of $\vp$ in $G_5$. The stabiliser of $\vp$ is the maximal compact subgroup $K_5$ of $G_5$ if
$p_1>0$, whereas it is a non-compact real form $K_5^* \subset G_5$ if $p_1<0$. The non-compact real
form $K_5^*$ is the divisor group that would define the pseudo-Riemannian scalar manifold $K_5^* \backslash G_5$ of the
theory obtained by time-like reduction of a genuine six-dimensional theory, so we shall write its
maximal compact subgroup as $K_6\subset K_5^*$. \footnote{We write this group $K_6$ because it is also the maximal compact subgroup of the six-dimensional theory duality group for magic supergravity theories. For the infinite series of axion-dilaton theories, $K_6\cong SO(n) $ is  the compact group acting on the $n$ vector multiplets coupled to gravity and one tensor multiplet  in six dimensions.} In the special case $p_1=0$, the stabiliser subgroup $I\hspace{-0.6mm}K_6$ is the contracted form of $K_5$ interpolating between the two real forms $K_5$ and $K_5^*$.  For example, in the exceptional theory with K\"{a}hler geometry $(U(1) \times E_{6(-78)}) \backslash E_{7(-25)}$, one has 
\begin{equation} K_5 \cong F_{4(-52)} \ , \quad I\hspace{-0.6mm}K_6 \cong Spin(9) \ltimes \mathds{R}^{16} \ , \quad K_5^* \cong F_{4(-20)} \ . \end{equation} 
Note however that this configuration is not generic, because the common stabiliser in $E_{7(-25)}$ of two independent generic vectors $\Gamma_1$ and $\Gamma_2$ is either $Spin(8)$ or $Spin(1,7)$ in general, as shown in appendix \ref{app:stabilizer}. Note that when $Spin(8)$ stabilizes both $\vp$ and $\vq$, one can always use $E_{6(-26)}$ to diagonalise them, such that they can be realised within the STU truncation, whereas this is not possible when their common stabilizer is $Spin(1,7)$. In the latter situation, both $\vp$ and $\vq$ are negative, and are linearly independent. 

One can easily convince oneself from our analysis that such configurations indeed exist. The positivity condition on $\vl_{\Gamma_2}$ and $\vk_{\Gamma_2}-\vk_{\Gamma_1}$ are identical upon substituting $\vp - 2 \ve \times \vq$ to $\vp$ and adding to $\vk_{\Gamma_2}-\vk_{\Gamma_1}$ a term linear in $\vq$. If one consider a situation in which $\vq$ is very small compare to the other charges ($\vq <\hspace{-1mm}< \vp$), it is clear that one can find solutions as deformations of the D0-D4-D6 ones. This does not require any particular property of $\vq$ with respect to $\vp$ apart from being very small, and so one can clearly find regular solutions for charge of common stabilizer $Spin(8)$ or $Spin(1,7)$.  

The two-centre solutions therefore admit drastically different sets of flat directions,  from zero to sixteen in the exceptional theory, \eg 
\begin{equation} \{ \mathds{1} \} \subset Spin(7) \backslash Spin(1,7) \subset \dots \subset  Spin(9) \backslash F_{4(-20)} \ . \end{equation} 
Although the flat directions associated to the individual charges seem to play an important role in the physical properties of the solution through their link to the vector $\ve$ parametrising the fake superpotential, it is not clear at this stage if the common flat directions of the two charges carry similar properties. We did not find specific differences between the solutions carrying or not flat directions.

\section{Derivation of composite non-BPS solutions}
\label{sec:derive-sols}

In this section, we present the detailed analysis of the duality covariant form
of the composite non-BPS system, as defined in \cite{Bossard:2013oga}. This leads
to a characterisation of solutions in terms of harmonic functions in an arbitrary
symplectic basis, leading to the results already presented in section \ref{sec:def-sys}
in a convenient basis. After a short summary of the system as defined in \cite{Bossard:2013oga}
in section \ref{sec:summ-5}, we discuss the general single-centre
solution of the multi-centre system in section \ref{sec:single-5}. This turns out to be
slightly more complicated than the purely single-centre system of
\cite{Bossard:2012xsa}, but leads to exactly the same physical results.
We then turn to the analysis of the multi-centre configurations in section
\ref{sec:multi-5}, where we present the general solution of the system in an
arbitrary frame and give the duality covariant constraints on the allowed
charges and distances between centres.

\subsection{The composite non-BPS system in a general basis}
\label{sec:summ-5}

Given the ansatze for the metric and gauge fields in
\eqref{metricMltc}-\eqref{gauge-decop}, the first order flow equation for the
composite non-BPS system is given as
\begin{align} \label{dw-summ}
\star dw =- \left(d -2\,\dTgp_{\scriptscriptstyle \cK} \right)
\left(2\,\mbox{Im}(e^{-U-i\alpha}\cV) - \frac12\, V\,\hat{R} -\frac{M}{V}\,\hat{R}^*\right)
\,.
\end{align}
Here, $M$, $V$, are functions to be specified below, while $\hat{R}$ and
$\hat{R}^*$ are a constant and a non-constant very small vector respectively,
where $\Iprod{\hat R}{\hat{R}^*}= 4$. The non-constant $\hat{R}^*$ is related to a
constant very small vector, $\Rstz$, by
\begin{equation} \label{Rstar0-real}
 \hat{R}^*= \exp[\Tgp_\csK]\Rstz\,,
\end{equation}
which also satisfies $\Iprod{\hat R}{\Rstz}= 4$. In this and all equations
in this section, $\Tgp_{\csK}$ is a generator of the
T-dualities leaving $\hat{R}$ invariant, parametrised by a vector of
harmonic functions, $\cK$. The vector of parameters $\cK$ lies in the grade
$(-1)$ component of the vector space according to the decomposition
\eqref{eq:vec-decomp} implied by the T-duality, \ie a
three-charge vector satisfying
\begin{equation}\label{grade-1}
\frac{1}{2} I_4^\prime(\hat{R},\Rstz, \cK ) = -\Iprod{\hat R}{\Rstz}\, \cK \,,
\end{equation}
which indeed specifies a vector of $n_v$ degrees of freedom. Without loss of
generality, we will consider the $\cK$ to asymptote to zero, \ie all the harmonic
functions contained in this vector have no constant parts. This choice
identifies the asymptotic value of $\hat{R}^*$ with the
constant vector $\Rstz$ and can be changed by passing to a different
$\Rstz$ by a constant T-duality. Note that this choice is convenient for
discussing the general properties of the system, but not necessarily for
constructing explicit solutions. Indeed, we use the freedom of
reintroducing asymptotic values for $\cK$ in the discussion of the explicit
representation of solutions in section \ref{sec:explicit-sols}.

The solutions to the flow equation \eqref{dw-summ} are simplified by
introducing a vector, $\cHz$, of grade $(-1)\oplus(+3)$, \ie satisfying
\begin{equation}
 \frac{1}{2} I_4^\prime(\hat{R},\Rstz, \cHz ) =
-\Iprod{\hat R}{\Rstz}\,\cHz + 3\, \Iprod{\cHz}{\Rstz} \hat{R} \ .
\label{sympl-constr-summ}
\end{equation}
Note that \eqref{grade-1} is trivially a solution of the last equation,
found by setting the grade $(+3)$ component, $\Iprod{\Rstz}{\cHz}$,
to vanish. In practice, once a basis is chosen, as was done in section
\ref{sec:system}, the constraints \eqref{grade-1} and \eqref{sympl-constr-summ}
determine $n_v$ and $n_v+1$ allowed components for the two vectors, $\cK$
and $\cHz$ respectively (cf. \eqref{V-L-def}).

The equations resulting from \eqref{dw-summ} take the form
\begin{gather}
 2\,e^{-U}\mbox{Im}(e^{-i\alpha}\cV)
=-\exp[\Tgp_{\scriptscriptstyle \cK}]
\left(\cHz - \frac12\, V\,\hat{R} - \frac{M}{V}\,\Rstz \right) \,,
\label{scals-5}
\\
 \star dw = \exp[\Tgp_{\scriptscriptstyle \cK}]
 \left( d\cHz - \dTgp_{\scriptscriptstyle \cK}\cHz\right)\,,
\label{dw-fin-summ}
\end{gather}
where $V$ is now identified with the grade $(+3)$ component of $\cHz$,
as $V=\Iprod{\cHz}{\Rstz}$. The compatibility relation for the last
relations leads to the field equation for $\cHz$, given by
\begin{gather}
 d\star d \cHz = \dTgp_{\scriptscriptstyle \cK}\! \wedge \star \dTgp_{\scriptscriptstyle \cK}\, \cHz
=-\frac1{64}\, I_4(d\cK,\star d\cK,\cHz,\hat{R})\,\hat{R}
\,. \label{Lapl-fin-summ}
\end{gather}
As the right hand side of this relation is only along $\hat{R}$, it follows that
all grade $(-1)$ components of $\cHz$ are harmonic, whereas $V$ is not, leading
to
\begin{equation}\label{V-Poiss}
 d\star d V = -\frac1{16}\,I_4(d\cK, \star d\cK, \cHz, \hat{R}) \,,
\end{equation}
by taking the inner product of \eqref{Lapl-fin-summ} with $\Rstz$.
Note that this is a linear equation for $V$, since the grade $(+3)$
component of $\cHz$ drops out from the right hand side
(cf. \eqref{V-Poiss-b} in a specific basis).
The final dynamical equation required is the one for the function $M$
in \eqref{dw-fin-summ} and the angular momentum vector $\omega$, both
of which are conveniently given as
\begin{align}\label{eq:dom-5}
\star d \omega - d M =
\Iprod{\cHz}{d\cHz - \dTgp_{\scriptscriptstyle \cK}\,\cHz}=
\frac1{16}\,I_4(d\cK ,\cHz, \cHz, \hat{R})\,.
\end{align}
Taking the divergence of this equation, one obtains
a Poisson equation for $M$, as
\begin{equation} \label{M-Poisson-5}
 d \star d M =
-\frac1{8}\,I_4(d\cK ,\star d\cHz, \cHz, \hat{R})\,
\end{equation}
whose solution can be used back in \eqref{eq:dom-5} to obtain
the angular momentum one-form, $\omega$.

These equations can be seen to be equivalent to the formulation given
in section \ref{sec:system}, by choosing the constant vectors
$\hat R$ and $\Rstz$ as in \eqref{SimilarlyR}-\eqref{RsK}. 
Similarly, one can verify that the formulations of the composite non-BPS
system given in a fixed duality frame in \cite{Bossard:2011kz, Bossard:2012ge}
can be also obtained from the above equations. The relevant choice for
comparing with \cite{Bossard:2011kz} is
\begin{equation}\label{fixedframe}
\hat{R}=\left( \begin{array}{c} 0 \\ 0 \\ 0 \\ - 2 \sqrt{2}  \end{array} \right) \ ,
\qquad
\Rstz = \left( \begin{array}{c} \sqrt{2} \\ 0 \\ 0 \\ 0  \end{array} \right) \ ,
\end{equation}
while \cite{Bossard:2012ge} uses the base obtained by an S-duality on the choice
above. Note that \eqref{fixedframe} are very similar to \eqref{SimilarlyR}-\eqref{RsK},
but do not include the arbitrary overall T-dualities that allow to cover all frames.
It then follows that one can only describe a restricted set of charges using
\eqref{fixedframe}, contrary to the system in section \ref{sec:system}.

\subsection{Single centre flows}
\label{sec:single-5}

As a first application of the covariant system defined in this section, we now
consider single-centre flows, i.e. the explicit solution when only one centre
is involved. While this case was treated in detail in \cite{Bossard:2012xsa}, we
find it illuminating to solve the general equations in this case, since they are
still nontrivial even though they lead to the same physical results as in
a purely single-centre treatment. Additionally, the structure of the solution
near each centre in the multi-centre case is necessarily of the type discussed
here and the precise embedding of the single-centre attractor in a multicentre
solution is of particular importance for later applications.

We therefore assume that all functions depend only on the coordinates relative
to one point, which represents the single horizon, and which we take to be the
origin of $\mathbb{R}^3$. In this case, \eqref{Lapl-fin-summ} determines $\cHz$
as
\begin{gather}
  \cHz  = \cH_s -\frac14\, V_n \hat{R}\,,
\CR
\cH_s = \mathrm{h} + \frac{\gz}{r}\,, \qquad
V_n= \frac1{32}\, \frac{1}{r^2} I_4(\kd, \kd, \mathrm{h} + \frac{\gz}{3r}, \hat{R}) \,,
\label{H-exp-hor-5}
\end{gather}
where $\mathrm{h}$ and $\gz$ are two constant vectors satisfying the same constraints
as $\cHz$, that correspond to the harmonic parts that remain arbitrary, while $\kd$
are the poles of the vector $\cK$, defined as
\begin{equation} \cK= \frac{\kd}{r}\,. \label{K-decomp-1}\end{equation} %
As noted below \eqref{Rstar0-real}, a possible constant part in $\cK$ can be
absorbed in $\Rstz$ and can be disregarded.
The non-harmonic terms in \eqref{H-exp-hor-5} arise by solving \eqref{V-Poiss}
for the only nontrivial component, $V$, by
\begin{gather}
 d\star d V 
 = -\frac1{16}\,\frac1{r^4}\,I_4(\kd, \kd, \cHz, \hat{R})
\quad \Rightarrow \quad
% \CR
V \equiv -\Iprod{\Rstz}{\cHz} =  V_s - V_n
\,.
\end{gather}
The harmonic function $V_s$ is the grade $(+3)$ component of the ones in \eqref{H-exp-hor-5},
which are naturally decomposed as
\begin{equation}\label{harm-decom-5}
 \mathrm{h} + \frac{\gz}{r} = \mathrm{h}^{\ord{-1}} + \frac{\gz^{\ord{-1}}}{r} + \frac14\, V_s\, \hat{R}\,,
\qquad
V_s = \Iprod{\mathrm{h}}{\Rstz} + \frac{\Iprod{\gz}{\Rstz}}{r}\,.
\end{equation}

We may now relate the integration constants in $\cHz$ to the physical charges, by using \eqref{dw-fin-summ}
and the definitions above, to obtain the following equation
\begin{equation}
  dw = -\frac1{r^2}\, \left(\gz - \Tgp_{\kd}\mathrm{h} \right) \star dr\,, 
\end{equation}
so that the charge vector at the given pole is given by
\begin{equation}\label{ch-attr-5}
 \Gamma =  \gz - \Tgp_{\kd}\mathrm{h} \,. 
\end{equation}
Note that the presence of nontrivial T-dualities implies that the poles of $\cHz$ are different
than the charges, which explicitly involve the constant parts of $\cHz$, through $\mathrm{h}$.
Note that once a given set of charges $\Gamma$ is chosen, one can use the fact
that the poles of $\cHz$ and $\Tgp_{\kd}\mathrm{h}$ lie in independent Lagrangian
submanifolds to determine them explicitly.

The final equation to be solved is \eqref{eq:dom-5},
which leads to the solution
\begin{align}\label{M-sol0}
M =  M_{\zer} -\frac1{16}\,I_4(\cK ,\mathrm{h}, \mathrm{h}, \hat{R})
 - \frac1{16}\frac1{r^2}\,I_4(\kd ,\mathrm{h}, \gz, \hat{R})
 - \frac1{48}\frac1{r^3}\,I_4(\kd ,\gz, \gz, \hat{R})
\,.
\end{align}
Here, $M_{\zer}$ stands for an dipole harmonic function describing rotation
through
\begin{equation}\label{single-M}
 \star d \omega = d M_{\zer} \,, \qquad M_{\zer}= m + \J\,\frac{\cos{\theta}}{r^2}\,,
\end{equation}
where $\J$ is the angular momentum along the axis $\theta=0$, as is conventionally
chosen for a single-centre solution.

We now consider the metric starting with the expression for the scale factor, given
in the standard way by
\begin{equation}\label{e4U-sc-5}
 e^{-4U} = -I_4(\cHz) - M^2\,,
\end{equation}
where $\cHz$ and $M$ are given by the expressions above.
Firstly, the quartic invariant can be expanded in the possible combinations of the different
terms in \eqref{H-exp-hor-5}, leading to
\begin{align}
 I_4(\cHz)=&
\frac1{r^4}\,I_4(\gz)
   - \frac1{24\, r^3}\,V_n\, I_4(\gz, \gz, \gz, \hat{R})
\nonumber\\
& - \frac1{8\, r^2}\,V_n\, I_4(\mathrm{h}, \gz, \gz, \hat{R})
- \frac1{8\, r}\,V_n\, I_4(\mathrm{h}, \mathrm{h}, \gz, \hat{R})
+ \mathcal{O}(r^{-3})
\,,\label{I4-decomp-5}
\end{align}
where we omitted the terms of lower order in $r$, that can however be straightforwardly
computed. Using the expression for $V_n$ as given in \eqref{H-exp-hor-5}, one finds poles
of order higher than $4$ in this expression, which lead to unphysical behaviour
near the horizon and must not be present for a physical solution.

We now note that only the grade $(-1)$ components of $\mathrm{h}$ and $\gz$ appear
in the full expression \eqref{I4-decomp-5}, since one can verify that the components
along $\hat{R}$ in \eqref{harm-decom-5} drop out due to the presence of $\hat{R}$ in all
terms involving the quartic invariant. Using the solution \eqref{M-sol0} for the function
$M$, one finds that the terms of order $r^{-6}$ and $r^{-5}$ in \eqref{e4U-sc-5} are
skew-symmetric forms in $\kd$ and $\gz^{\ord{-1}}$, so that they vanish only if the condition
\begin{gather}\label{keql-5}
\kd=\gamma\, \gz^{\ord{-1}}\,,
\end{gather}
is imposed on the poles of the harmonic functions, where $\gamma$ is an arbitrary constant.
Imposing this condition, we obtain the expression
\begin{align}\label{single-metric-5}
 e^{-4U} =& -I_4(\cHz) - M^2
\nonumber\\
=&-\frac1{r^4}\,I_4(\gz)
  - \frac1{192}\, \frac{\gamma^2}{r^4}\,I_4(\gz, \gz, \hat{R}, \gz)
I_4(\mathrm{h}, \mathrm{h}, \gz, \hat{R})
\nonumber\\
&-M_{\zer} \left(M_{\zer} -\frac14\frac{\gamma}{r^2}\,I_4(\gz ,\mathrm{h}, \gz, \hat{R})
 - \frac1{12}\frac{\gamma}{r^3}\,I_4(\gz ,\gz, \gz, \hat{R})\right)
+ \mathcal{O}(r^{-3})\,.
\end{align}
This still contains an unwanted pole of order $5$, proportional to
$I_4(\gz ,\gz, \gz, \hat{R})$, whenever the angular momentum is
nonzero, i.e. in the presence of a dipole harmonic term in $M_{\zer}$. This term can be
easily canceled by adding a dipole harmonic piece in the function $V$, which amounts
to shifting
\begin{equation}\label{H-shift0}
\cHz \rightarrow \cHz
+ \frac1{4}\,\gamma\, \frac{\J\,\cos{\theta}}{r^2}\,\hat{R} \,.
 \end{equation}

Note that in solving all non-harmonic equations above, we did not use the freedom of adding fixed
harmonic pieces in all functions, as the one in \eqref{H-shift0}. Indeed, adding such terms not
only simplifies expressions significantly, but also leads to a simpler identification of charges.
We therefore modify the solution \eqref{H-exp-hor-5} and \eqref{M-sol0} to
\begin{gather}
  \cHz = \cH_s + \frac14\,\gamma\,M_{\zer}\, \hat{R}
-\frac1{384}\, \gamma^2 I_4(\cH_s, \cH_s, \cH_s, \hat{R})\, \hat{R} \,,
\CR
M= M_{\zer} -\frac1{48}\, \gamma I_4(\cH_s, \cH_s, \cH_s, \hat{R})\,,\label{HM-5-str}
\end{gather}
which, by repeating the steps above, leads to the metric function
\begin{align} \label{metr-sing}
 e^{-4U} =& -I_4(\cHz) - M^2 = - I_4(\cH_s) -M_{\zer}^2  \,,
\end{align}
where we have computed the full expression rather than the leading term.
Observe that this is identical to the corresponding single-centre expression
for charges equal to $\gz$, if the poles of $\cH_s$ are taken to be equal to the
charges \eqref{ch-attr-5} computed above.

Using this solution, it is now possible to completely fix the integration
constants in terms of the charges, starting from \eqref{ch-attr-5}, which
is now modified to
\begin{equation}\label{ch-5-n}
  \Gamma =
\gz - \Tgp_{\kd}\mathrm{h} -\frac1{128}\, \gamma^2 I_4(\mathrm{h}, \mathrm{h}, \gz, \hat{R})\, \hat{R} \,,
\end{equation}
after taking into account the additional harmonic term in \eqref{HM-5-str}.
Now, we can use \eqref{keql-5} to rearrange the second term as
\begin{equation}
\Tgp_{\kd}\mathrm{h}= \gamma\,\Tgp_{\mathrm{h}^{\ord{-1}}}\gz\,,
\end{equation}
where we used the linearity of T-dualities and the fact that the grade $(+3)$ component
of $\mathrm{h}$ and $\gz$ drops out from both sides. Combining this with \eqref{Tp2},
the expression \eqref{ch-5-n} takes the rather simple form
\begin{equation}\label{ch-5-fin}
  \Gamma = \exp[\Tgp_{-\gamma\,\mathrm{h}^{\ord{-1}}}]\gz \,,
\end{equation}
so that the charges are indeed given by $\gz$, up to an overall finite T-duality with
parameter $-\gamma\,\mathrm{h}^{\ord{-1}}$.

In order to show that the equivalence with the single-centre solutions is complete, one needs
to show that not only the metric in \eqref{metr-sing}, but also the scalar fields are driven
by a single-centre flow with charges as in \eqref{ch-5-fin}. This computation involves a
local K\"ahler transformation governed by the non harmonic part of $\cHz$ above and leaving
the physical moduli invariant. Such gauge transformations were recently discussed in
\cite{Galli:2012ji}. The interested reader can find an outline of this computation
in appendix \ref{app:kahler-trans}, where we show that an appropriate K\"ahler transformation
indeed brings the section \eqref{scals-5} with $\cHz$ as in \eqref{HM-5-str} to exactly the
single-centre form. The action on the various functions is given in \eqref{kah-trans-VM}-
\eqref{kah-sec} in the general case and in \eqref{sing-complicated} in the basis used in
sections \ref{sec:system}-\ref{sec:explicit-sols}.

A final point worthwhile discussing is the inversion of \eqref{ch-5-fin} to find
the asymptotic constants $ -\gamma\,\mathrm{h}^{\ord{-1}}$ in terms of the charges.
Since this equation is based on a finite T-duality that leaves $\hat{R}$ invariant and acts
nontrivially on $\Rstz$ by definition, one needs to relate a combination of the charges
to the vector $\Rstz$. This can be easily done starting from the expression
\begin{equation}\label{pre-d}
 I_4^\prime(\gz) = \frac1{24}\, I_4(\gz,\gz,\gz,\hat R)\, \Rstz
  + \frac18\, \Iprod{\gz}{\Rstz} I_4^\prime(\gz,\gz,\hat R)\,,
\end{equation}
which is simply the decomposition of the Freudenthal dual of a charge $\gz$ as in
\eqref{sympl-constr-summ}, in its grade $(-3)$ and $(+1)$ components, from which one
also derives
\begin{equation}\label{dec-I4}
  I_4(\Gamma) = I_4(\gz)
= \frac16\, I_4(\gz,\gz,\gz,\hat R)\, \Iprod{\gz}{\Rstz}
= \frac16\, I_4(\Gamma,\Gamma,\Gamma,\hat R)\, \Iprod{\gz}{\Rstz} \,,
\end{equation}
by contracting with the charges. Using the manifest covariance of \eqref{pre-d}, we
can boost $\gz$ to the physical charge $\Gamma$ according to \eqref{ch-5-fin},
to obtain
\begin{gather}\label{d-expr}
\exp[\Tgp_{ -\gamma\mathrm{h}}] \Rstz =
\frac{6\,\Iprod{\hat R}{\Rstz}}{I_4(\Gamma,\Gamma,\Gamma,\hat R)}
\left(I_4^\prime(\Gamma)
  - \frac{3\,I_4(\Gamma)}{ I_4(\Gamma,\Gamma,\Gamma,\hat R) } I_4^\prime(\Gamma,\Gamma,\hat R)\right)
\,,
\end{gather}
where we also used \eqref{dec-I4} to express $\Iprod{\gz}{\Rstz}$ in terms of charges.
The last expression gives the combination $ -\gamma\,\mathrm{h}^{\ord{-1}}$
in terms of the charges in any given frame, once the vectors $R$ and $\Rstz$ are determined.
It follows that once these vectors and the total charge are chosen, one can directly invert
\eqref{ch-5-fin} to obtain the charge $\gz$ that governs the flow. Note that this does
not imply that the above combination is fixed in terms of charges only, since in practice
the constant vector $\Rstz$ depends on asymptotic scalars, as in the single-centre
case discussed in \cite{Bossard:2012xsa}. In terms of an explicit basis, the value
$-\gamma\mathrm{h}$ for the relevant T-duality parameter can be seen explicitly by the
shift the harmonic functions $\vK$ in \eqref{sing-complicated}.

\subsection{Multi-centre flows}
\label{sec:multi-5}
In view of the solutions presented in the previous section on single-centre flows, one can consider
solutions involving multiple centres. In this setting, one has to superpose a set single-centre
black holes, as described above, by allowing for the various functions to have poles in all allowed
centres. Of central importance in this respect is the fact that all centres in a given solution must
be compatible with a single pair of vectors $\hat{R}$ and $\Rstz$, which poses a strong constraint on the
allowed structures.

\subsubsection*{Local structure}
In the composite non-BPS system, all centres carry non-BPS charge vectors and their near horizon
regions are of the type described in the previous section. Considering a multi-centre flow, all but one
function in $\cHz$ continue to be harmonic, in addition to the $\cK$ describing the T-dualities. Therefore,
they take a form similar to \eqref{H-exp-hor-5}, as
\begin{gather}
   \cHz = \cH_s - \frac14\, V_n \, \hat R \,,
\nonumber\\
\cH_s= \mathrm{h} + \sum_\pA\frac{\gzA}{r_\pA}\,,
\qquad
  \cK = \sum_\pA\frac{\kd_\pA}{r_\pA}\,,
\label{hars-multi-5}
\end{gather}
where $V_n$ contains the non-harmonic part of the function $V$, to be
determined below. In order to have a regular solution near the centres,
we need to impose the restrictions found for the single-centre case above,
and in particular \eqref{keql-5}, so that the poles of $\cH_s$ and $\cK$
must be colinear at every centre, i.e.
\begin{gather}\label{keql-5-multi}
\kd_\pA= \gamma_\pA\, \gzA^{\ord{-1}}\,,
\end{gather}
where $\gamma_\pA$ are a set of constants.

We can now use these expressions to obtain the non-harmonic functions $V_n$
and $M$, by solving \eqref{V-Poiss} and \eqref{M-Poisson-5} respectively.
As shown in \cite{Bossard:2012ge}, it is possible to find the exact solutions
for these functions in terms of the function $F_{\pA\pB,\pC}$, defined as the
everywhere regular solution to the Poisson equation
\begin{equation}\label{F-triple-def}
 d\star d F_{\pA\pB,\pC}=\frac{1}{r_\pC}\,d\star d\left( \frac{1}{r_\pA} \frac{1}{r_\pB} \right)\,.
\end{equation}
While the existence and regularity of this function was shown in
\cite{Bossard:2012ge}, it cannot be expressed by elementary functions generically,
but only in the special case when all three centres $A,\,B,\,C$ are aligned.
Additionally, in view of the discussion in the previous section, we find it
convenient to include a harmonic part in each of these functions, both to impose
regularity of the metric at each pole, as well as to simplify some expressions.
The complete expressions for these two functions are as follows
\begin{align}\label{V-gen}
 V_n =&\,  \sum_\pA\gamma_\pA\,\J_{i\,\pA} \,\frac{r^i_\pA}{r^3_\pA} 
      -\frac1{96}\sum_{\pA} \gamma^2_\pA
I_4\Scal{\frac{\gzA}{r_\pA}, \frac{\gzA}{r_\pA}, 3\,\mathrm{h} + \frac{\gzA}{r_\pA}, \hat R}
\CR
 &\,
-\frac1{32}\,\sum_{\pA\neq \pB} \gamma_\pA\gamma_\pB \, 
I_4\Scal{\frac{\gzA}{r_\pA}, \frac{\gzB}{r_\pB}, \mathrm{h} +\!\!\! \sum_{\pC\neq\{\pA,\pB\}} \frac{\gzC}{r_\pC} , \hat R}
\nonumber\\
&\,
-\frac1{32}\,\sum_{\pA\neq \pB} \gamma_\pA\, I_4(\gzA, \gzA, \gzB, \hat R)
\left( \frac{  \gamma_\pB}{\rc{A}^2 \rc{B}}
     + \frac{ \gamma_\pA -  \gamma_\pB}{\RABsq} \left(\frac{\rc{B}}{\rc{A}^2} - \frac{1}{\rc{B}} \right)\right) \,
\nonumber\\
&\,
+\frac1{16}\,\sum_{\pA\neq \pB\neq \pC}  \gamma_\pA  \gamma_\pC
\, I_4(\gzA, \gzB, \gzC, \hat R)
 \,\left( F_{\pAB,\pC} + \frac1{\RAB \RBC} \frac1{\rc{B}} \right)
\,,
\end{align}
\begin{align}\label{M-gen}
M =&\, m + \sum_\pA \J_{i\,\pA} \,\frac{r^i_\pA}{r^3_\pA}
% \nonumber\\
% &\,
- \frac1{48}\,\sum_\pA \gamma_\pA\,
I_4\Scal{\mathrm{h}+ \frac{\gzA}{r_\pA}, \mathrm{h}+ \frac{\gzA}{r_\pA}, \mathrm{h}+ \frac{\gzA}{r_\pA}, \hat R}
\nonumber\\
&\,
-\frac1{16}\,\sum_{A\neq B} \gamma_\pA\, I_4(\gzA, \gzB, \mathrm{h}, \hat R)
\left( \frac{1}{\rc{A}} \frac{1}{\rc{B}} + \frac1{\RAB} \frac{1}{\rc{A}}
 - \frac1{\RAB} \frac{1}{\rc{B}} \right)
\nonumber\\
&\,
-\frac1{32}\,\sum_{\pA\neq \pB} I_4(\gzA, \gzA, \gzB, \hat R)
\left( \frac{ \gamma_\pA +  \gamma_\pB}{\rc{A}^2 \rc{B}}
     + \frac{ \gamma_\pA -  \gamma_\pB}{\RABsq} \left(\frac{\rc{B}}{\rc{A}^2} - \frac{1}{\rc{B}} \right)\right) \,
\nonumber\\
&\,
-\frac1{16}\,\sum_{\pA\neq \pB\neq \pC}  \gamma_\pA
\, I_4(\gzA, \gzB, \gzC, \hat R)
 \,\left( F_{\pA\pB,\pC} + \frac1{\RAB \RBC} \frac1{\rc{B}} \right)
\,,
\end{align}
where $\J_{i\,\pA}$ is the angular momentum vector associated to the centre $A$.
Note that both these expressions contain harmonic parts, chosen so that the appropriate
behaviour near each centre is obtained,
in direct analogy with \eqref{HM-5-str}. Upon specifying to the frame \eqref{fixedframe},
one easily recovers the results of \cite{Bossard:2012ge}, while the general choice
\eqref{SimilarlyR}-\eqref{RsK} similarly leads to \eqref{V-frame}-\eqref{M-frame}.

Using these expressions, one can show that the expression for the charges
at each centre takes a form very similar to the single-centre result
\eqref{ch-5-fin}, as
\begin{equation}\label{ch-full-5}
 \Gamma_\pA = \exp[\Tgp_{\fd_\pA}] \gzA\,.
\end{equation}
Here and in the following, we use the combination
\begin{align}\label{comp-const}
\fd_\pA= \sum_{\pB\neq \pA}\frac{\kd_\pB}{\RAB}- \gamma_\pA\mathrm{h}^{\ord{-1}}_\pA
       = \sum_{\pB\neq \pA}( \gamma_\pB -  \gamma_\pA)\frac{\gzB^{\ord{-1}}}{\RAB} -  \gamma_\pA\mathrm{h}^{\ord{-1}}\,,
\end{align}
which is given in terms of the constant parts of the harmonic functions
$\cH_s$ and $\cK$ at each pole.
Note that there are multiple ways of casting \eqref{ch-full-5}, in particular one
can use the relations \eqref{Tp2} to find the expansion
\begin{equation}\label{ch-full-5-alt}
\Gamma_\pA = \gzA
-\frac18\, I_4^{\prime}(\fd_\pA, \gzA, \hat{R})
-\frac1{64}\, I_4 (\fd_\pA, \fd_\pA , \gzA, \hat{R})\,\hat{R}
\,,
\end{equation}
which will prove useful in what follows.
Finally, note that following the arguments in \eqref{pre-d} - \eqref{d-expr}
we can solve for the combination of asymptotic constants
$\fd_\pA$ as
\begin{gather}\label{d-expr-mult}
\exp[\Tgp_{\fd_\pA}] \Rstz =
\frac{\Iprod{\hat R}{\Rstz}}{I_4(\Gamma_\pA,\Gamma_\pA,\Gamma_\pA,\hat R)} I_4^\prime(\Gamma_\pA)
  - \frac{\Iprod{\hat R}{\Rstz}\,I_4(\Gamma_\pA)}
  {\left(I_4(\Gamma_\pA,\Gamma_\pA,\Gamma_\pA,\hat R)\right)^2} I_4^\prime(\Gamma_\pA,\Gamma_\pA,\hat R)
\,,
\end{gather}
which must hold at each centre independently. Similar to \eqref{d-expr}, one can
write the solution as
\begin{equation} 
 \fd_\pA =
\frac{\Iprod{\hat R}{\Rstz}}{I_4(\Gamma_\pA,\Gamma_\pA,\Gamma_\pA,\hat R)}
\mathsf{P}(I_4^\prime(\Gamma_\pA))^{\ord{-1}}\,.
\end{equation}
Note that this relation explicitly contains the distances between the various
centres, through \eqref{comp-const}. This property will be used in the following,
in the discussion of the global structure of solutions.
Once again, writing the formal expression \eqref{d-expr-mult} in terms of an explicit
basis leads to a simple identification of T-duality parameters, as in \eqref{BubbleTframe}.

\subsubsection*{Global structure}
Turning to the global features of multi-centre solutions, we consider the total
angular momentum. Inserting \eqref{hars-multi-5} and \eqref{M-gen} in \eqref{eq:dom-5},
we obtain the full expression for the total angular momentum one-form as
\begin{align}
\omega =  &\,
\sum_\pA \varepsilon_{ijk} \frac{ \J_{\pA}^i\, r_\pA^j\, dx^k}{ r_\pA^3}
-\frac18\! \sum_{\pA\ne \pB\ne \pC}\!  \gamma_\pC\, I_4(\gzA, \gzB, \gzC, \hat R) \; \omega_{\pA\pB,\pC}
\nonumber\\
&\,
-\frac18\, \sum_{\pA\ne \pB} \frac{  (  \gamma_\pA- \gamma_\pB)\, I_4(\gzA, \gzB, \mathrm{h}, \hat R)\,
 \, \varepsilon_{ijk}\, \RABin\, r_\pB^j\, dx^k }{\RAB\, r_\pA\, r_\pB\, \scal{ r_\pA + r_\pB + \RAB}}
\nonumber\\
&\,
-\frac18\,\sum_{\pA \ne \pB} (  \gamma_\pA- \gamma_\pB)
\frac{ I_4(\gzA, \gzB, \gzC, \hat R)\,
\varepsilon_{ijk}\, \RABin\, r_\pA^j\, dx^k }{\RABsq\, r_\pA^2\, r_\pB}\,,
\label{omega-gen}
\end{align}
where $\omega_{AB,C}$ is defined as the solution to
 \begin{equation}
\star d \omega_{\pA\pB,\pC} = d \Scal{ F_{(\pA,\pB)\pC} + \frac{1}{\RAC \, \RBC \, r_\pC } }
  - \frac{1}{r_\pA\,r_\pB } d \frac{1}{r_\pC}\,.
\end{equation}
In order to extract the angular momentum, we expand in the asymptotic region, using
the asymptotic expansion for the function $F_{\pA\pB,\pC}$ as given in
\cite{Bossard:2012ge}, along with the corresponding contribution to the
angular momentum one-form through \eqref{eq:dom-5}. One can verify that the
resulting expression for the asymptotic total angular momentum is
 \begin{eqnarray}
\J^i_{\text{t}} &=&  \sum_\pA \J_\pA^i
+\frac18\,\sum_{A>B} \frac{\RABin}{\RAB} \biggl(
 I_4(\gzA, \gzB, \fd_\pA, \hat R)
 - I_4(\gzA, \gzB, \fd_\pB, \hat R)
 \biggr)
\CR &&
+\frac18\, \sum_{\pA\ne \pB\ne \pC}  \gamma_\pC \,
I_4(\gzA, \gzB, \gzC, \hat R)
\frac{ \RAB ^2 \, R^i_\pBC - \RAB \hspace{-1mm}\cdot \hspace{-1mm} \RBC\, R^i_\pAB }{ \RAB\, \RAC\, \RBC\,  \scal{ \RAB  + \RAC  + \RBC  }}  \CR
 &=&
\sum_\pA \J_\pA^i
-   \sum_{\pA>\pB} \, \Iprod{\Gamma_\pA}{\Gamma_\pB}\, \frac{\RABin}{\RAB}
 \hspace{35mm}
\CR &&
+\frac18\, \sum_{\pA\ne \pB\ne \pC}  \gamma_\pC \,
I_4(\Gamma_\pA, \Gamma_\pB, \Gamma_\pC, \hat R)
\frac{ \RAB ^2 \, R^i_\pBC - \RAB \hspace{-1mm}\cdot \hspace{-1mm} \RBC\, R^i_\pAB }{ \RAB\, \RAC\, \RBC\,  \scal{ \RAB  + \RAC  + \RBC  }} \,,
\label{NonBPSJ}
 \end{eqnarray}
where in the second equality we used \eqref{ch-full-5-alt} to rewrite
the second sum as the inner product of charges at all the centres, and the property that  $I_4(\gzA, \gzB, \gzC, \hat R)=I_4(\Gamma_\pA, \Gamma_\pB, \Gamma_\pC, \hat R)$ because of the grading.
Note that, while the first two terms are standard, the third term in this
expression, resulting from the asymptotic expansion of $\omega_{\pA\pB,\pC}$,
is rather non-standard and appears only when not all centres lie on a line.

The final step is to consider the spatial structure of the solution,
by fixing the distances between the centres through the interactions
described above. To this end, we define the antisymmetric combination
\begin{equation}
\fd_{\pA\pB}\equiv \fd_\pA - \fd_\pB\,,
\end{equation}
which is fixed in terms of charges by \eqref{d-expr-mult}. Using
\eqref{comp-const}, this can be written as
\begin{align}\label{distan-5}
\fd_{\pA\pB}= &\,
\sum_{\pC\neq \pA}( \gamma_\pC-  \gamma_\pA)\,\frac{\gzC^{\ord{-1}}}{\RAC}
  -\sum_{\pC\neq \pB}( \gamma_\pC-  \gamma_\pB)\,\frac{\gzC^{\ord{-1}}}{\RBC}
 -( \gamma_\pA -  \gamma_\pB)\,\mathrm{h}^{\ord{-1}}\,,
\end{align}
which is the covariant version of the relation \eqref{BubbleTframe}, as given in the
explicit basis of section \ref{sec:system}. 
Note that due to the presence of all distances between all $N$ centres,
one can use \eqref{distan-5} to constrain their values.

To obtain explicit solutions to \eqref{distan-5} however, one has to appreciate the fact
that while the $\mathrm{h}$ parametrise (some of) the asymptotic scalars, all other
terms are fixed in terms of charges at each centre by \eqref{d-expr-mult}, so that
this set of equations is overconstrained. This was shown in detail in section
\ref{sec:explicit-sols}, where we saw the emergence of hypersurfaces on which the
composite solutions are constrained to exist. It would be interesting to investigate
whether particular contractions of \eqref{distan-5} can be used to study these properties
directly, \ie without going to an explicit basis.

\section{Conclusion}
\label{sec:concl}

In this paper, we presented the first detailed analysis of the properties of non-BPS
black hole bound states in extended supergravity, which is not attached to a specific duality frame. This permitted us to study the domain of existence of these solutions  in moduli space for fixed electromagnetic charges. In particular, we showed explicitly
the existence of walls of marginal stability where the binding energy vanishes. Moreover we define the notion of a non-BPS attractor flow tree as a criterion of existence for these solutions. This was done for a relatively simple subclass of non-BPS extremal solutions, corresponding to
the so-called composite non-BPS system of black hole solutions. These correspond to a
system of black holes each carrying a non-BPS charge (of negative quartic invariant) and an angular momentum that is
bounded above by the charges.

The results derived in section \ref{sec:explicit-sols} show that all the features familiar
from the study of multi-centre BPS solutions appear for non-BPS composites as well. The
only crucial difference is that, while BPS solutions a priori exist on codimension zero
subspaces of moduli space, non-BPS solutions can only exist on specific hypersurfaces in moduli space,
depending on the charges involved \cite{Bossard:2011kz, Bossard:2012ge}. One of our main technical result is to prove that the composite solutions always carry a non-trivial binding energy between the constituents, exhibiting that they indeed define bound states (with the exception of the solutions of the $T^3$ model). In order to arrive to this conclusion we showed that the fake superpotential, defined in \cite{Ceresole:2009vp} as a function of auxiliary parameters associated to the flat directions of the individual centres, defines the mass of a single-centre non-BPS black hole at its global maximum. Using the property that the mass of a composite solution is defined from the same function at different values of these auxiliary parameters, one concludes that the binding energy is always positive.

As for the BPS solutions, the distances between the centres are determined in terms of the individual charges and the moduli. We show explicitly that the distance between two centres (or two clusters of centres) diverges for finite values of the asymptotic scalars. Moreover, we prove that the corresponding binding energy vanishes in this limit, exhibiting that it defines a wall of marginal stability in moduli space. These domains indeed define codimension one boundaries of the hypersurface on which the solution exists in moduli space.

As it turns out, the auxiliary parameters associated to flat directions play a very similar
role to that of the K\"ahler phase of the central charge for BPS solutions, leading to a natural
notion of attractor flow tree. Indeed, as explained in section \ref{sec:attr-tree}, a solution
may only exist if a wall of marginal stability exists, on which the values for the auxiliary parameters
for the constituents are the same as for the bound state. It then follows that any solution
can be assembled in this way, so that one can associate an attractor flow tree to any such
composite solution. It is therefore tempting to conjecture that the reverse would be true, \ie that
the existence of such an attractor flow tree would imply the existence of a solution, as proposed
in \cite{Denef:2000nb} for BPS solutions.

Note that the property that these solutions only exist on a hypersurface in moduli space may be an artifact of the composite non-BPS system we are solving, rather than a physical property. The system somehow forces us to restrict ourselves to a hypersurface without boundaries in moduli space, on which we can identify a boundary carrying all physical properties of a wall of marginal stability. If we know that there is no deformation of our solutions in the normal directions to this hypersurface within the composite non-BPS system of equations, there may exist more general extremal solutions that would extend the domain of existence to a codimension zero domain in moduli space. This would require to give up some special properties of these solutions, as for example the condition that the three-dimensional Euclidean base is flat.

There are a number of future directions related to the above developments.
First, it would be interesting to extend the analysis to more general systems of black
hole composites. The most obvious such example is to allow for BPS charges as well, so
that a BPS/non-BPS system of charges may arise. This is described by the almost-BPS system
\cite{Goldstein:2008fq, Bena:2009ev, Bena:2009en, Bossard:2013oga}, which can be treated
in a very similar fashion. Further extensions may involve solutions that do not admit a flat
three-dimensional base space, which are however much less understood and there is no known
system of equations to describe them systematically. Finally, it would be very interesting
to understand the possible higher dimensional origin of the hypersurfaces in moduli space
on which the non-BPS solutions are constrained to exist.

%%%%%%%%%%%%%%%%%%%%%%%%%%%%%%%%%%%%%%%%%%%%%%%%%%%%%%%%%%%%%%%%
\section*{Acknowledgement}
%%%%%%%%%%%%%%%%%%%%%%%%%%%%%%%%%%%%%%%%%%%%%%%%%%%%%%%%%%%%%%%%
This work was supported by the French ANR contract 05-BLAN-NT09-573739, the ERC
Advanced Grant no. 226371 and the ITN programme PITN-GA-2009-237920. The work of
SK was supported in part by the ANR grant 08-JCJC-0001-0, and by the ERC Starting
Independent Researcher Grant 240210-String-QCD-BH.
%%%%%%%%%%%%%%%%%%%%%%%%%%%%%%%%%%%%%%%%%%%%%%%%%%%%%%%%%%%%%%%%
%%%%%%%%%%%%%%%%%%%%%%%%%%%%%%%%%%%%%%%%%%%%%%%%%%%%%%%%%%%%%%%%

\begin{appendix}

\section{T-dualities}
\label{sec:T-duality}

In this appendix, we discuss in some detail the properties of T-duality operators, following \cite{Bossard:2013oga},
and indicate how to obtain explicit parametrisations for their action. The discussion is based on two very small
vectors, $R$ and $R^*$, which are ultimately identified to the two vectors used to describe composite non-BPS
solutions in \eqref{eq:R-gen} and \eqref{RsK}.

The description of T-dualities is based on the grading of the symplectic vector space
according to the eigenspaces of the generator
\begin{equation}
{\bf h}_T\, \Gamma \equiv \Iprod{R}{R^*}^{-1} \Scal{  \frac{1}{2} I_4^{\prime}( R,R^* ,\Gamma)  + \Iprod{\Gamma}{R^*} R  -R^* \Iprod{R}{\Gamma}} \  , \label{hTReal}
\end{equation}
in terms of its eigenvalues, $\pm 1$, $\pm 3$, as
\begin{equation} \label{eq:vec-decomp}
\mathds{R}^{2n_v+2} \cong \mathds{R}^\ord{-3} \oplus  ({\mathds{R}}^{n_v})^\ord{-1}  \oplus  ({\mathds{R}}^{n_v})^\ord{1} \oplus \mathds{R}^\ord{3} \ .
\end{equation}
The corresponding projectors to each of the four eigenspaces are given by
\begin{eqnarray}
 \Gamma^\ord{3}  &=& \Iprod{R}{R^*}^{-1} \Iprod{\Gamma}{R^*} R\ ,  \CR
 \Gamma^\ord{1} &=& \frac{1}{2} \Gamma+ \frac{1}{2}\Iprod{R}{R^*}^{-1}  \Scal{  \frac{1}{2} I_4^{\prime}( R,R^* ,\Gamma)  - 3 \Iprod{\Gamma}{R^*} R  +R^* \Iprod{R}{\Gamma}} \ , \CR
  \Gammamun &=& \frac{1}{2} \Gamma- \frac{1}{2}\Iprod{R}{R^*}^{-1}  \Scal{  \frac{1}{2} I_4^{\prime}( R,R^* ,\Gamma)  -  \Iprod{\Gamma}{R^*} R  +3R^* \Iprod{R}{\Gamma}} \ , \CR
 \Gammamtrois  &=& \Iprod{R}{R^*}^{-1} \Iprod{R}{\Gamma} R^* \ . \label{RProjector}
\end{eqnarray}
This construction allows for practical simplifications, since all inner products must
respect the grading. For instance, the grading implies that
\begin{equation}
I^\prime(\Gammamun,\Gammamun,R^*)
=  0 \ , \qquad I^\prime(\Gamma^\ord{1},\Gamma^\ord{1},R)=  0  \ ,
\end{equation}
since there is no vector of weight $\pm 5$ that these cubic terms could be equal to.
Similar considerations apply to scalar products, which necessarily vanish unless the
sum of grades of the vectors involved is zero.

As shown explicitly in \cite{Bossard:2013oga}, one may consider any grade $-1$ vector of
parameters $\kmun$ to define the grade 2 T-duality generators as
\begin{equation}
 \Tgp_k \Gamma = \Iprod{R}{R^*}^{-1}  \Scal{  \kmun \Iprod{R}{\Gamma^\ord{-3}} -   \frac{1}{4} I_4^{\prime}( R,\kmun ,\Gammamun)    - \Iprod{\Gamma^\ord{1}}{\kmun} R } \  , \label{Tp1}
\end{equation}
where $\Gamma$ is a generic symplectic vector and $\Gamma^\ord{\pm 3}$, $\Gamma^\ord{\pm 1}$
are its components of each respective grade.
All these generators clearly commute between themselves for different $\kmun$'s.
Similarly, one defines the grade $-2$ generator in terms of a grade $1$ vector $k^\ord{1}$
\begin{eqnarray}
\Tgm_k \Gamma &\equiv&
\Iprod{R}{R^*}^{-1}  \Scal{k^\ord{1} \Iprod{\Gamma^\ord{3}}{R^*}  + \frac{1}{4} I_4^{\prime}( R^*,k^\ord{1} ,\Gamma^\ord{1})  - \Iprod{k^\ord{1}}{\Gammamun} R^*  } \  . \label{Tm1}
\end{eqnarray}
The normalisations we have chosen are such that
\begin{equation}\label{t-dual-params}
 \Tgp_k R^* = \kmun \ , \qquad \Tgm_k R = k^\ord{1} \ ,
\end{equation}
while one easily computes that
\begin{equation}\label{R-Rst-inv}
\Tgp_k R = 0 \ , \qquad \Tgm_k R^* = 0 \ .
\end{equation}
Conversely, any grade $\pm 1$ vectors can be re-expressed in terms of a T-duality
acting as in \eqref{t-dual-params}, while the T-dualities can be defined by specifying
the invariant very small vectors, as in \eqref{R-Rst-inv}.
In this form, one easily computes that these generators are nilpotent of order 4, as
\begin{equation}\label{eq:nil-cond-T}
 (\Tgpm_k)^{4} \Gamma = 0 \ ,
\end{equation}
consistent with the grading \eqref{eq:vec-decomp}, which only allows for four eigenspaces.
Explicitly, we find the following expressions for the two sets of generators
\begin{eqnarray}
 (\Tgp_k)^{ 2} \Gamma&=& - \frac{1}{4}  \Iprod{R}{R^*}^{-2} \Scal{ I_4^\prime(R,\kmun,\kmun) \Iprod{R}{\Gamma} + I_4(R,\kmun,\kmun,\Gamma) R } \  , \label{Tp2} \\
(\Tgp_k)^{3} \Gamma &=&- \frac{1}{4}  \Iprod{R}{R^*}^{-3}  I_4(R,\kmun,\kmun,\kmun) \Iprod{R}{\Gamma} R \ , \label{Tp3}\\
(\Tgm_k)^{ 2} \Gamma&=&  \frac{1}{4}  \Iprod{R}{R^*}^{-2} \Scal{ I_4^\prime(R^*,k^\ord{1},k^\ord{1}) \Iprod{\Gamma}{R^*} - I_4(R^*,k^\ord{1},k^\ord{1},\Gamma) R^* } \  , \label{Tm2}\\
(\Tgm_k)^{3} \Gamma &=&-\frac{1}{4}  \Iprod{R}{R^*}^{-3}  I_4(R^*,k^\ord{1},k^\ord{1},k^\ord{1}) \Iprod{\Gamma}{R^*} R^* \ . \label{Tm3}
\end{eqnarray}
Finally, a finite T-duality is defined by the exponential of $\Tgpm$, as
\begin{equation}\label{expTdual}
\exp[\Tgpm_k] = 1 + \Tgpm_k + \tfrac12\, (\Tgpm_k)^2 + \tfrac16\, (\Tgpm_k)^3\,,
\end{equation}
where we used \eqref{eq:nil-cond-T}. In the main text, we always use \eqref{expTdual}
in combination with \eqref{Tp1}, \eqref{Tm1} and \eqref{Tp2}-\eqref{Tm3} to compute
the action of a general T-duality on an arbitrary vector in a general frame.

In order to construct particular black hole solutions, it is necessary to give a
representation of $\Tgpm$ explicitly, which is simplified by observing that the
variety of very small vectors, such as $R$ and $R^*$, can be generated by action
of any T-duality on any very small vector that is not invariant under
it\footnote{Note that a given
parametrisation does not generically cover all possible very small vectors,
but it is always possible to find a parametrisation that is non-singular for a
given vector.}. Therefore, we can choose any distinguished pair of T-dualities,
such as the spectral flows $\cTgpm$ in \eqref{Tminus0}-\eqref{Tplus0}, to obtain an explicit
representation of all T-dualities.

More concretely, the most general transformation that brings the charge along
$q_0$ to the most general vector $R$ is given by
$\exp\Scal{ \cTgm_{ k^{-}_0 } } \exp\Scal{ \cTgp_{ k^{+}_0 } } $, so that
\begin{equation}\label{eq:R-rot}
 R_{k^{-}_0} = \exp\Scal{ \cTgm_{ k^{-}_0 } } \exp\Scal{ \cTgp_{ k^{+}_0 } } \overset{\,\circ}{R}
         = \exp\Scal{ \cTgm_{ k^{-}_0 } } \overset{\,\circ}{R} \,,
\end{equation}
where  $k^\pm_0$ are arbitrary parameters of grade $(\mp 1)$. Here,
$\overset{\,\circ}{R}$ is the vector along $q_0$ and we used that all T-dualities $\cTgp$
are defined as leaving $\overset{\,\circ}{R}$ invariant. The associated $R^*$ then takes the form
\begin{equation}\label{eq:R_st-rot}
\qquad
 R^*_{k^{-}_0, k^{+}_0} = \exp\Scal{ \cTgm_{ k^{-}_0 } } \exp\Scal{ \cTgp_{ k^{+}_0 } } \overset{\,\circ}{R}{}^*\,,
\end{equation}
where $\overset{\,\circ}{R}{}^*$ is the vector along $p^0$ and we stress the fact that the
new $R^*$ depends on both $k^{\pm}_0$.

It now follows that all T-dualities can be obtained by conjugating
the simple spectral flows \eqref{Tminus0} by duality transformations above,
so that, \eg
\begin{align}\label{eq:T-conj}
 \Tgp_{k^{+}} = &\,
\exp\Scal{ \cTgm_{ k^{-}_0 } } \exp\Scal{ \cTgp_{ k^{+}_0 } }
\cTgp_{k^{+}}
\exp\Scal{ \cTgp_{ - k^{+}_0 } }\exp\Scal{ \cTgm_{ - k^{-}_0 } }
\CR
= &\,
\exp\Scal{ \cTgm_{ k^{-}_0 } }  \cTgp_{k^{+}}  \exp\Scal{ \cTgm_{ - k^{-}_0 } }  \ .
\end{align}
The representation for the dual T-dualities $\Tgm$, can be easily obtained
from \eqref{eq:T-conj}, as
\begin{equation}\label{eq:T-conj-dual}
  \Tgm_{k^{-}} =
\exp\Scal{ \cTgm_{ k^{-}_0 } } \exp\Scal{ \cTgp_{ k^{+}_0 } }
\cTgm_{k^{-}}
\exp\Scal{ \cTgp_{ - k^{+}_0 } }\exp\Scal{ \cTgm_{ - k^{-}_0 } } \ .
\end{equation}
These operators leave $R^*_{k^{-}_0, k^{+}_0}$invariant by construction,
but are not useful for the discussion of composite non-BPS solutions.
In the main text, we use the representation \eqref{eq:T-conj} to do explicit
computations of T-dualities in a duality covariant setting, using only the
simple spectral flows \eqref{Tminus0}.

\section{Physical moduli and local K\"ahler transformations}
\label{app:kahler-trans}

We summarise some of the relevant formulae for computing the physical
moduli in terms of the components of the symplectic section. Our starting
point is the expression
\begin{equation}\label{i-sec-app}
 2\, \mbox{Im}( e^{-U-i\alpha} \cV ) = - \cH + \frac{1}{2}\, V\, \hat R  + \frac{M}{V} \, \hat{R}^*  \ ,
\end{equation}
which gives the scalars in all solutions discussed in this paper, up
to an overall T-duality, which can be applied on the final moduli. From
this, one can compute the real part of the section by a straightforward
evaluation of the general solution, given by
\begin{align} 
e^{-4U} = &\, I_4\Scal{- \cH + \frac{1}{2}\, V\, \hat R  + \frac{M}{V} \, \hat{R}^*}
 = - I_4(\cH) -M^2\,,
\CR
 2\,\mbox{Re}( e^{-U-i\alpha} \cV ) = &\,
-\frac{1}{2}\,e^{2U} I_4^\prime\Scal{- \cH + \frac{1}{2}\, V\, \hat R  + \frac{M}{V} \,\hat{R}^* }\,.
\end{align}
Expanding the last expression in components along each grade, we obtain
the following result
\begin{align}\label{r-sec-app}
 2\,e^{-2U} \mbox{Re}( e^{-U-i\alpha} \cV ) = &\,
\frac{I_4(\cH)}{2 V}\, \hat{R}^*
-\frac{1}{16}\,V\,I_4^\prime(\cH, \cH , \hat R)
+ \frac{M^2}{V} \, \hat{R}^* -  M \,\cH\,.
\end{align}
We can now write the complete expression for the section as
\begin{align}
 2\, e^{-U-i\alpha} \cV = &\,
\frac{1}{2V}e^{-2U}\left( e^{2U} M + i \right)^2\, \hat{R}^*
-\left( e^{2U} M + i\right)\,\cH
\CR
&\,
-\frac{1}{16}\,e^{2U} V\,I_4^\prime(\cH, \cH , \hat R)
+ i\,\frac12\,V \, \hat{R} \,, \label{ComplexSection}
\end{align}
from which follows the solution for the physical moduli.

In this paper we have shown that a single-centre solution can be described within the composite non-BPS system with non-trivial functions $\mathcal{K}$. This rewriting of the single-centre solution can be reabsorbed in a T-duality, which requires a modification of the K\"{a}hler phase defining the system in order to be identified with the single-centre solution in its standard form. To prove this, we will show the existence of a phase $\alpha_\zer$ such that 
\begin{align}\label{kah-sec}
  2\, \mbox{Im}(  e^{-U-i\alpha_\zer} \cV ) =&\,
\exp\Bigl[-\gamma\,\Tgp_{\cH^{{\mbox{\fontsize{3.35pt}{3pt}\selectfont(-1)\fontsize{12.35pt}{12pt}\selectfont}}}}\Bigr]\,
\left(- \cH^\ord{-1} + \frac{1}{4}\, V_{\zer} \, \hat R  + \frac{M_{\zer}}{V_{\zer}} \, \hat{R}^* \right)\,,
\end{align}
where the functions $V_\zer$ and $M_\zer$ are related to the original ones through
\begin{align}\label{kah-trans-VM}
 V=&\, V_{\zer} + \gamma\,M_{\zer} +\frac{\gamma^2}{4}\frac{e^{-4U} + M_\zer^2 }{V_{\zer}}\,,
\CR
 M=&\, M_{\zer} +\frac{\gamma}{2}\frac{e^{-4U} + M_\zer^2}{V_{\zer}}\,,
\end{align}
for some constant real parameter $\gamma$. Here, we used the grade $-1$ component $\cH^\ord{-1}$
for convenience, noting that one can straightforwardly define a new vector as
$\cH^\ord{-1} + \frac{1}{4}\, V_{\zer} \, \hat R$, to match with the standard form \eqref{i-sec-app}.
One can easily verify that $e^{-4U}$ is invariant under these transformations, and one computes that \eqref{kah-sec} is indeed satisfied for $\cV$ defined as \eqref{ComplexSection} and
\begin{equation}
 e^{i(\alpha-\alpha_\zer)} = 1 -\frac{\gamma^2}{2}\frac{1}{V\,V_{\zer}} e^{-4U}
 + i\,\frac{\gamma}{V\,V_{\zer}}\,e^{-2U}\left(V_{\zer} + \frac12\,\gamma\,M_{\zer}\right)\, .
\end{equation}

\section{Stabilizer of two charges}
\label{app:stabilizer}

In this appendix we briefly discuss the stabiliser of an electric and a magnetic vector of charges in five
dimensional supergravity coupled to a symmetric scalar manifold. This stabiliser defines the possible
flat directions of the example two-centre solution in four dimensions constructed in section
\ref{sec:two-cent-ex}. Indeed, since one of the centres is chosen to carry a D0-D6 charge, the possible
flat directions are classified by exactly the stabiliser of the $\vp$, $\vq$ charges at the second
centre, and is therefore identical to a five dimensional computation.

Let us consider the exceptional theory for which the five dimensional duality group is $E_{6(-26)}$.
From this example it is completely straightforward to extend the results to all other symmetric theories
with a cubic prepotential, since the computation would go exactly the same way for the three other magic
supergravity theories, and is even simpler for the infinite series of axion-dilaton theories. It is
convenient to consider the following graded decomposition of $E_{6(-26)}$, which arises by viewing
the five dimensional theory as the Kaluza-Klein reduction of a six dimensional theory of
duality group $Spin(1,9)$ 
\begin{equation} \mathfrak{e}_{6(-26)} \cong \overline{\bf 16}^\ord{-3} \oplus \scal{ \mathfrak{gl}_1 \oplus \mathfrak{so}(1,9)}^\ord{0} \oplus {\bf 16}^\ord{3}\,, \end{equation}
with respect to which the fundamental representation decomposes as 
\begin{equation}\label{5d-ch-dec}
 {\bf 27} \cong {\bf 1}^\ord{-4} \oplus {\bf 16}^\ord{-1} \oplus {\bf 10}^\ord{2} \ .  
\end{equation}
In the corresponding decomposition of the five dimensional vector fields in terms of the six-dimensional
field components, the singlet comes from the six-dimensional metric, the spinor from the six-dimensional
1-forms, and the vector from the six-dimensional 2-forms. Let us write the electric and magnetic charges
according to \eqref{5d-ch-dec} as 
\begin{equation}
 \vq = (q_1, \chi, q^a)\,, \qquad \vp = ( p_1, \psi , p^a)\,, 
\end{equation}
respectively, where $q_1$, $p_1$, are real numbers, $\chi$, $\psi$ are commuting $Spin(1,9)$ Majorana--Weyl spinors  of
opposite chirality and $q^a$, $p^a$ are vectors. One then obtains that 
\begin{gather}
 \det \vp = p_1 p_a p^a - 2\, p_a \bar \psi \gamma^a \psi \ , \quad 
\det \vq = q_1 q_a q^a + 2\, q_a \bar \chi \gamma^a \chi\,,
\CR
\tr \vp \vq = p_1 q_1 + 4\, \bar \psi \chi + 2\, p_a q^a \ .  
\end{gather}
The action of the $GL(1) \times Spin(1,9)$ subgroup on these components is manifest, so we shall
only display the transformations associated to the other generators, parameterised by spinor
parameters $\Lambda_+,\Lambda_-$ of opposite chirality, as
\begin{equation}
\begin{split} \delta p_1 &= 4\, \bar \Lambda_- \psi\,, \\
\delta \psi &= p_1 \Lambda_+ - p_a \gamma^a \Lambda_-\,, \\
\delta p^a &= 2\, \bar \Lambda_+ \gamma^a \psi \,,
\end{split}\hspace{10mm}\begin{split}
\delta q_1 &=- 4\, \bar \Lambda_+ \chi \,, \\
\delta \chi &= q_1 \Lambda_- - q_a \gamma^a \Lambda_+ \,, \\
\delta q^a &= -2\, \bar \Lambda_- \gamma^a \chi \,.
\end{split}\end{equation}
Using these generators, one can always set the spinor component of $\vp$ to zero. The stabilizer
of a generic magnetic charge with $\det \vp \ne 0$ contains therefore the stabilizer of the
non-null vector $p^a$ in $Spin(1,9)$, and the elements generated by the spinor generators
satisfying to 
\begin{equation} \Lambda_+ = \frac{ p_a}{p_1} \gamma^a \Lambda_- \ . \label{SpinStab} \end{equation}
For $p^a$ time-like, these generators are compact if $p_1p^a$ is a positive energy vector (\ie $p_1 p^0 > 0$),
and non-compact otherwise. Accordingly, one finds that the stabilizer of $\vp$ is $F_{4(-52)}$, with
\begin{equation} \mathfrak{f}_{4(-52)} \cong \mathfrak{so}(9) \oplus {\bf 16} \ , \label{F4-SO9} \end{equation} 
if all the eigen values of $\vp$ have the same sign, and $F_{4(-20)}$ otherwise (in which case the
spinor generators in \eqref{F4-SO9} are non-compact). If $p^a$ is space-like, the generators
\eqref{SpinStab} decompose into 8 compact plus 8 non-compact generators, such that the stabilizer
subgroup is also $F_{4(-20)}$. This reproduces the results derived in \cite{Ferrara:2006xx}.

If the stabilizer of $\vp$ is compact, one can always use it to rotate the second charge $\vq$ to
a basis in which its spinor component vanishes as well. However, this is not always possible when
the stabilizer of $\vp$ is non-compact. Nevertheless, we will see that it is enough to consider
an example with vanishing spinor component to get all possible stabilizers of generic charges.
Assuming that the spinor component of $\vq$ vanishes, the constraint that a spinor generator
leaves it invariant gives
\begin{equation} ( q_1 p_1 - q_a \gamma^a p_b \gamma^b )  \Lambda_-  = 0 \ . \end{equation}
Consistency requires that $\Lambda_-$ can only be non-zero if 
\begin{equation} q_a q^a \, p_b p^b - 2 q_1 p_1 q_a p^a + q_1^{\; 2} q_1^{\; 2} = 0 \ , \end{equation}
which is not the case for generic charges. It follows that for generic charges (without spinor
components), the stabilizer of $\vp$ and $\vq$ in $E_{6(-26)}$ is identified with the stabilizer
of $p^a$ and $q^a$ in $Spin(1,9)$. Being generic, these vectors are linearly independent. Their common stabilizer is therefore $Spin(8)$, unless they are both space-like and $q_a q^a p_b p^b > ( q_a p^a )^2$, in which case it is $Spin(1,7)$. 

Note that altogether with the four invariants
\begin{equation} \det \vp \ , \quad \tr \vq \vp \ , \quad \tr \vq \times \vq \, \vp \times \vp \ , \quad \det \vq \ , \end{equation}
the angles of the homogeneous spaces $Spin(8) \backslash E_{6(-26)}$ and  $Spin(1,7) \backslash E_{6(-26)}$,
provide the $2\times 27$ parameters of the two charges. The same counting applies for two four-dimensional
charges in the ${\bf 56}$ with stabilizer $Spin(8)$ or $Spin(1,7)$ and their seven $E_{7(-25)}$ invariants
defined in \cite{Andrianopoli:2011gy}. 

Although we did not consider generic charge configurations, for which one cannot remove the spinor components
of both the electric and magnetic charges $\vq$, $\vp$, the stabilizer of the two charges must also be a real
form of the same complex group $D_4$ in this case, since it is always possible to remove the spinor component
by a complex $F_4$ rotation. However, there is no other real form of $D_4$ than $Spin(8)$ and $Spin(1,7)$ that
one can embed in $E_{6(-26)}$, and the result above is therefore general.

A similar analysis shows that the stabilizer of two four-dimensional charges can only be $Spin(8)$
when one charge is BPS, since $Spin(8)$ is the only real form of $D_4$ inside $E_{6(-78)}$. Instead, the
stabilizer of a non-BPS charge of positive quartic invariant is $E_{6(-14)}$ which includes $Spin(2,8)$ and
therefore $Spin(8)$, $Spin(1,7)$ and $Spin(2,6)$. If the two charges are both of this type, the stabilizer
can be any of these three groups, as can be checked explicitly in the $SL(2) \times SO(2,10)$ truncation of
the theory. These stabilizers are discussed in \cite{Andrianopoli:2011gy}.

\end{appendix}

\bibliography{PaperG} \bibliographystyle{JHEP}

\end{document}